\documentclass[11pt,a4paper]{article}
\pdfoutput=1  

\usepackage{jheppub}
\usepackage[T1]{fontenc} 

\usepackage{upgreek}
\usepackage{slashed}
\usepackage{subfigure}

\newcommand{\ud}{\mathrm d}
\allowdisplaybreaks[4]

\graphicspath{{figures/}} 

\title{\boldmath Relativistic effects in the semileptonic $B_c$ decays to charmonium with the Bethe-Salpeter method}

\author[a]{Zi-Kan Geng,}
\author[a]{Tianhong Wang,}
\author[a]{Yue Jiang,}
\author[a]{Geng Li,}
\author[a]{Xiao-Ze Tan}
\author[a,1]{and Guo-Li Wang\note{Corresponding author.}}

\affiliation[a]{Department of Physics, Harbin Institute of Technology,\\Harbin, 150001, China}

\emailAdd{zikangeng@hit.edu.cn}
\emailAdd{thwang@hit.edu.cn}
\emailAdd{jiangure@hit.edu.cn}
\emailAdd{karlisle@hit.edu.cn}
\emailAdd{xz.tan@hit.edu.cn}
\emailAdd{gl\_wang@hit.edu.cn}

\abstract{Relativistic effects are important in the rigorous study of heavy quarks. In this paper, we study the relativistic corrections of semileptonic $B_c$ decays to charmonium with the instantaneous Bethe-Salpeter method. Within the Bethe-Salpeter framework, we use two methods to study the relativistic effects. One of them is to expand the transition amplitude in powers of $\vec{q}$ which is the relative momentum between the quark and antiquark, and the other is to expand the amplitude base on the wave functions. In the level of decay width, the results show that, for the transition of $B_c\to \eta_c$, the relativistic correction is about $22\%$; for $B_c\to J/\psi$, it is about $19\%$; the relativistic effects of $1P$ final states are about $14\sim46\%$ larger than those of $1S$ final states; for $2S$ final states, they are about $19\sim 28\%$ larger than those of $1S$ final states; for $3S$ final states, they are about $12\sim13\%$ larger than those of $2S$ final states; for $2P$ final states, they are about $10\sim14\%$ larger than those of $1P$ final states; for $3P$ final states, they are about $7\sim12\%$ larger than those of $2P$ final states. We conclude that the relativistic corrections of the $B_c$ decays to the orbitally or radially excited charmonium (2S, 3S, 1P, 2P, 3P) are quite large.}

\keywords{relativistic correction, heavy meson}

\begin{document}
\maketitle
\flushbottom



\section{Introduction}

When studying the properties of heavy-light mesons, we always pay attention to the relativistic effect of light quark but ignore that of heavy quark. However, the careful investigation of the relativistic corrections to heavy quark is important, especially the charm quark. For example, recently, the authors of ref.~\cite{Zhu:2017lqu} have calculated the relativistic corrections to the form factors of the $B_c$ decays to S-wave charmonium by non-relativistic QCD (NRQCD) approach, with the heavy quark relative velocities $\vec{v}^{\:2}_{J/\psi}=0.267$ and $\vec{v}^{\:2}_{B_c}=0.186$. They pointed out that the relativistic corrections can bring about additional $15\sim27\%$ contributions. The authors of ref.~\cite{Wang:2017bgv} have calculated the $\mathcal{O}(v^2)$ corrections to twist-2 light-cone distribution amplitudes (LCDAs) of S-wave $B_c$ mesons. They pointed out that the relativistic corrections are sizable, and comparable with the next-to-leading order radiative corrections. Another famous example, the leading order NRQCD predictions \cite{Braaten:2002fi,Liu:2002wq} of the production $e^+ + e^- \to J/\psi + \eta_c$ are at least 5 times smaller than the experimental measurements \cite{Abe:2002rb,Aubert:2005tj}. Later, people found that the relativistic corrections increase the results to 2 times as much as the non-relativistic predictions \cite{Bodwin:2006ke,Bodwin:2007ga}. Therefore, the relativistic effect of heavy quark are important and need to be studied carefully.

The $B_c^+$ meson consists of $c$ quark and $\bar b$ antiquark, and carries two different flavors. It only decays via weak interactions, thus the $B_c$ meson has attracted a lot of attentions both in theories and experiments \cite{Patrignani:2016xqp}. Recently the cross section of the $B_c$ meson is expected to reach the level of $\rm{\upmu b}$ via the proton-nucleus and the nucleus-nucleus collision modes at the Large Hadron Collider \cite{Chen:2018obq}. The LHCb experiment can produce and reconstruct a large number of the $B_c$ meson events, and it provides a solid platform to study the properties of the $B_c$ meson precisely.

A great deal of work has been done on various $B_c$ decays under different approaches, such as NRQCD approach \cite{Zhu:2017lqu,Qiao:2012hp,Shen:2014msa,Zhu:2017lwi}, the perturbative QCD approach (PQCD) \cite{Du:1988ws,Sun:2008ew,Wen-Fei:2013uea,Rui:2014tpa}, the relativistic quark model (RQM) \cite{Nobes:2000pm,Ebert:2003cn,Ivanov:2005fd,Ebert:2010zu}, Light-cone sum rules (LCSR) \cite{Huang:2007kb}, the non-relativistic constituent quark model (NCQM) \cite{Hernandez:2006gt} and QCD sum rules (QCDSR) \cite{Colangelo:1992cx,Kiselev:1993ea,Kiselev:1999sc,Azizi:2009ny}.

In previous study, according to the numerical wave function which is the solution of instantaneous Bethe-Salpeter (BS) equation (also called Salpeter equation) \cite{Salpeter:1951sz,Salpeter:1952ib}, we have qualitatively pointed out that the relativistic correction of a excited state is larger than that of the corresponding ground state. The relativistic correction can not be ignored, so a relativistic model is needed to deal with the problems including excited states \cite{Wang:2007nb}. In this paper, we give quantitatively study of this topic and choose the semileptonic $B_c$ decays to charmonium by using the instantaneous BS method. This method has a comparatively solid foundation because both the BS equation and the Mandelstam formula \cite{Mandelstam:1955sd} are established on relativistic quantum field theory. We have solved the full Salpeter equations for different $J^{P(C)}$ states \cite{Cvetic:2004qg,Chang:2004im,Wang:2007av}, and deduced the transition amplitude formula by performing the instantaneous approach to the Mandelstam formula. In these processes, the corresponding quark and antiquark are charm and bottom quarks which both are heavy. The instantaneous approximation is reasonable, and we can provide a relatively rigorous relativistic calculation.

The paper is organized as follows. In section~\ref{sec:formula}, we give the useful formulas for the $B_c$ decays to charmonium. In section~\ref{sec:wavefunction}, we give the relativistic wave function in the the instantaneous BS method. In section~\ref{sec:method1}, we give a method to separate the relativistic corrections. In section~\ref{sec:method2}, we give another method to calculate the relativistic corrections. In section~\ref{sec:result}, we give the numerical results and discussions. We summarize and conclude in section~\ref{sec:conclude}, and put the wave functions and Salpeter equation in the appendix~\ref{appendix}.

\section{\label{sec:formula} Form factors and semileptonic decay width }

\begin{figure}[tbp]
\centering
\includegraphics[width=0.45\textwidth]{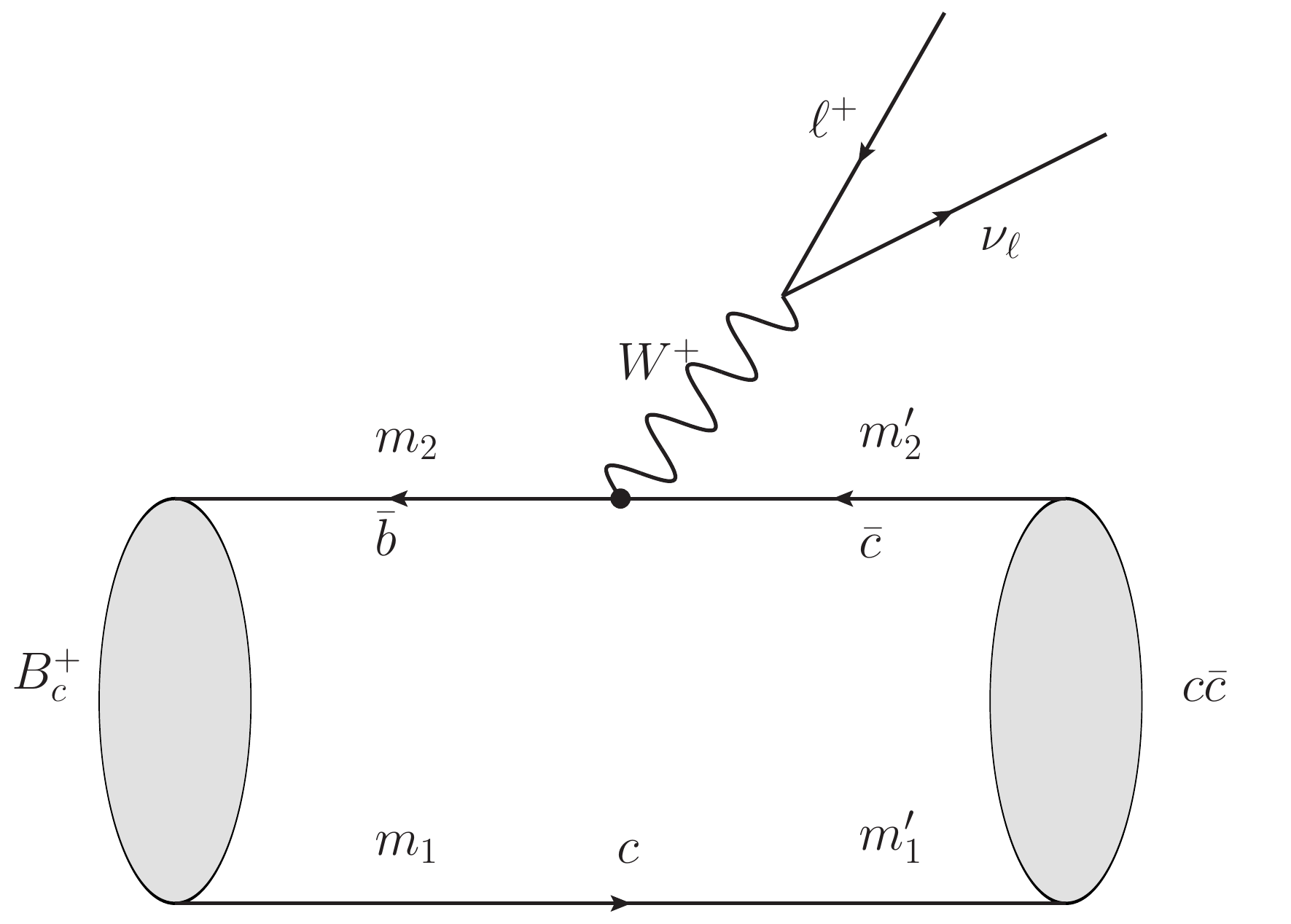}
\caption{Feynman diagram corresponding to the semileptonic decays $B_c^+\to (c\bar c)\ell^+\nu_\ell$.}\label{fig:feymanw}
\end{figure}

For the $B_c^+\to (c\bar c)\ell^+\nu_\ell$ processes shown in figure~\ref{fig:feymanw}, the transition amplitude element  reads
\begin{equation}
\begin{aligned}
T=\frac{G_F}{\sqrt 2}V_{cb}\bar{u}_{\nu_\ell}\gamma^\mu(1-\gamma_5)v_\ell\left\langle (c\bar c)(P_f)|J_\mu|B_c^+(P)\right\rangle,
\end{aligned}
\end{equation}
where $(c\bar c)$ denotes charmonium; $V_{cb}$ is the Cabibbo-Kobayashi-Maskawa (CKM) matrix element; $J_\mu\equiv V_\mu-A_\mu$ is the charged current responsible for the decays; $P$ and $P_f$ are the momenta of the initial $B_c^+$ and the final charmonium, respectively.

Taking $\eta_c$ meson as an example, the hadronic transition element can be written as the overlapping integral over the initial and final relativistic BS wave functions within Mandelstam formalism. We would not solve the full BS equation, but the instantaneous one, namely, the full Salpeter equation. We perform the instantaneous approximation to the transition element \cite{Chang:2006tc} and write it as
\begin{equation}
\begin{aligned}
\langle\eta_c| \bar b\gamma^\mu(1-\gamma^5)c|B_c^+\rangle = \int\frac{\ud\vec q}{(2\pi)^3} {\rm Tr}\bigg[\overline\varphi_{P_f}^{++}(\vec q\:')\frac{\slashed P}{M}\varphi_P^{++}(\vec q\:)\gamma^\mu(1-\gamma^5)\bigg],\label{eq:mandelstam}
\end{aligned}
\end{equation}
where $\varphi_P^{++}$ denotes the positive energy component of the instantaneous BS wave function of the initial state; $\overline\varphi_{P_f}^{++}\equiv\gamma^0\varphi_{P_f}^{++}\gamma^0$ is the Dirac conjugate of the positive energy component of the final state; $m'_1$ and $m'_2$ are the masses of quark and antiquark in the final state, respectively, and $\vec q\:'=\vec q-\frac{m'_1}{m'_1+m'_2} \vec P_f$ is the relative momentum between them. In this paper, we keep only the positive energy component $\varphi^{++}$ of the relativistic wave functions, because the contributions from other components are much smaller than $1\%$ in transition of $B_c \to (c\bar c)$ \cite{Wang:2016enc}.

For $B_c^+\to P\ell^+\nu_\ell$ (here $P$ denotes $\eta_c$ or $\chi_{c0}$), the hadronic matrix element can be written as
\begin{equation}
\begin{aligned}
&\left\langle P|\bar b\gamma^\mu(1-\gamma^5)c|B_c^+\right\rangle\\
=&\int\frac{\ud \vec q}{(2\pi)^3}\mathrm{Tr}\left[\overline\varphi_{P_f}^{++}(\vec q\:')\frac{\slashed P}{M}\varphi_P^{++}(\vec q\:)\gamma^\mu(1-\gamma^5)\right]\\
=&S_+(P+P_f)^\mu+S_-(P-P_f)^\mu\label{eq:0-ff},
\end{aligned}
\end{equation}
where $S_{+}$ and $S_{-}$ are the form factors.

For $B_c^+\to V\ell^+\nu_\ell$ (here $V$ denotes $J/\psi$, $h_c$ or $\chi_{c1}$), the hadronic matrix element can be written as
\begin{equation}
\begin{aligned}
&\langle V|\bar b\gamma^\mu(1-\gamma^5)c|B_c^+\rangle\\
=&\int\frac{\ud \vec q}{(2\pi)^3}\mathrm{Tr}\left[\overline\varphi_{P_f}^{++}(\vec q\:')\frac{\slashed P}{M}\varphi_P^{++}(\vec q\:)\gamma^\mu(1-\gamma^5)\right]\\
=&(t_1 P^\mu+ t_2 P_f^\mu)\frac{\epsilon\cdot P}{M}+t_3(M+M_f)\epsilon^{\mu}+\frac{2t_4}{M+M_f}\mathrm i\varepsilon^{\mu\nu\sigma\delta}\epsilon_{\nu} P_\sigma P_{f\delta},
\end{aligned}
\end{equation}
where $\epsilon_\nu$ is the polarization vector of the final vector meson; $t_1$, $t_2$, $t_3$ and $t_4$ are the form factors.

The summation formulas for polarization of the final vector meson used in this paper are
\begin{equation}
\begin{aligned}
\epsilon_\mu^{(\lambda)}(P_f)P_f^\mu&=0,\\
\sum_{\lambda}\epsilon_\mu^{(\lambda)}(P_f)\epsilon_\nu^{\dagger(\lambda)}(P_f)&=-g_{\mu\nu}+\frac{P_{f\mu}P_{f\nu}}{M_f^2}.
\end{aligned}
\end{equation}

Finally, the semileptonic decay width can be expressed as
\begin{equation}
\Gamma=\frac{1}{8M(2\pi)^3}\int\frac{|\vec P_\ell|}{E_\ell}\ud |\vec P_\ell|\int\frac{|\vec P_f|}{E_f}\ud |\vec P_f|\sum_{\lambda}|T|^2\label{eq:width},
\end{equation}
where $\vec P_\ell$ is the three-dimensional momentum of the final lepton, and $\vec P_f$ is the three-dimensional momentum of the final meson.

\section{\label{sec:wavefunction} Relativistic wave function}

Usually, the non-relativistic wave function for a pseudoscalar is written as \cite{Chang:1992pt}
\begin{equation}
\Psi_P(\vec{q}\:)=(\slashed P+M)\gamma_5 f(\vec{q}\:)\label{eq:schrodinger},
\end{equation}
where $M$ and $P$ are the mass and momentum of the meson, respectively; $\vec{q}$ is the relative momentum between the quark and antiquark in the meson, and the radial wave function $f(\vec{q})$ can be obtained numerically by solving the Schrodinger equation.

But in our method, we solve the full Salpeter equation. The form of wave function is relativistic and depends on the $J^{P(C)}$ quantum number of the corresponding meson. For a pseudoscalar, the relativistic wave function can be written as the four items constructed by $P$, $q_{\perp}$ and $\gamma$-matrices \cite{Kim:2003ny}
\begin{equation}
\begin{aligned}
\varphi_{0^-}(q_\perp)&=M\left[\frac{\slashed{P}}{M}f_1(q_\perp)
+f_2(q_\perp)+\frac{\slashed{q}_\perp}{M}f_3(q_\perp)+\frac{\slashed{P}\slashed{q}_\perp}{M^2}f_4(q_\perp)\right]\gamma_5,
\end{aligned}\label{eq:BS wave function}
\end{equation}
where $q=p_1-\alpha_1 P=\alpha_2 P-p_2$ is the relative momentum between quark (with momentum $p_1$ and mass $m_1$) and antiquark (momentum $p_2$ and mass $m_2$), $\alpha_1=\frac{m_1}{m_1+m_2}$, $\alpha_2=\frac{m_2}{m_1+m_2}$;  $q_{\perp}=q-\frac{P\cdot q}{M^2}P$, in the rest frame of the meson, $q_{\perp}=(0,\vec{q})$.

All the items in the wave function eq. (\ref{eq:BS wave function}) have the quantum number of $0^-$. Because of instantaneous approximation, the items with $P\cdot q_{\perp}$ disappear, so this wave function is a general relativistic form for a pseudoscalar with the instantaneous approximation. If we set the items with $f_3$ and $f_4$ to zero, and set $f_1=f_2$, the relativistic wave function is reduced to the  Schrodinger wave function eq.~(\ref{eq:schrodinger}).

Taking into account the last two equations in eq. (\ref{eq:phi}), we obtain the relations
\begin{equation}
\begin{aligned}
f_3(q_\perp)&=\frac{M(\omega_2-\omega_1)}{(m_1\omega_2+m_2\omega_1)}f_1, \\
f_4(q_\perp)&=-\frac{M(\omega_1+\omega_2)}{(m_1\omega_2+m_2\omega_1)}f_2,
\end{aligned}
\end{equation}
where the quark energy $\omega_i=\sqrt{m^2_i-q_{\perp}^2}=\sqrt{m^2_i+\vec{q}^{\:2}}$ ($i=1,2$).
The wave function corresponding to the positive energy projection has the form
\begin{equation}
\begin{aligned}
\varphi_{0^-}^{++}(q_{\perp})=\left[A_1(q_{\perp})+\frac{\slashed P}{M}A_2(q_{\perp})+\frac{\slashed q_{\perp}}{M}A_3(q_{\perp})+\frac{\slashed P\slashed q_{\perp}}{M^2}A_4(q_{\perp})\right]\gamma^5\label{eq:++wave},
\end{aligned}
\end{equation}
where
\begin{equation}
\begin{aligned}
A_1&=\frac{M}{2}\left[\frac{\omega_1+\omega_2}{m_1+m_2}f_1+f_2\right],\qquad A_3=-\frac{M(\omega_1-\omega_2)}{m_1\omega_2+m_2\omega_1}A_1,\\
A_2&=\frac{M}{2}\left[f_1+\frac{m_1+m_2}{\omega_1+\omega_2}f_2\right],\qquad A_4=-\frac{M(m_1+m_2)}{m_1\omega_2+m_2\omega_1}A_1\label{eq:A}.
\end{aligned}
\end{equation}

The normalization condition reads
\begin{equation}
\int\frac{\ud\vec q}{(2\pi)^3}4f_1f_2M^2\left\{\frac{m_1+m_2}{\omega_1+\omega_2}+\frac{\omega_1+\omega_2}{m_1+m_2}+\frac{2\vec q^{\:2}(m_1\omega_1+m_2\omega_2)}{(m_2\omega_1+m_1\omega_2)^2}\right\}=2M.\label{eq:norma}
\end{equation}

By solving the full Salpeter equation, the numerical values of wave functions $f_1$, $f_2$, $f_3$ and $f_4$ are obtained. The positive energy component eq.~(\ref{eq:++wave}) is brought into the Mandelstam formula eq.~(\ref{eq:mandelstam}). After the trace and integral are finished, the form factors $S_+$ and $S_-$ can be calculated numerically. Finally, the decay width of the semileptonic decay $B_c^+\to \eta_c \ell^+\nu_\ell$ can be obtained within the relativistic BS method. In this paper, besides the wave function for $0^-$ state, we also need the wave functions for the states of $1^{--}$ ($J/\psi$), $1^{+-}$ ($h_c$), $0^{++}$ ($\chi_{c0}$), $1^{++}$ ($\chi_{c1}$), etc., and we give them in the appendix~\ref{appendix}.

\section{\label{sec:method1} Method I of separating the relativistic corrections}

In this part, how to obtain the relativistic corrections is shown. The transition element is obtained by overlapping integral over the Schrodinger wave functions of the initial and final states in a non-relativistic model. The main difference between the relativistic and non-relativistic models comes from the wave functions. If we set the items  with $\vec{q}$ (or $q_{\perp}$) in eq.~(\ref{eq:++wave}) to zero and let $f_1=f_2$, the relativistic wave function is reduced to the non-relativistic one eq.~(\ref{eq:schrodinger}).

To see the relativistic corrections at each expansive order, we expand the amplitude in powers of $\vec{q}$. There are some reasons: (i) the quantity $\vec{q}$, which represents the relative momentum between the quark and the antiquark, is a kind of measure of relativistic effect of a meson; (ii) when $|\vec{q}\:|$ is small, the ratio $|\vec{q}\:|/M$ or $|\vec{q}\:|/m_i$ ($i=1,2$) is small and can be dealt as the power to expand the amplitude; when $|\vec{q}\:|$ is large, its contribution will be suppressed by the wave function $f_i(\vec{q}\:)$, especially for the ground state ($\eta_c$ and $J/\psi$); (iii) it has been investigated in NRQCD effective theory that the decay rates can be ordered in powers of the quark relative velocity $v$ \cite{Bodwin:1994jh}. The relative momentum $\vec{q}$ is related to relative velocity $\vec{v}$, and also can be used to perform the Taylor expansion of the amplitude, see refs. \cite{Zhu:2017lqu,Zhu:2017lwi}.

In the transition amplitude, the wave function of the final state is dependent on $\vec{q}\:'$. We use the relation $\vec q\:'=\vec q-\frac{m'_1}{m'_1+m'_2} \vec P_f$ during numerical calculation. But $\vec{q}\:'$ is treated as an independent variable to maintain covariance when we expand the amplitude. In other words, we perform the Taylor expansion of the amplitude eq. (\ref{eq:mandelstam}) (before doing the integrate over $\vec {q}$\:) in powers of relative momentum $\vec{q}$ and  $\vec{q}\:'$, where $\vec{q}$ and  $\vec{q}\:'$ are the relative momenta between quark and antiquark in the initial meson and final meson, respectively. The relativistic corrections to form factors can be given at each expansive order. According to the expansion in the transition amplitude, we can obtain the expansion in the level of decay width straightforwardly. We expand the amplitude (or form factors) to the third order of $\vec q$, but expand the decay width to the sixth order of $\vec q$. This difference results from the cross items in $|T|^2$, similarly as eq.~(\ref{eq:tt}).

In some non-relativistic methods, both the wave functions and the amplitude are non-relativistic. If we set the items with $\vec{q}$ (or $q_{\perp}$) in eq.~(\ref{eq:++wave}) to zero and set $f_1=f_2$ as mentioned above, and modify the normalization condition eq.~(\ref{eq:norma}) as
\begin{equation}
\int\frac{\ud\vec q}{(2\pi)^3}4f_1f_2M^2\times 2=2M,\label{eq:normaNR}
\end{equation}
the non-relativistic wave function can be obtained. Taking them into the leading order expansion of the amplitude, we can estimate the decay widths obtained by the non-relativistic (NR) methods. 

\section{\label{sec:method2} Method II of calculating the relativistic corrections}
Though the behavior of wave function $f_i(\vec{q}\:)$ will suppress the contribution of large $|\vec{q}|$, there is still problem of rapidity of convergence. This problem will be shown by numerical results later. We would like to provide another method to give the relativistic corrections.

The first two items in eq.~(\ref{eq:BS wave function}) are close to the non-relativistic wave function eq.~(\ref{eq:schrodinger}) because of $f_1\simeq f_2$ numerically. Base on that, we can treat them as the leading order, and the last two items as the relativistic corrections. Similarly, the positive energy wave function eq.~(\ref{eq:++wave}) can be decomposed into two components,
\begin{equation}
\begin{aligned}
\varphi_{0^-}^{++}(q_{\perp})=\varphi_{0}^{++}(q_{\perp})+\varphi_{1}^{++}(q_{\perp}),
\end{aligned}
\end{equation}
where $\varphi_{0}^{++}(q_{\perp})=\left[A_1(q_{\perp})+\frac{\slashed P}{M}A_2(q_{\perp})\right]\gamma^5$ is treated as the non-relativistic (NR) wave function, and $\varphi_{1}^{++}(q_{\perp})=\left[\frac{\slashed q_{\perp}}{M}A_3(q_{\perp})+\frac{\slashed P\slashed q_{\perp}}{M^2}A_4(q_{\perp})\right]\gamma^5$ is treated as the the relativistic corrections (RC) of the wave function.

If we use the approximate formula $\omega_i=m_i+\vec{q}^{\:2}/{2m_i}$ ($i=1,~2$) (which is valid in small $|\vec {q}|$, but the contribution from large $|\vec {q}|$ will be suppressed by the wave functions $f_i$), and set $f_1=f_2$, then  $\varphi_{0}^{++}(q_{\perp})=Mf_1
\left[\left(1+\frac{\vec{q}^{\:2}}{4m_1m_2}\right)+\frac{\slashed P}{M}\left(1-\frac{\vec{q}^{\:2}}{4m_1m_2}\right)\right]\gamma^5$. The difference between $\varphi_{0}^{++}$ and the NR wave function eq.~(\ref{eq:schrodinger}) is left to the second order of $\vec q$. Therefore $\varphi_{0}^{++}$ is approximated equivalent to the NR wave function. For other states,  we can reach the same conclusions.

The hadronic transition element can be decomposed into three components,
\begin{equation}
\begin{aligned}
&\quad\langle\eta_c|\bar b\gamma^\mu(1-\gamma^5)c|B_c^+\rangle \\
&= \int\frac{\ud\vec q}{(2\pi)^3} {\rm Tr}\bigg[\overline\varphi_{P_f}^{++}(\vec q')\frac{\slashed P}{M}\varphi_P^{++}(\vec q)\gamma^\mu(1-\gamma^5)\bigg]\\
&=\int\frac{\ud\vec q}{(2\pi)^3} {\rm Tr}\bigg[(\overline\varphi_{0}^{'++}+\overline\varphi_{1}^{'++})\frac{\slashed P}{M}(\varphi_0^{++}+\varphi_1^{++})\gamma^\mu(1-\gamma^5)\bigg]\\
&=\int\frac{\ud\vec q}{(2\pi)^3} {\rm Tr}\bigg[(\overline\varphi_{0}^{'++})\frac{\slashed P}{M}(\varphi_0^{++})\gamma^\mu(1-\gamma^5)\bigg]+\Leftrightarrow the\ leading\ order\ (LO)\\
&\quad+\int\frac{\ud\vec q}{(2\pi)^3} {\rm Tr}\bigg[(\overline\varphi_{1}^{'++})\frac{\slashed P}{M}(\varphi_0^{++})\gamma^\mu(1-\gamma^5)+(\overline\varphi_{0}^{'++})\frac{\slashed P}{M}(\varphi_1^{++})\gamma^\mu(1-\gamma^5)\bigg]+\\
&\qquad\qquad\qquad\qquad\qquad\Leftrightarrow the\ first\ order\ of\ relativistic\ correction\ (1stRC)\\
&\quad+\int\frac{\ud\vec q}{(2\pi)^3} {\rm Tr}\bigg[(\overline\varphi_{1}^{'++})\frac{\slashed P}{M}(\varphi_1^{++})\gamma^\mu(1-\gamma^5)\bigg]\\
&\qquad\qquad\qquad\qquad\qquad\Leftrightarrow the\ second\ order\ of\ relativistic\ correction\ (2ndRC).
\end{aligned}
\end{equation}

In the transition amplitude $T$, the leptonic component $\bar{u}_{\nu_\ell}\gamma^\mu(1-\gamma_5)v_\ell$ is independent of the relative momentum $\vec{q}$. The transition amplitude $T$ can also be decomposed into three components,
\begin{equation}
T=T_0+T_1+T_2,
\end{equation}
where $T_0$ denotes the leading order (LO); $T_1$ denotes the first order relativistic correction (1stRC), and $T_2$ denotes the second order relativistic correction (2ndRC). However, the decay width is related to the square of transition amplitude module, as shown in eq.~(\ref{eq:width}). The square of transition amplitude module can be decomposed into five components,
\begin{equation}
\begin{aligned}
|T|^2 &=(T_0+T_1+T_2)(T_0^*+T_1^*+T_2^*)\\
&=|T_0|^2+\Leftrightarrow LO\\
&\quad+T_0T_1^*+T_0^*T_1+\Leftrightarrow 1stRC\\
&\quad+|T_1|^2+(T_0T_2^*+T_0^*T_2)+\Leftrightarrow 2ndRC\\
&\quad+T_1T_2^*+T_1^*T_2+\Leftrightarrow 3rdRC\\
&\quad+|T_2|^2\Leftrightarrow 4thRC.
\end{aligned}\label{eq:tt}
\end{equation}
Taking an component into the phase-space integral eq.~(\ref{eq:width}), we can calculate the corresponding order of the decay width. In summary, we separate the positive energy wave function of the initial and final meson into two components, the NR wave function and the RC one. Then we compute each expansive orders of transition amplitude (or form factors) and finally we use them to obtain each expansive orders of decay width.

\section{\label{sec:result} Results and discussions}

The parameters used in this paper are: $\Gamma_{B_c}=1.298\times 10^{-12}~\mathrm{GeV}$, $G_F=1.166\times 10^{-5}~\mathrm{GeV^{-2}}$, $m_b=4.96~\mathrm{GeV}$, $m_c=1.62~\mathrm{GeV}$, $V_{cb}=40.5\times 10^{-3}$,  $M_{h_c(2P)}=3.887~\mathrm{GeV}$, $M_{\chi_{c0}(2P)}=3.862~\mathrm{GeV}$, $M_{\chi_{c1}(2P)}=3.872~\mathrm{GeV}$, $ M_{\eta_c(3S)}=3.949~\mathrm{GeV}$, $M_{\psi(3S)}=4.039~\mathrm{GeV}$, $M_{h_c(3P)}=4.242~\mathrm{GeV}$, $M_{\chi_{c0}(3P)}=4.140~\mathrm{GeV}$, $M_{\chi_{c1}(3P)}=4.229~\mathrm{GeV}$.

\begin{figure}[tbp]
\centering
\subfigure[$B_c$]{\label{fig:Bc}
			      \includegraphics[width=0.4\textwidth]{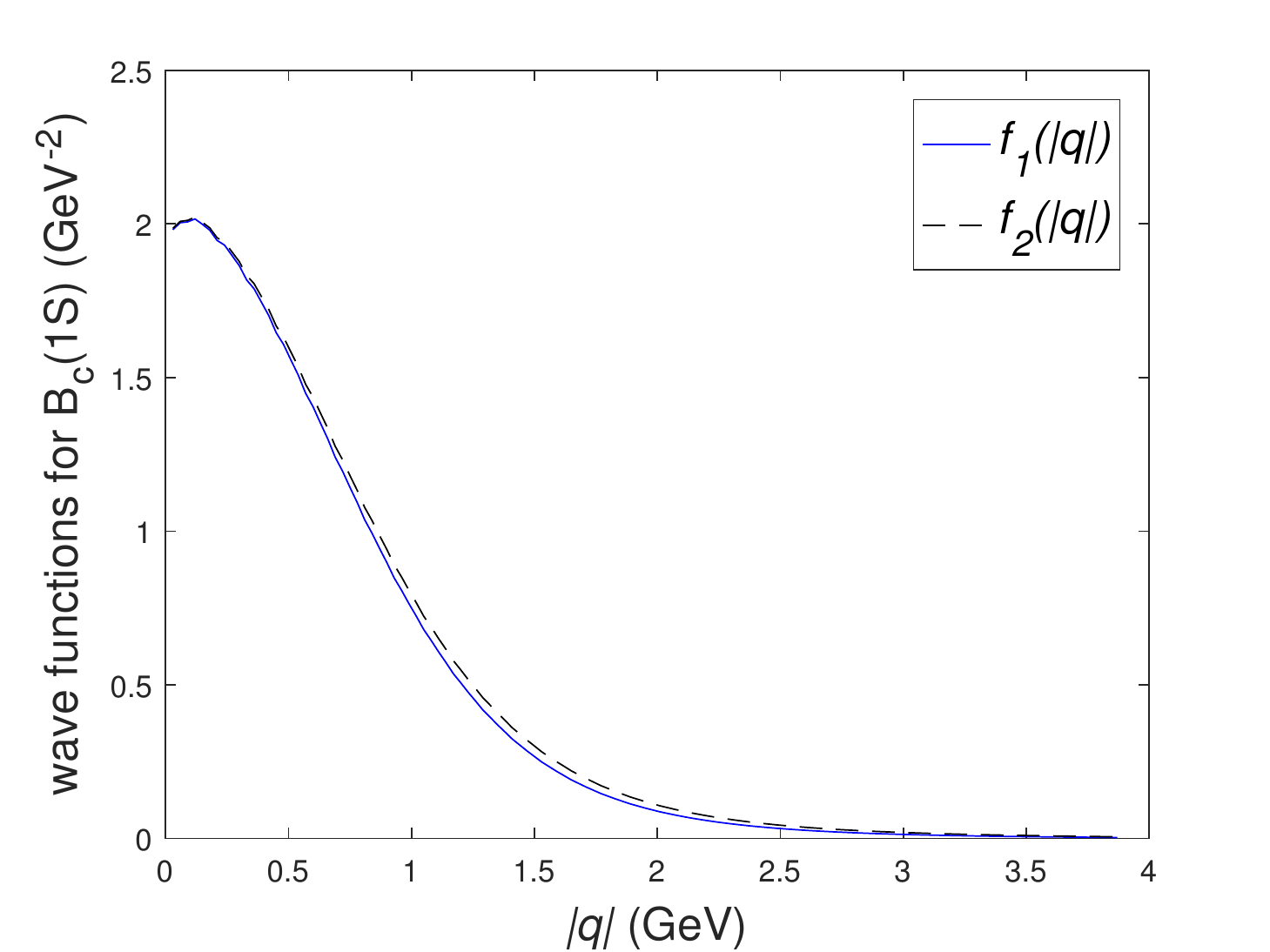}}
\subfigure[$\eta_c$]{\label{fig:eta_c}
			      \includegraphics[width=0.4\textwidth]{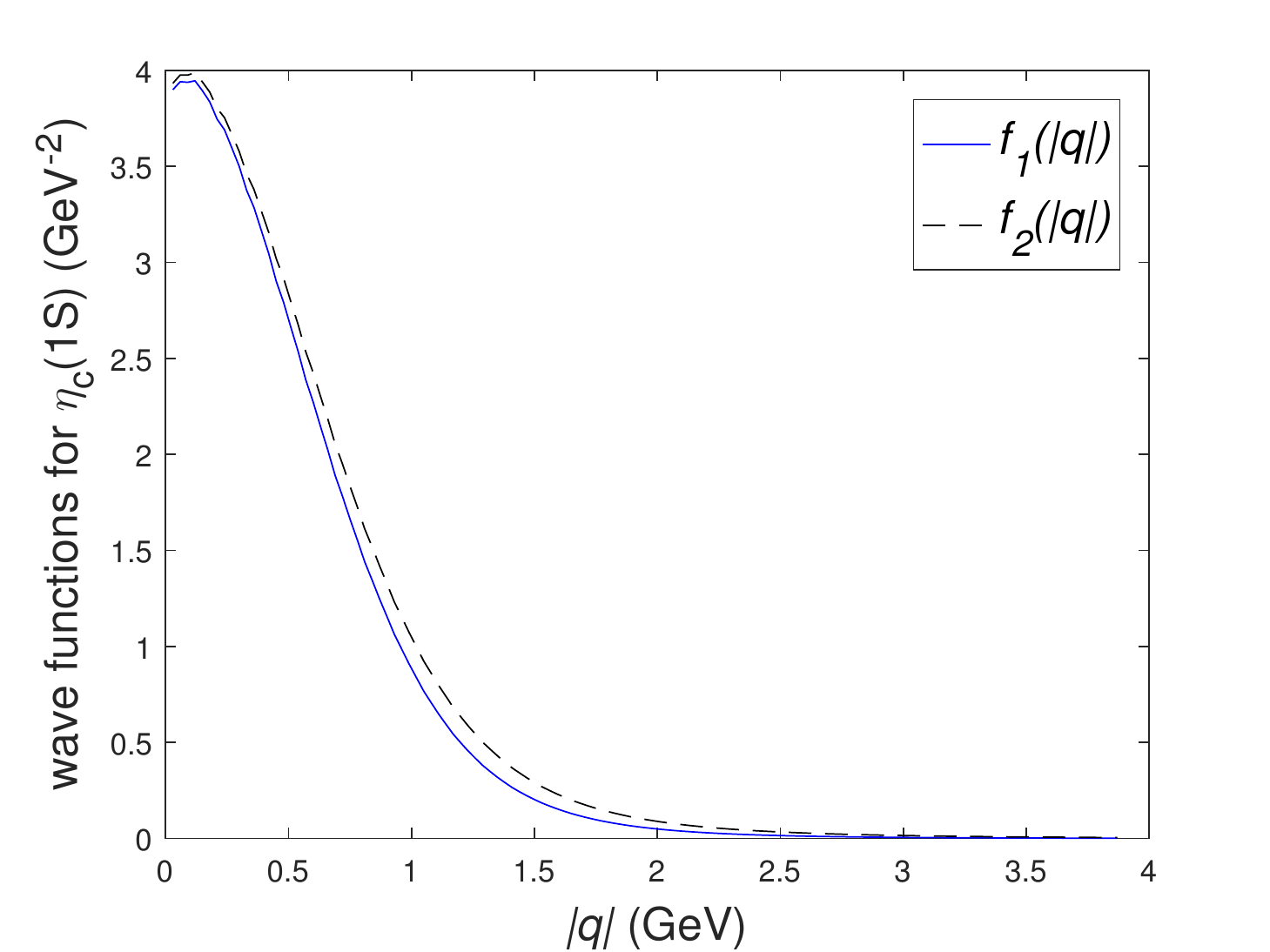}}
\subfigure[$J/\psi$]{\label{fig:J-psi}
			      \includegraphics[width=0.4\textwidth]{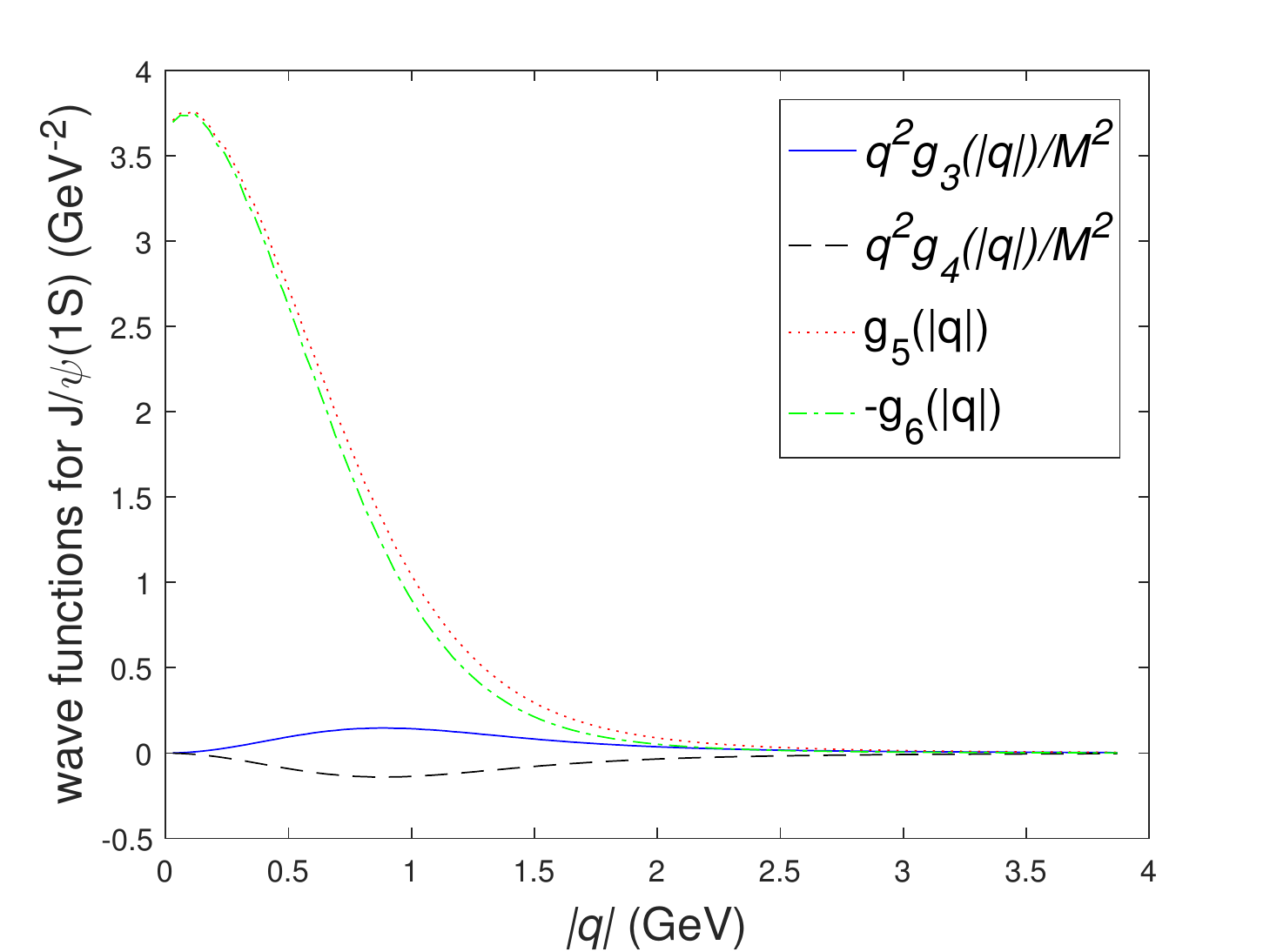}}
\subfigure[$h_c$]{\label{fig:hc}
			      \includegraphics[width=0.4\textwidth]{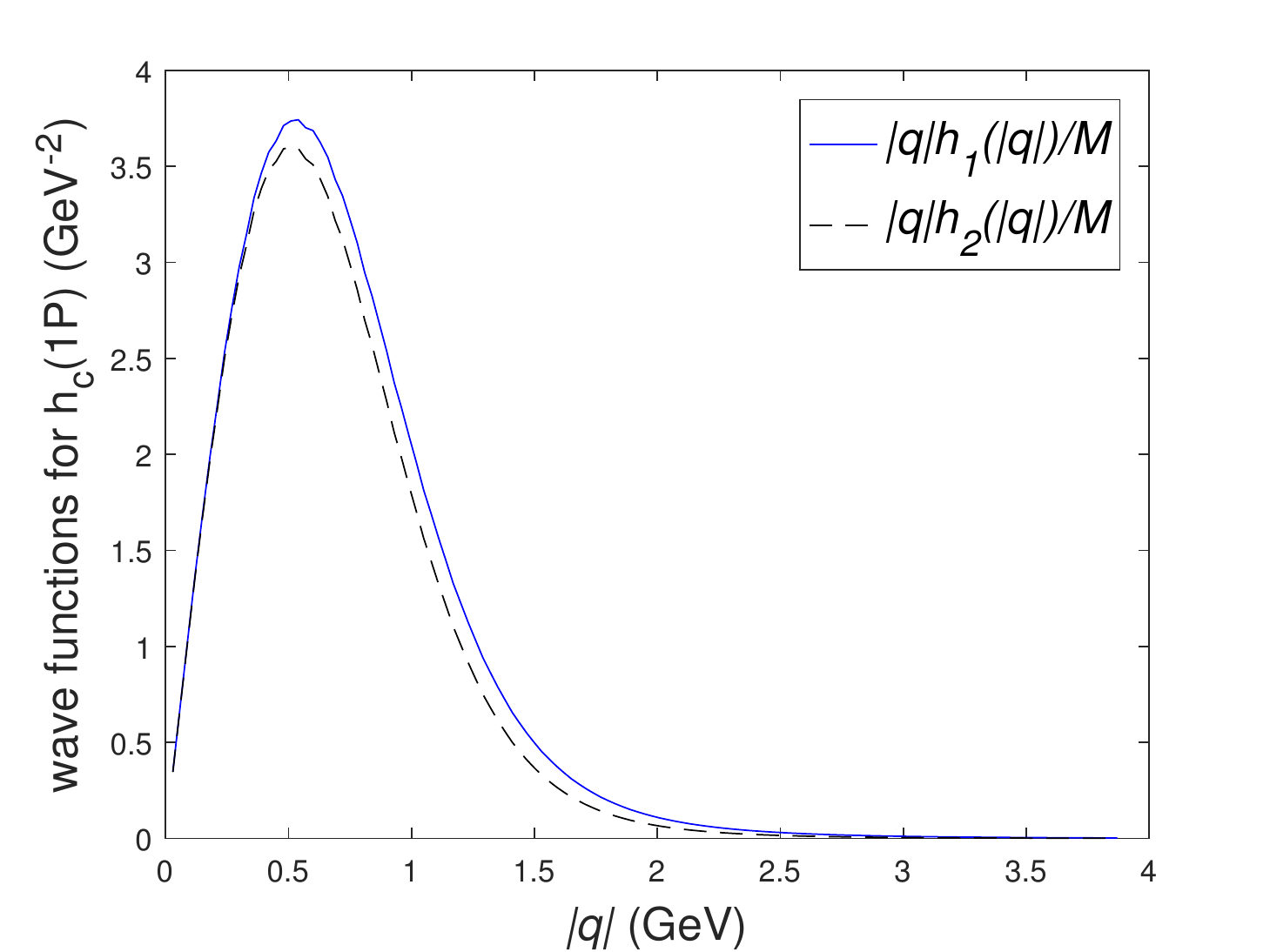}}
\subfigure[$\chi_{c0}$]{\label{fig:Xc0}
			      \includegraphics[width=0.4\textwidth]{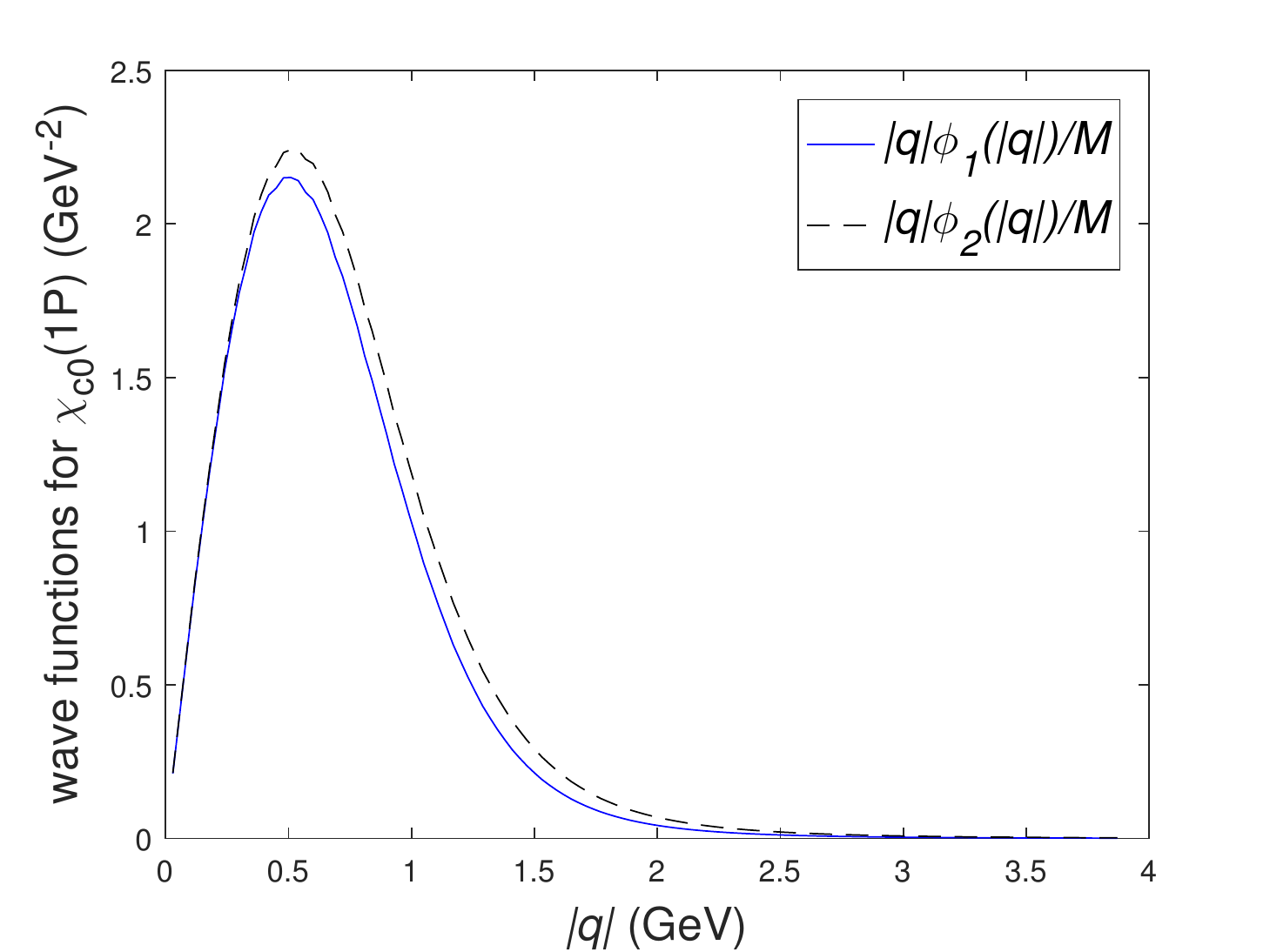}}
\subfigure[$\chi_{c1}$]{\label{fig:Xc1}
			      \includegraphics[width=0.4\textwidth]{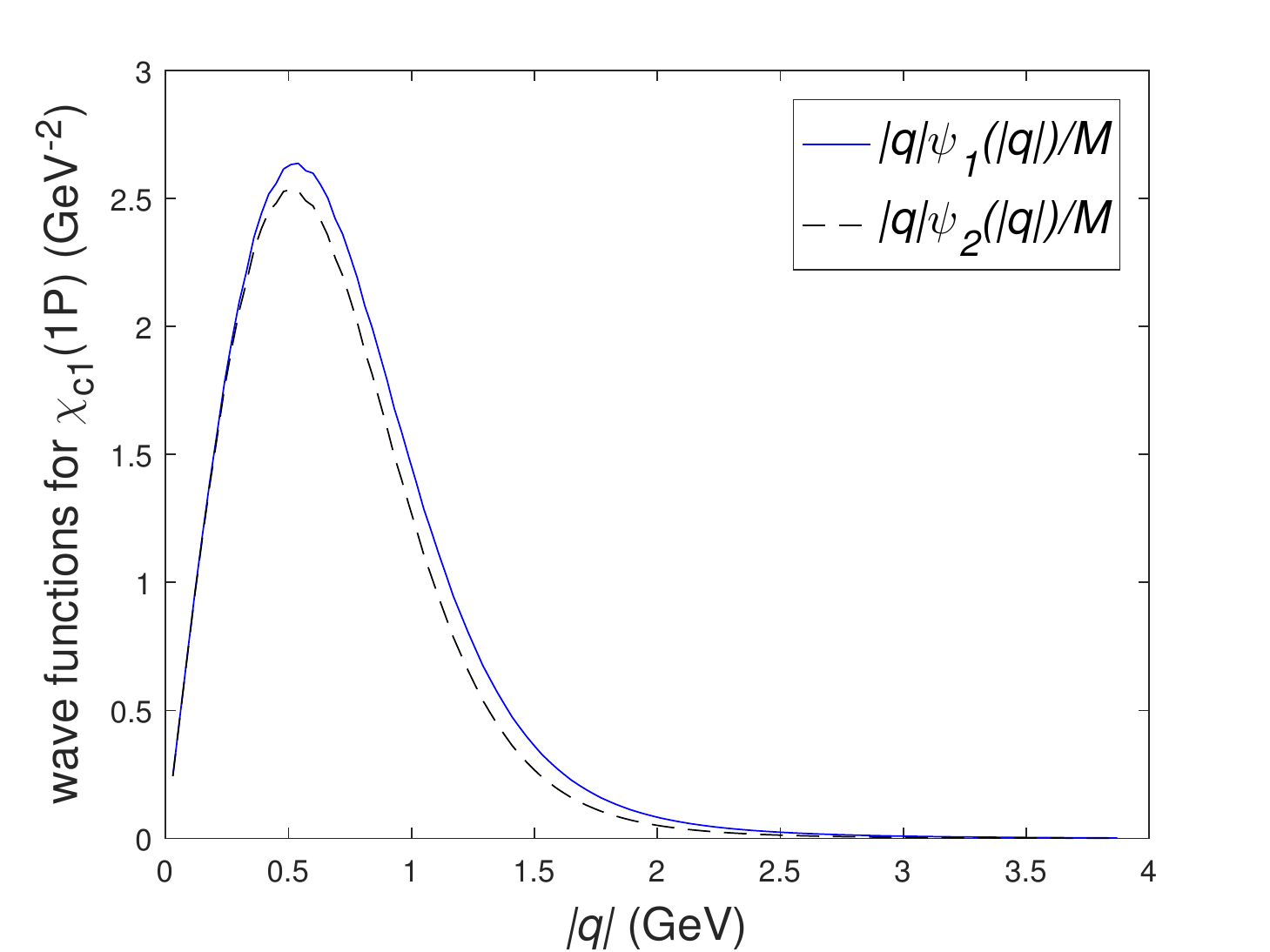}}
\caption{The wave functions for the ground state mesons.}\label{fig:wfs}
\end{figure}

\begin{figure}[tbp]
\centering
\subfigure[$\eta_c(2S)$]{\label{fig:eta_c2S}
			      \includegraphics[width=0.4\textwidth]{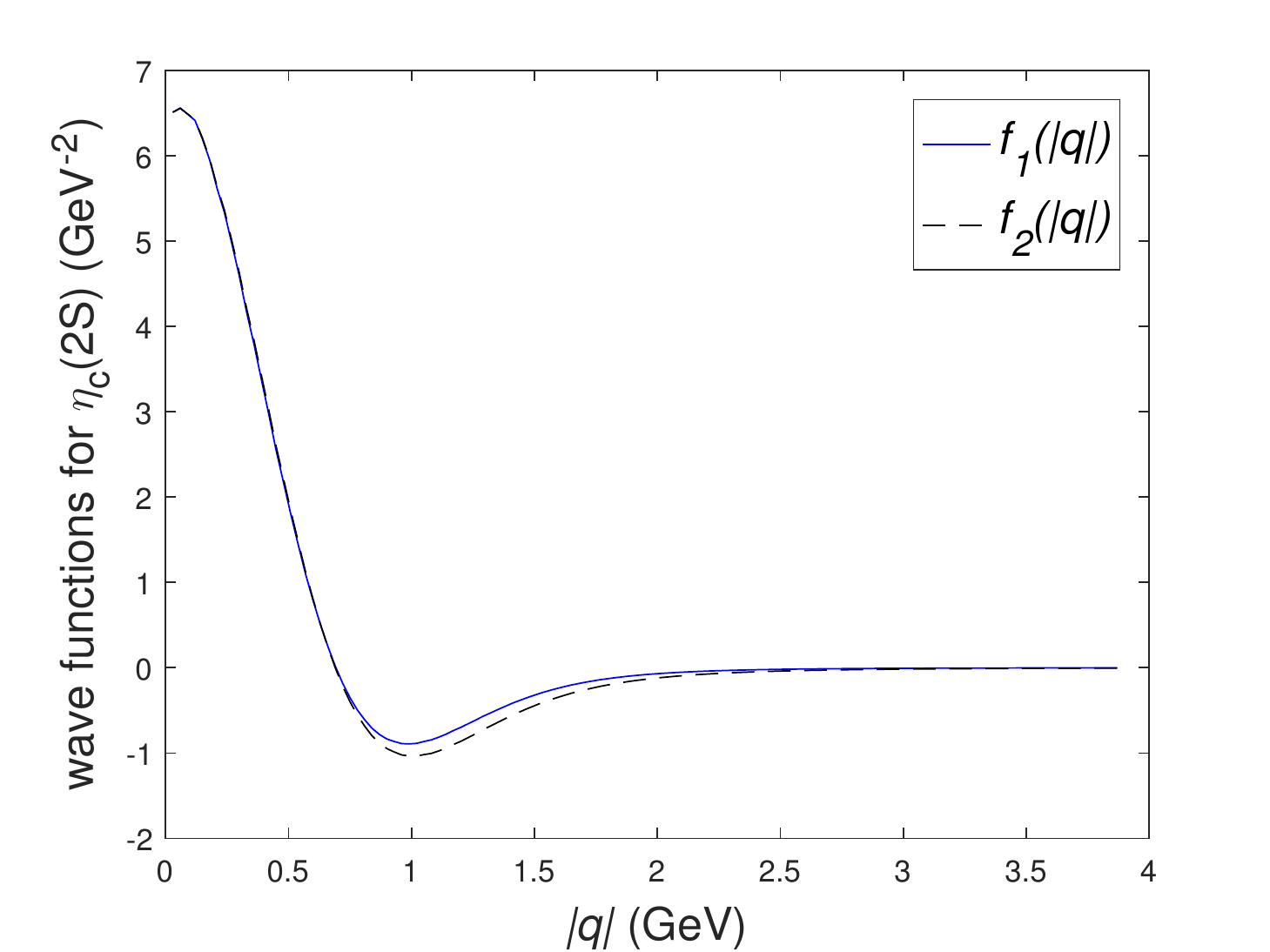}}
\subfigure[$\psi(2S)$]{\label{fig:J-psi}
			      \includegraphics[width=0.4\textwidth]{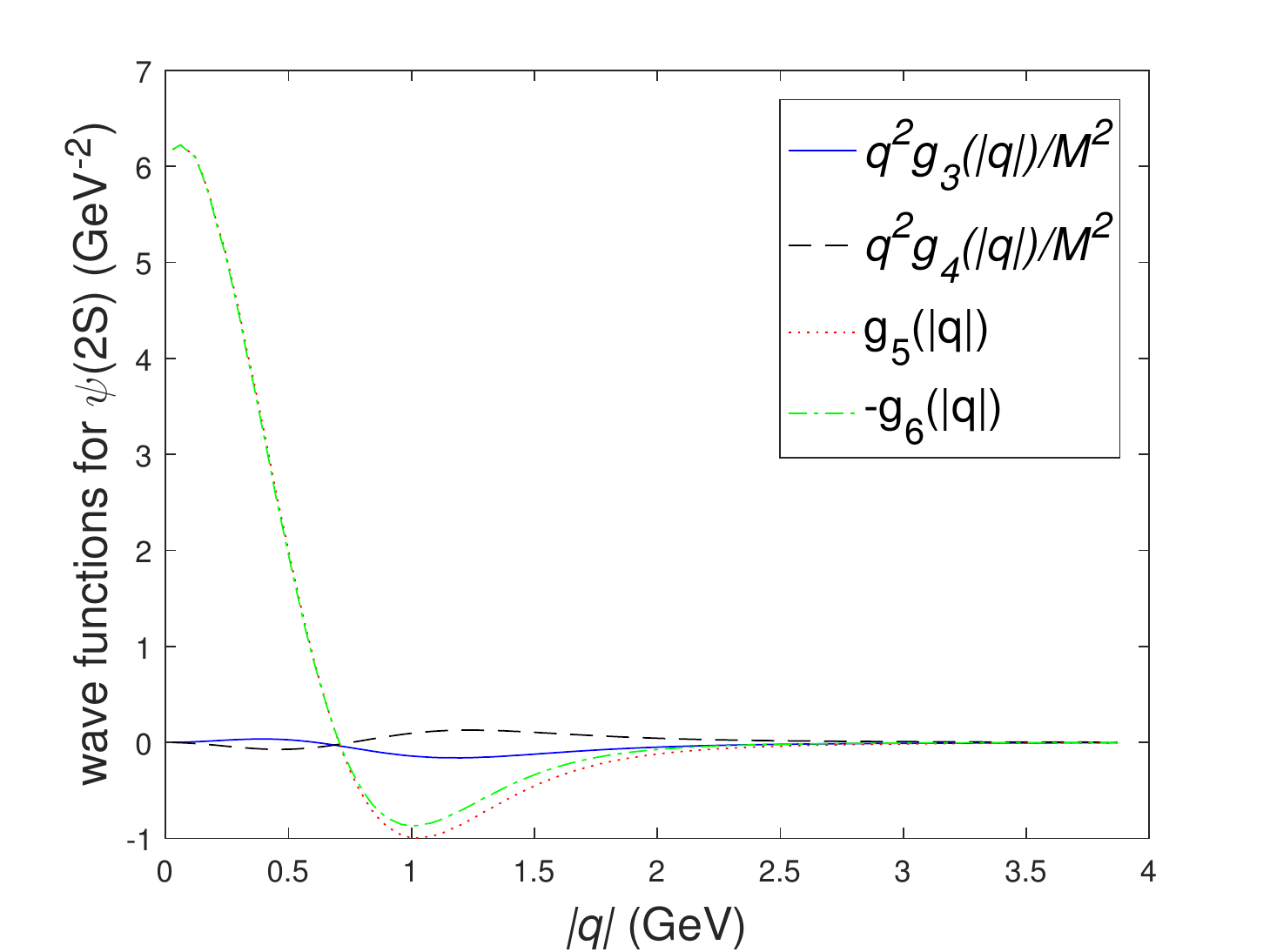}}
\subfigure[$h_c(2P)$]{\label{fig:hc}
			      \includegraphics[width=0.4\textwidth]{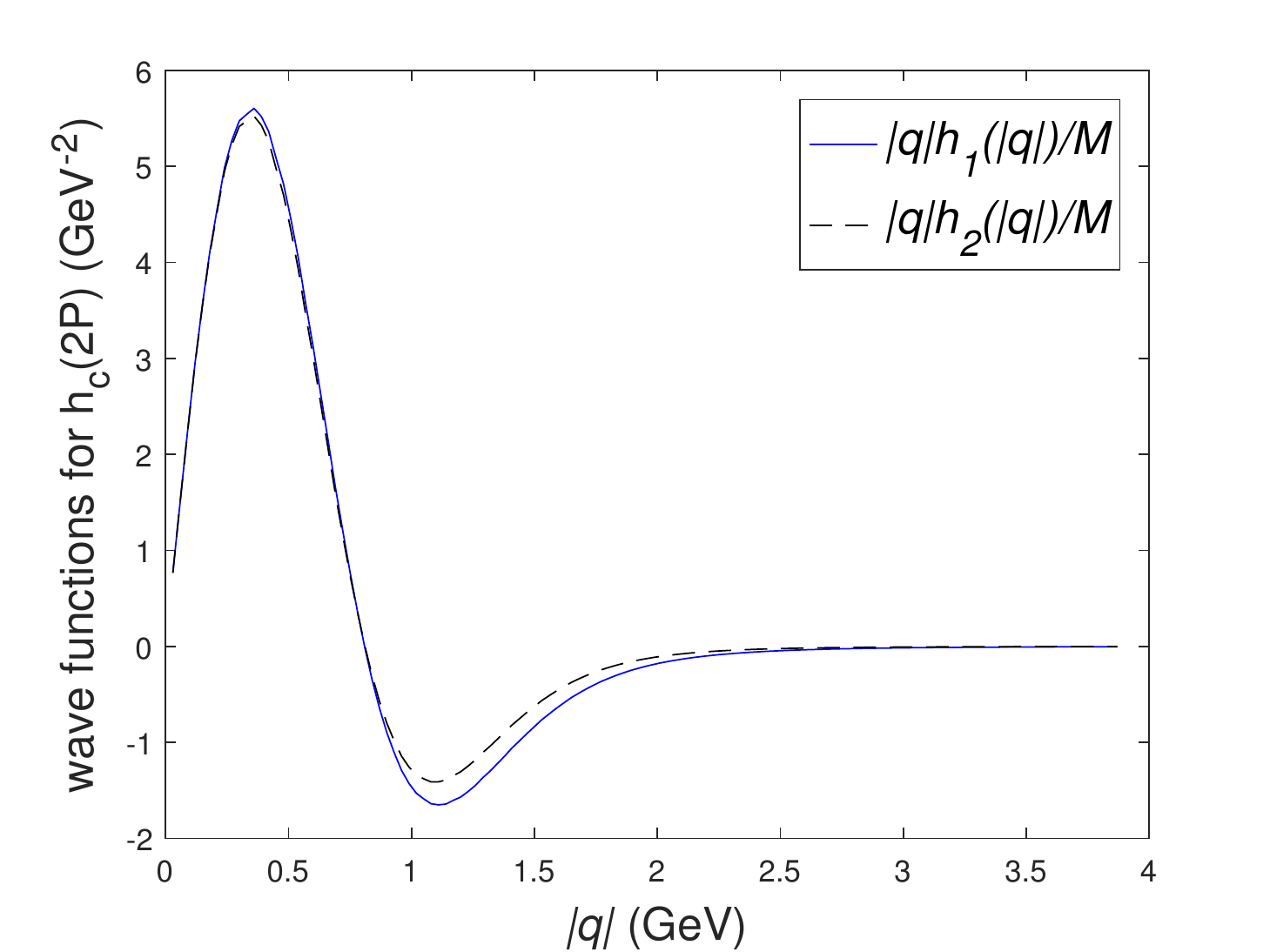}}
\subfigure[$\chi_{c0}(2P)$]{\label{fig:Xc0}
			      \includegraphics[width=0.4\textwidth]{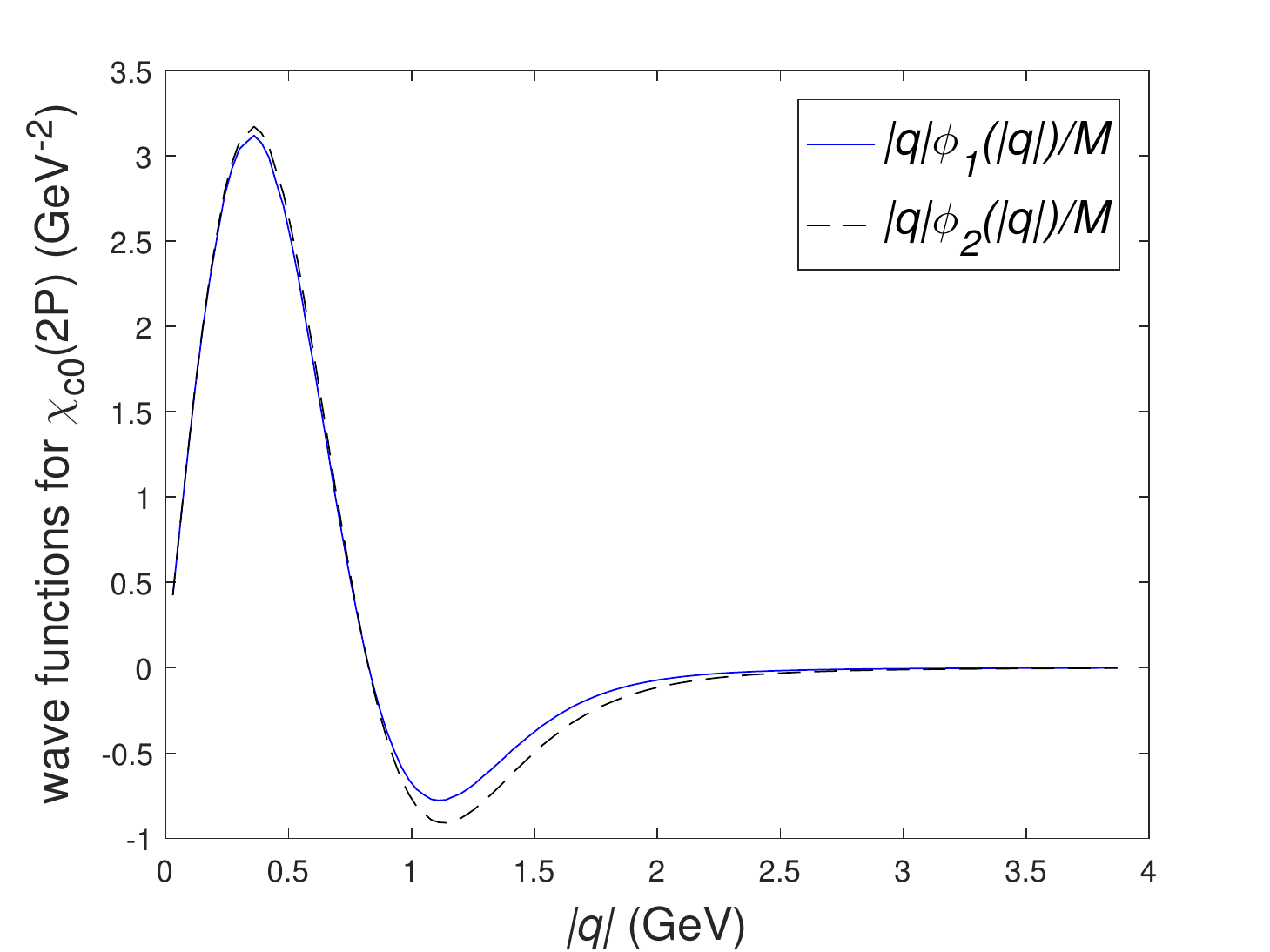}}
\subfigure[$\chi_{c1}(2P)$]{\label{fig:Xc1}
			      \includegraphics[width=0.4\textwidth]{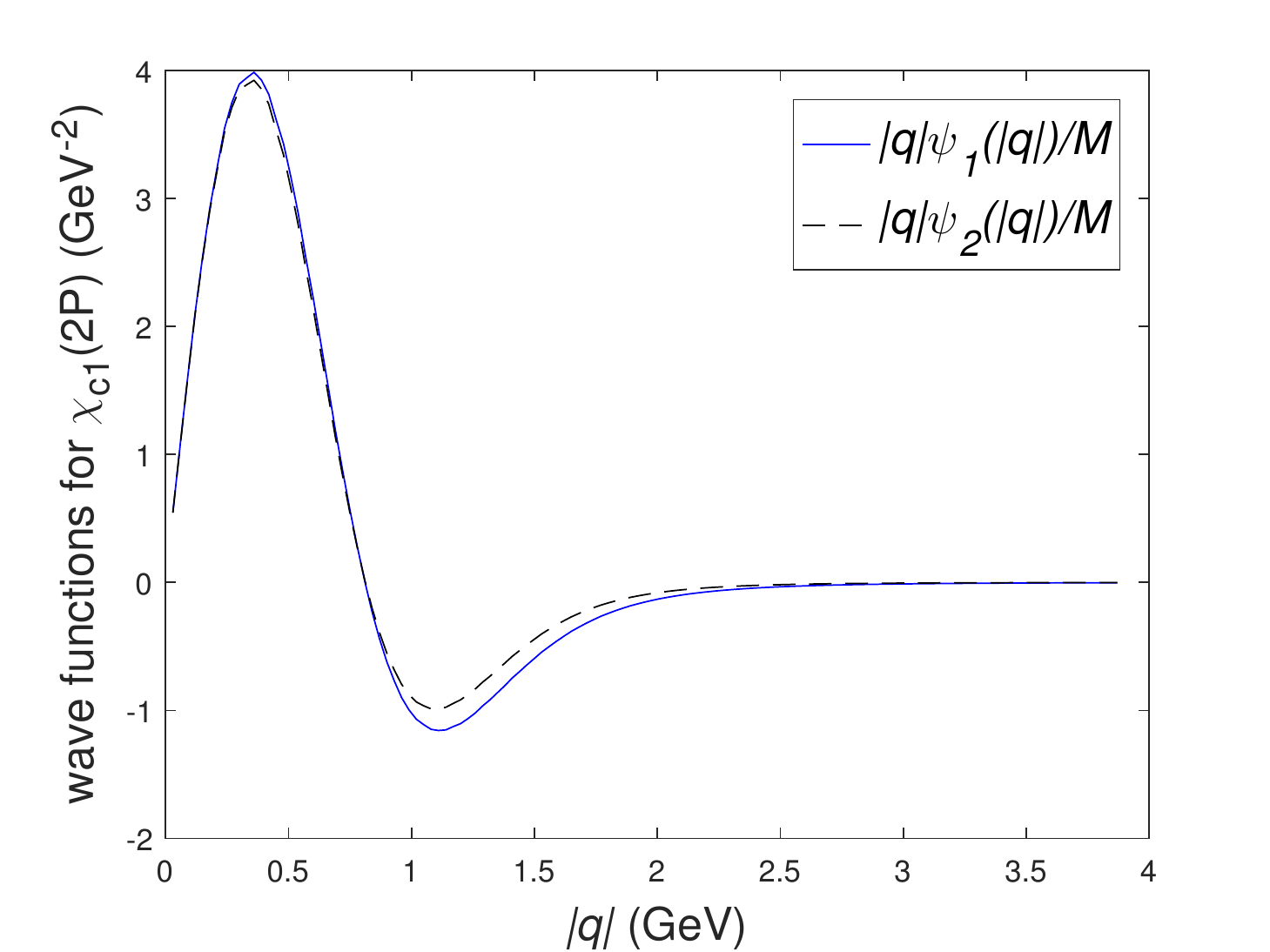}}
\caption{The wave functions for the radially excited state mesons.}\label{fig:wfs1}
\end{figure}

\begin{figure}[tbp]
\centering
\subfigure[$\eta_c(3S)$]{\label{fig:eta_c2S}
			      \includegraphics[width=0.4\textwidth]{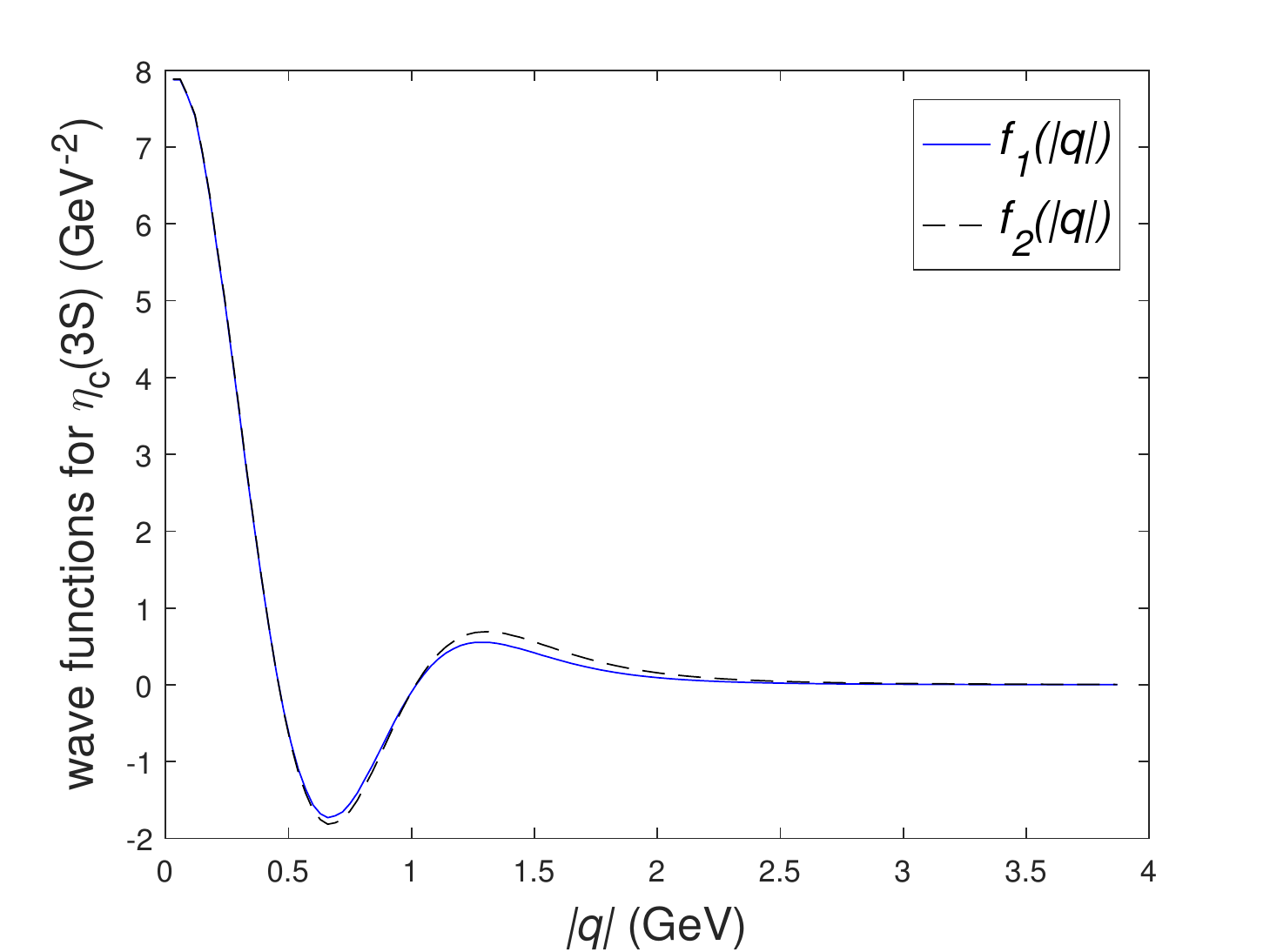}}
\subfigure[$\psi(3S)$]{\label{fig:psi3s}
			      \includegraphics[width=0.4\textwidth]{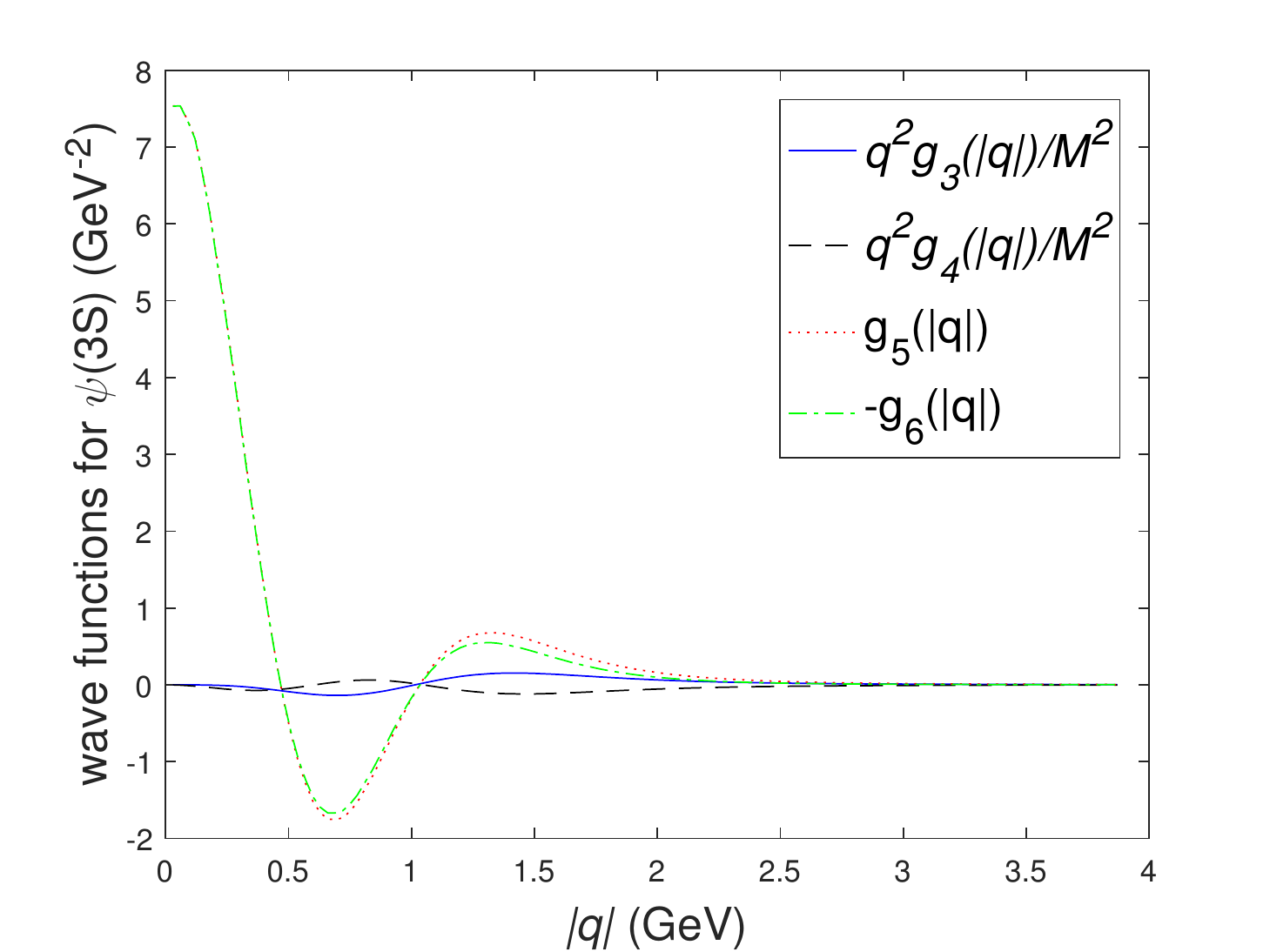}}
\subfigure[$h_c(3P)$]{\label{fig:hc}
			      \includegraphics[width=0.4\textwidth]{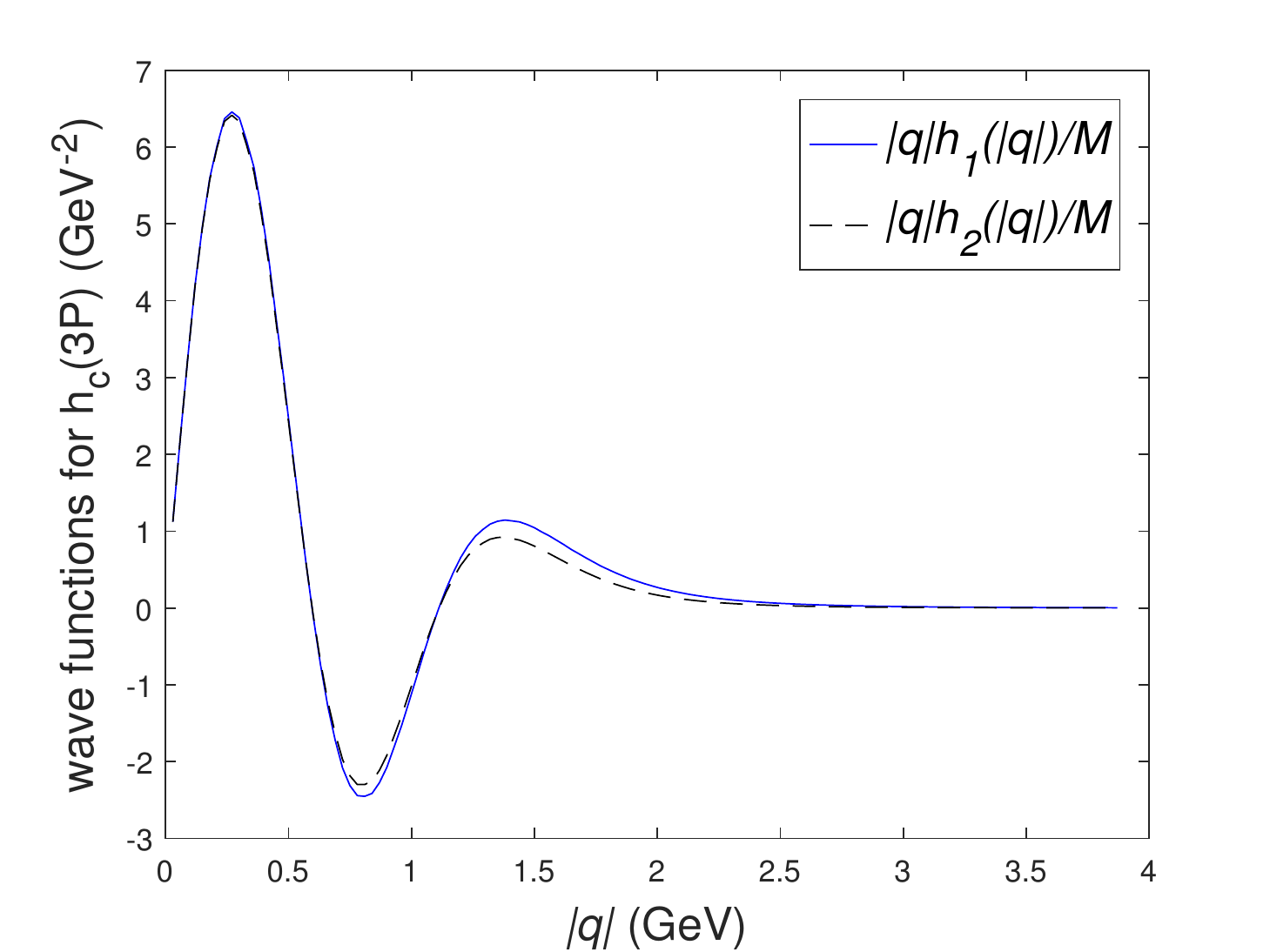}}
\subfigure[$\chi_{c0}(3P)$]{\label{fig:Xc0}
			      \includegraphics[width=0.4\textwidth]{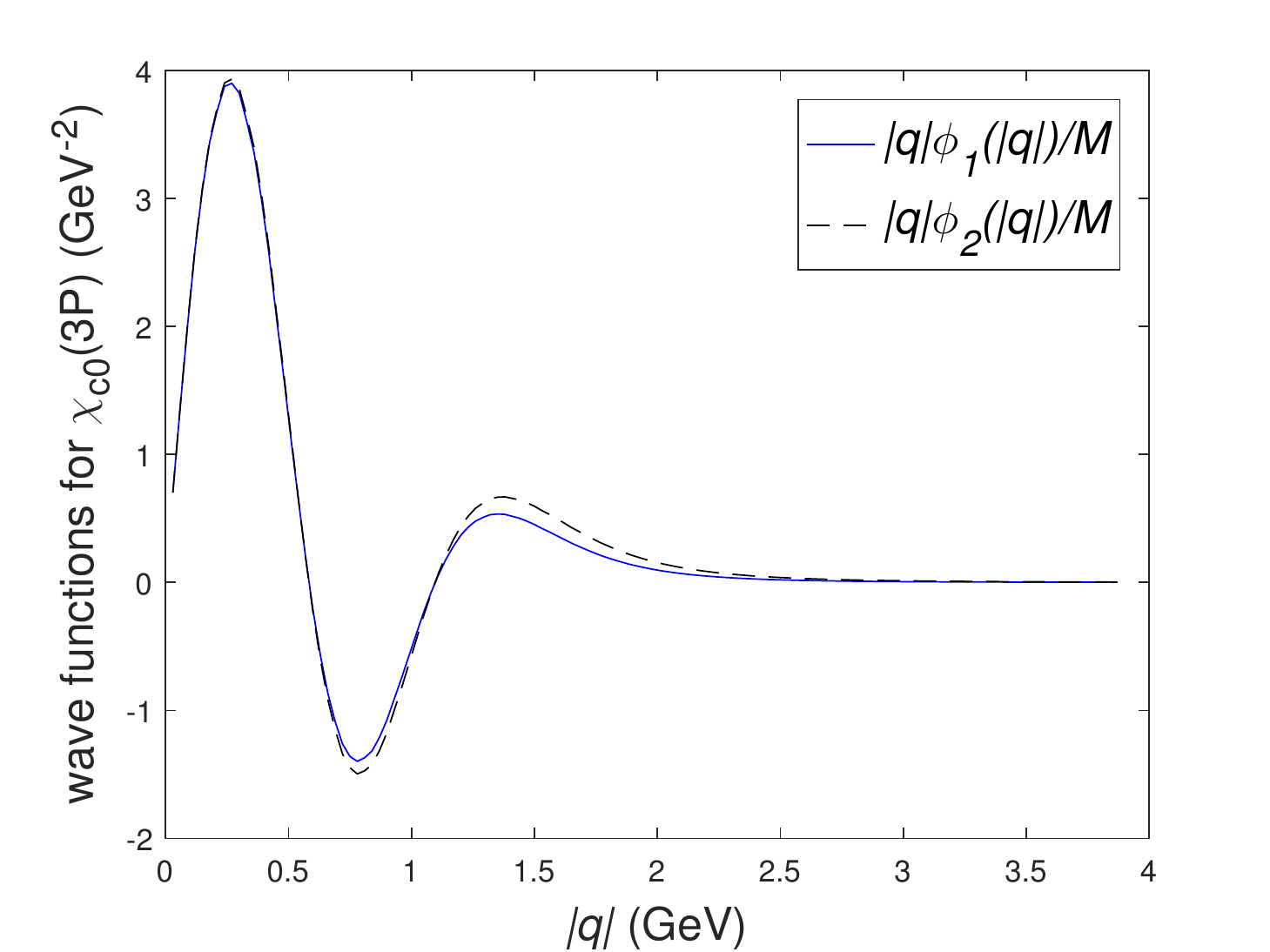}}
\subfigure[$\chi_{c1}(3P)$]{\label{fig:Xc1}
			      \includegraphics[width=0.4\textwidth]{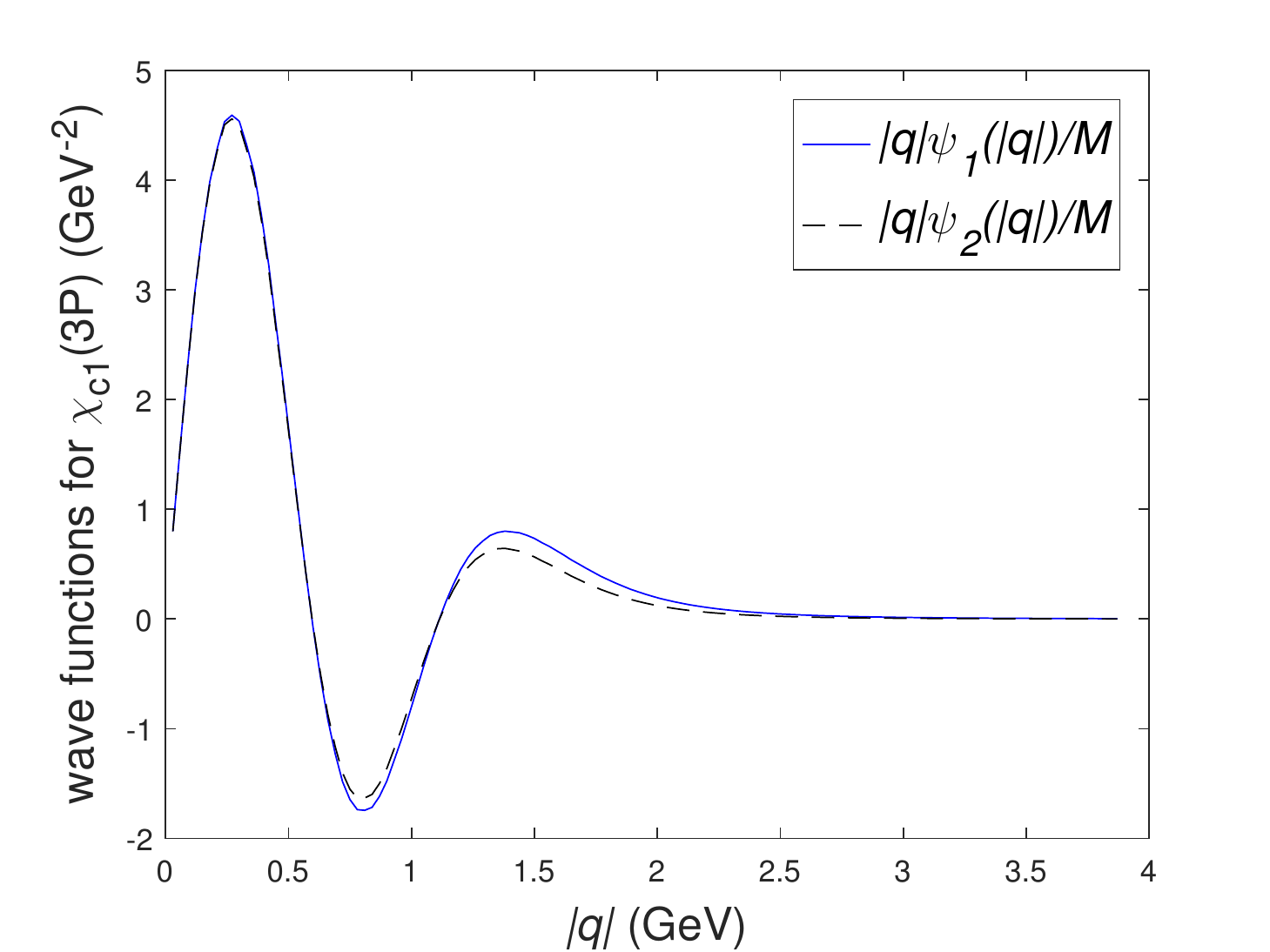}}
\caption{The wave functions for the radially excited state mesons.}\label{fig:wfs2}
\end{figure}

By solving the corresponding full Salpeter equations, we obtain the numerical results of wave function for different $J^{P(C)}$ states and show them in figures~\ref{fig:wfs}--\ref{fig:wfs2}. In some non-relativistic approaches, there is only one wave function. According to our results, the dominate two wave functions are almost equivalent, for example, $f_1\simeq f_2$ for $0^-$ state, $g_5\simeq -g_6$ for $1^{-}$ state, and $\phi_1\simeq \phi_2$ for $0^+$ state, etc. Therefore these non-relativistic approaches are reasonable in some cases.

\begin{figure}[tbp]
\centering
\subfigure[$B_c^+\to\eta_ce^+\nu_e$]{
			      \includegraphics[width=0.4\textwidth]{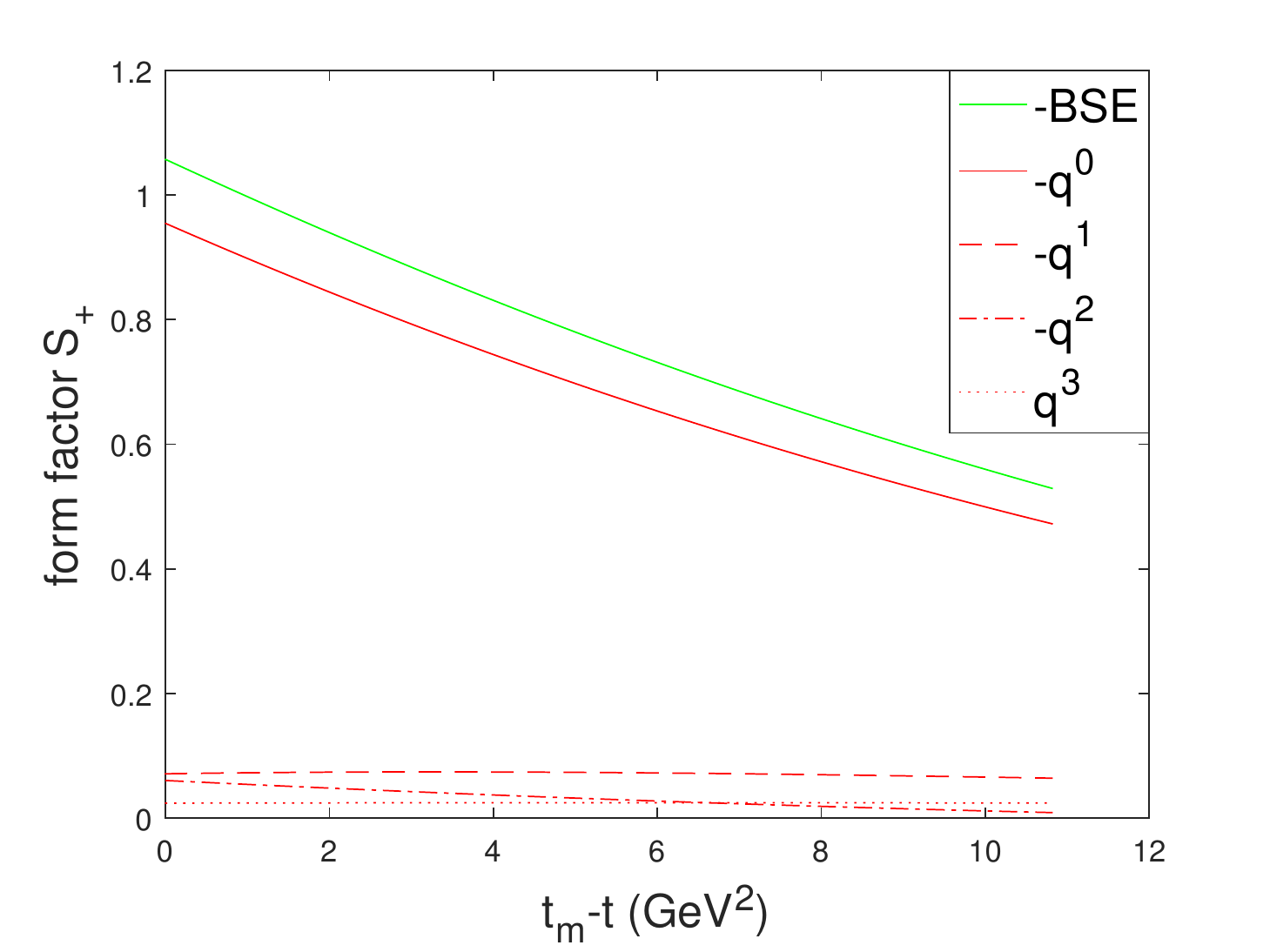}}
\subfigure[$B_c^+\to J/\psi e^+\nu_e$]{
			      \includegraphics[width=0.4\textwidth]{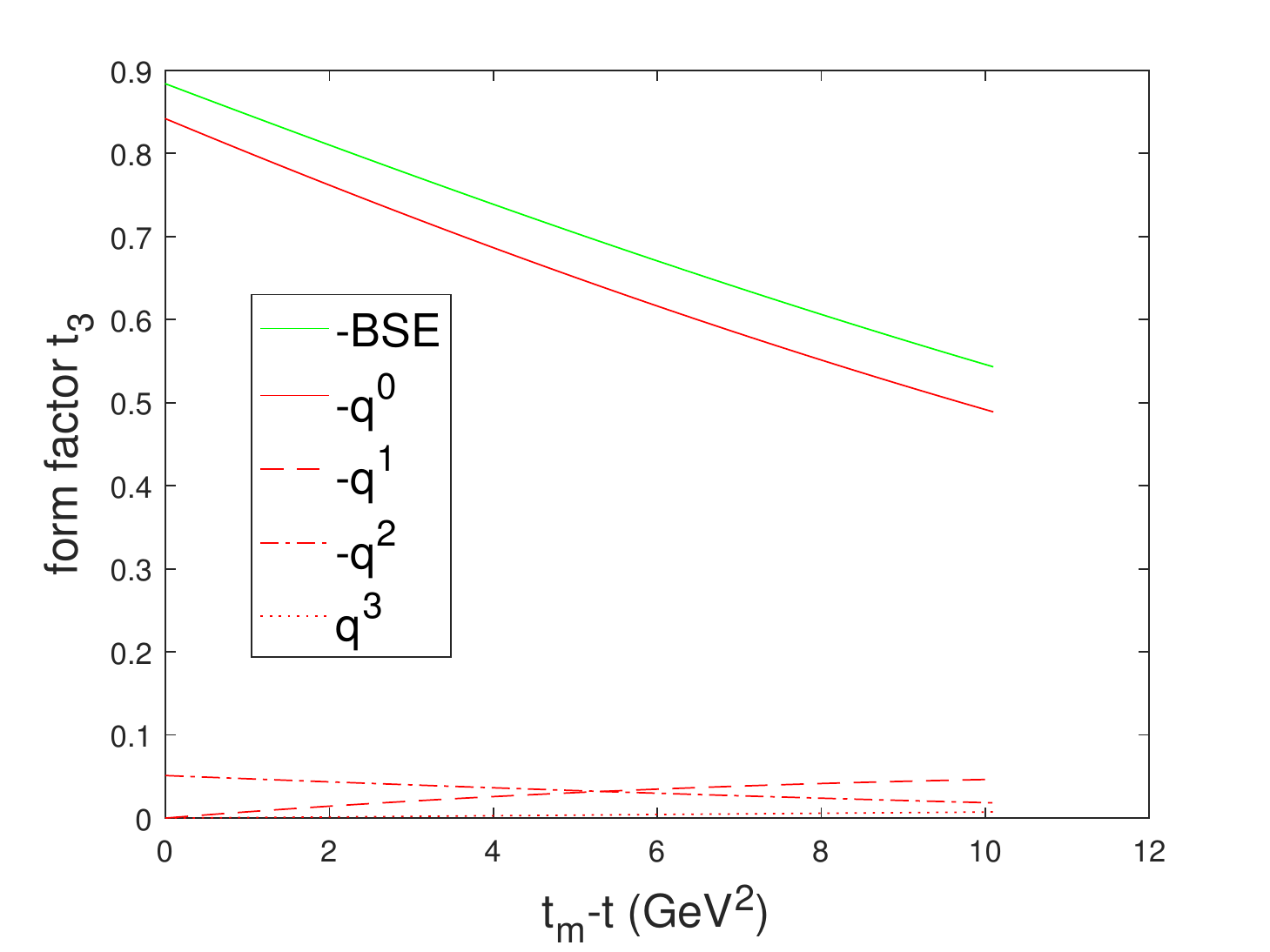}}
\subfigure[$B_c^+\to h_ce^+\nu_e$]{
			      \includegraphics[width=0.4\textwidth]{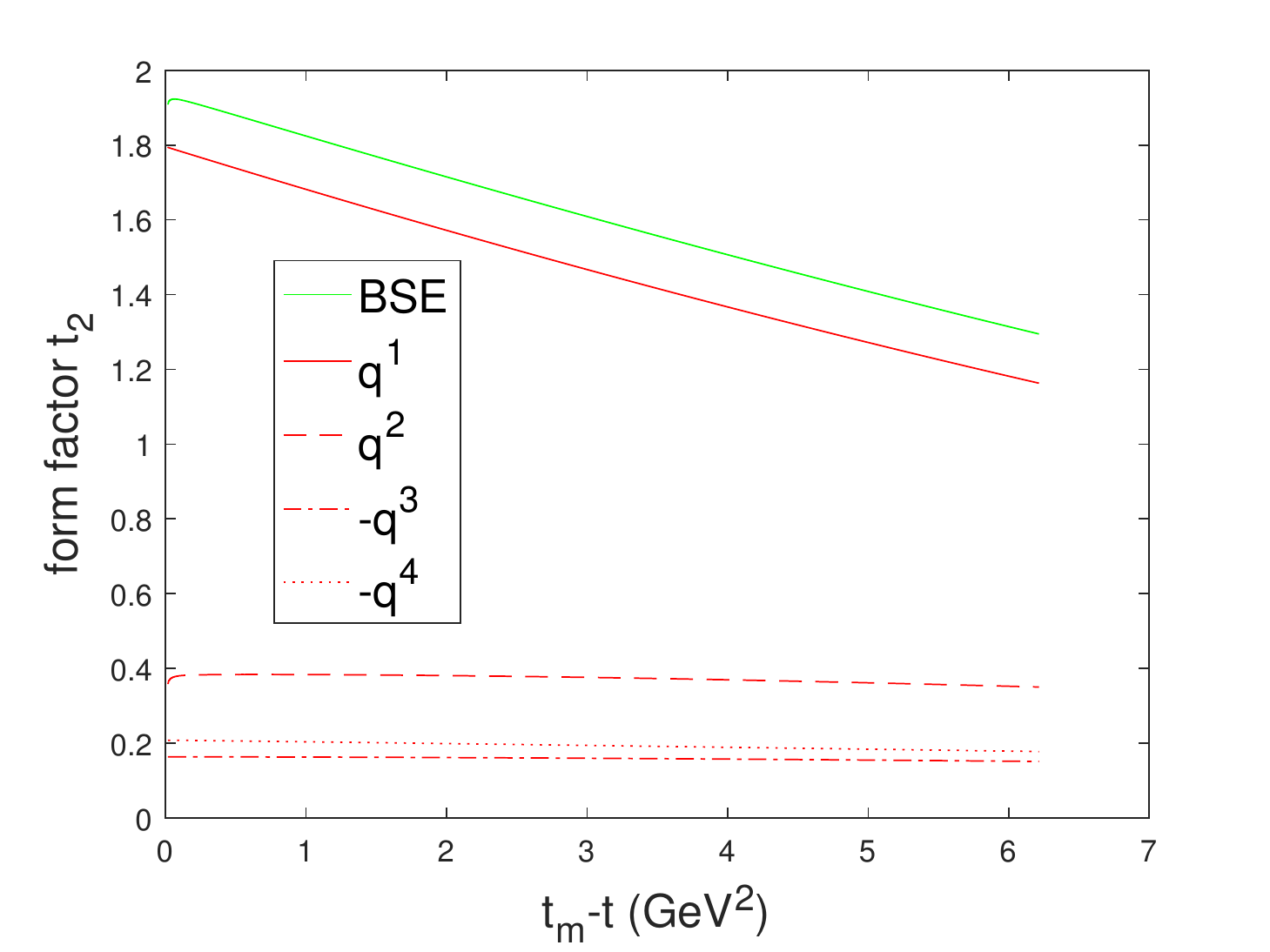}}
\subfigure[$B_c^+\to\chi_{c0}e^+\nu_e$]{
			      \includegraphics[width=0.4\textwidth]{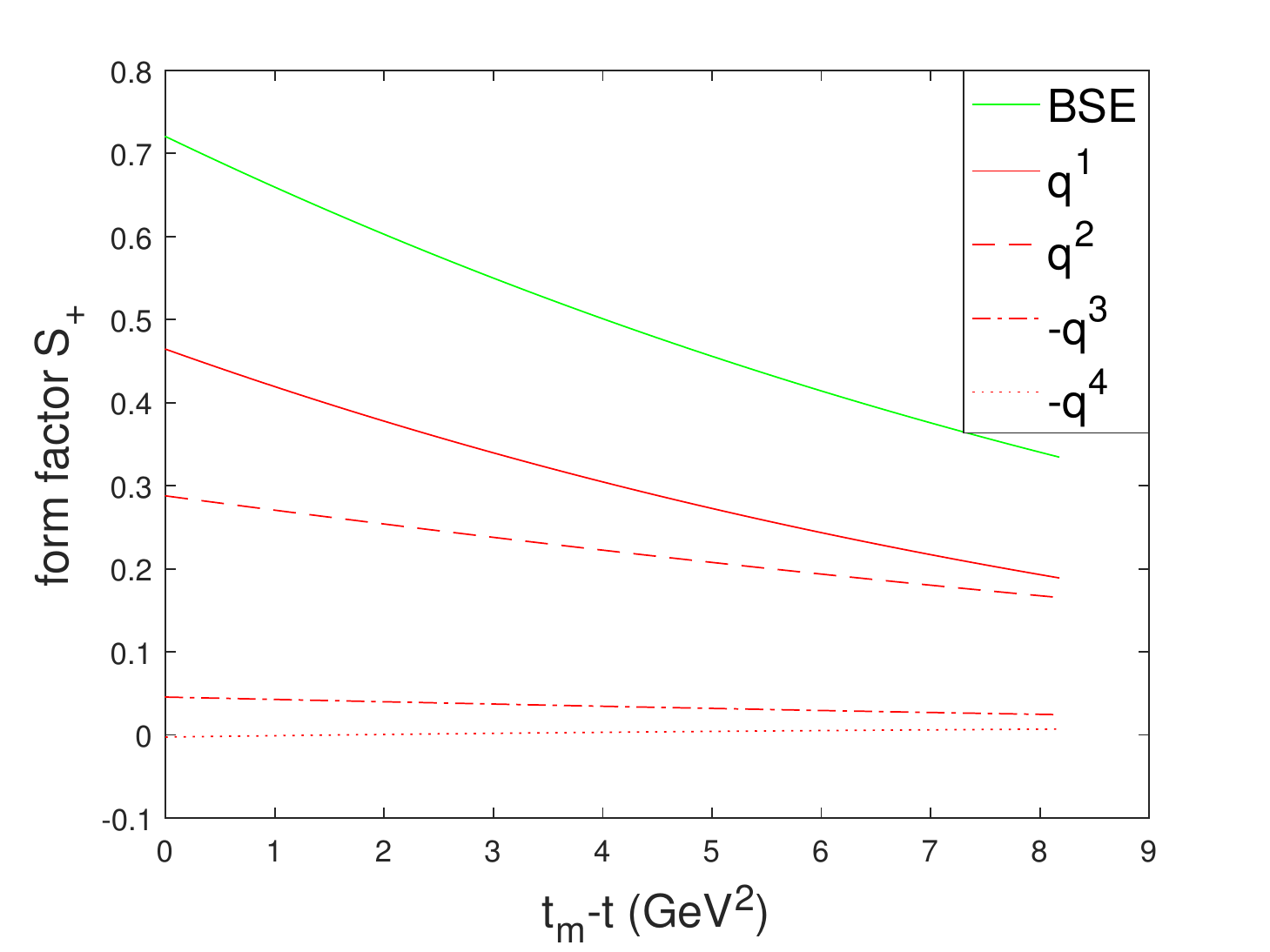}}
\subfigure[$B_c^+\to\chi_{c1}e^+\nu_e$]{
			      \includegraphics[width=0.4\textwidth]{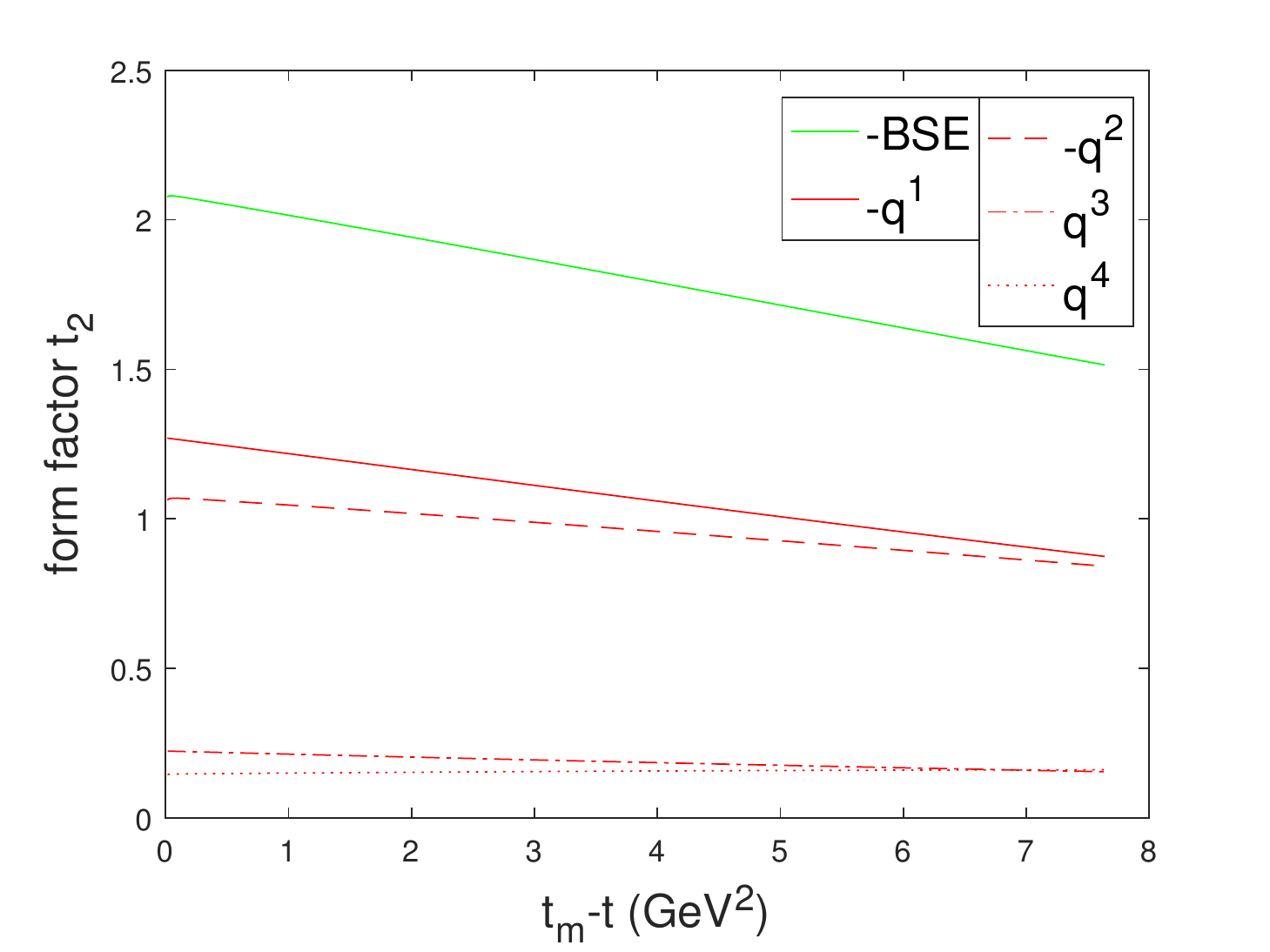}}

\caption{Form factors of the $B_c$ decays to $1S$ and $1P$ charmonium.}\label{fig:ffg}
\end{figure}

With these relativistic wave functions, the form factors and the corresponding relativistic corrections are obtained. With the non-relativistic wave functions (see eq.~(\ref{eq:normaNR})) and the leading order amplitude, the results obatained by those NR approches are estimated. For each process in Method I, we only show one of the form factors which makes the main contribution to the decay width. The corresponding results of the $B_c$ decays to $1S$-wave and $1P$-wave charmonium are shown in figure~\ref{fig:ffg}, where the green solid lines are the results of relativistic BS method (BSE) without expansion; the red solid lines are the leading order (non-relativistic) contributions to the form factors; dash lines are the first order relativistic corrections ($q^2$); dot-dash lines are the second order relativistic corrections ($q^3$), and dot lines are the third order relativistic corrections ($q^3$); $t\equiv(P-P_f)^2$, and $t_m$ is the maximum of $t$. The LO contributions in form factors are dominant. The 1st relativistic corrections for the $1S$ final state cases are negligible. In the cases of $1P$ final states, the 1st relativistic corrections provide sizable contributions, especially for $\chi_{c0}$ and $\chi_{c1}$. The ratio of 1st relativistic corrections to LO in $B_c\to h_c$ is around 15\%. But for $\chi_{c0}$ or $\chi_{c1}$ final states, this ratio can reach up to 80\%. Other higher order corrections are less than 10\% as much as the leading order and are negligible. Therefore we conclude that the relativistic corrections in the $B_c$ decays to P-wave charmonium have large contributions, even though both the initial state and final state are the double heavy mesons.

\begin{figure}[tbp]
\centering
\subfigure[$B_c^+\to\eta_c(2S)e^+\nu_e$]{
			      \includegraphics[width=0.4\textwidth]{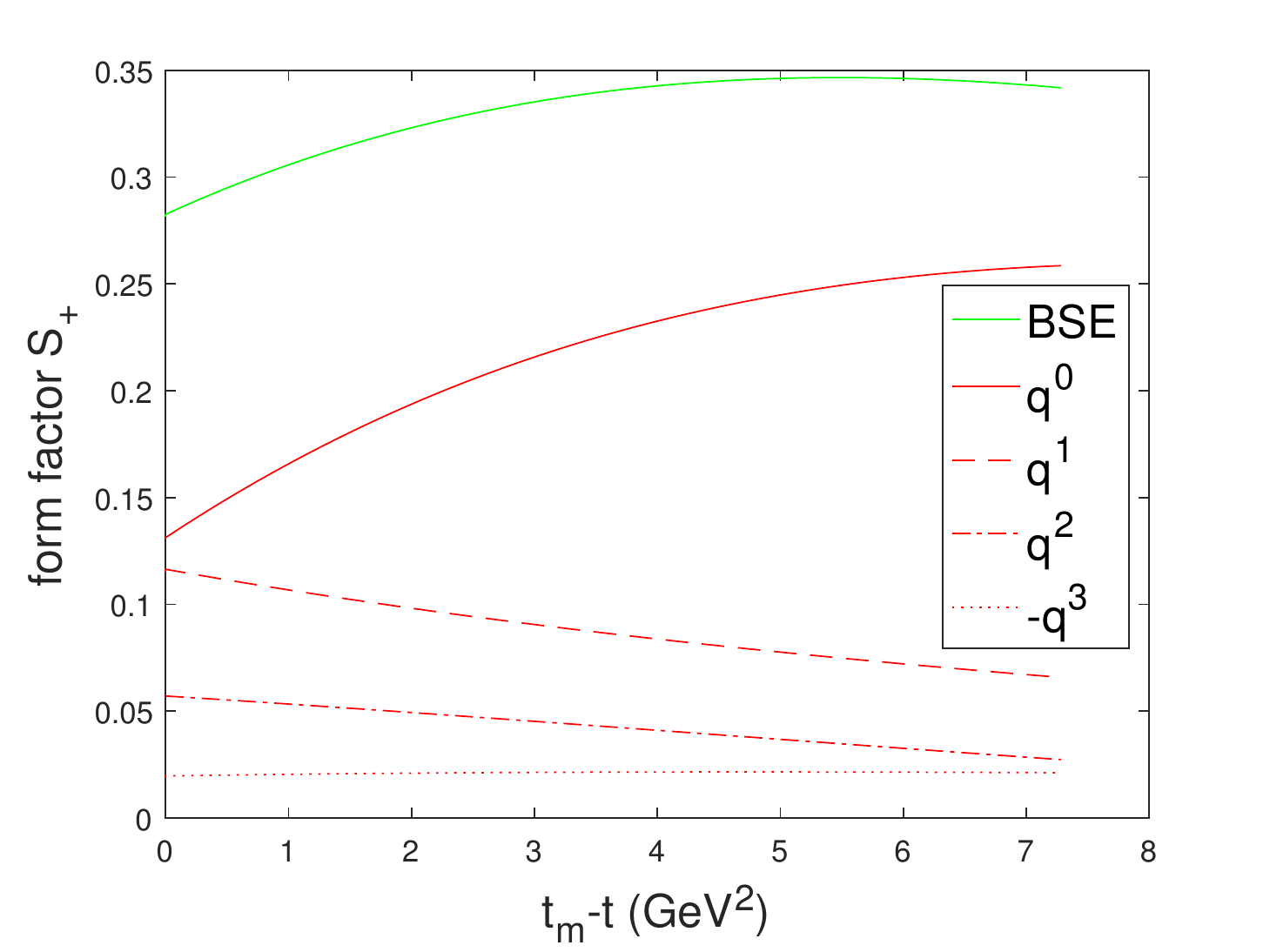}}
\subfigure[$B_c^+\to\psi(2S)e^+\nu_e$]{
			      \includegraphics[width=0.4\textwidth]{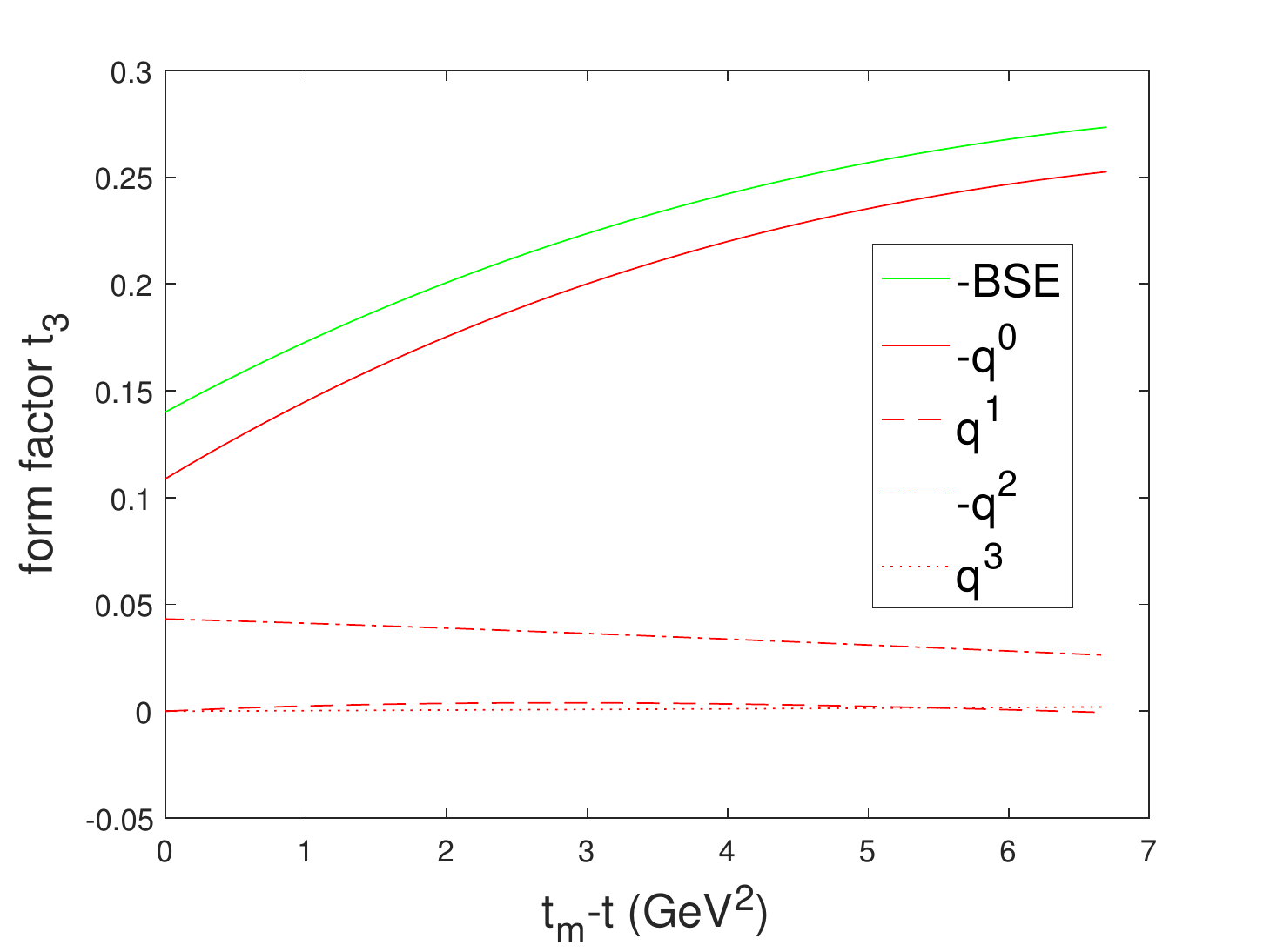}}
\subfigure[$B_c^+\to h_c(2P)e^+\nu_e$]{
			      \includegraphics[width=0.4\textwidth]{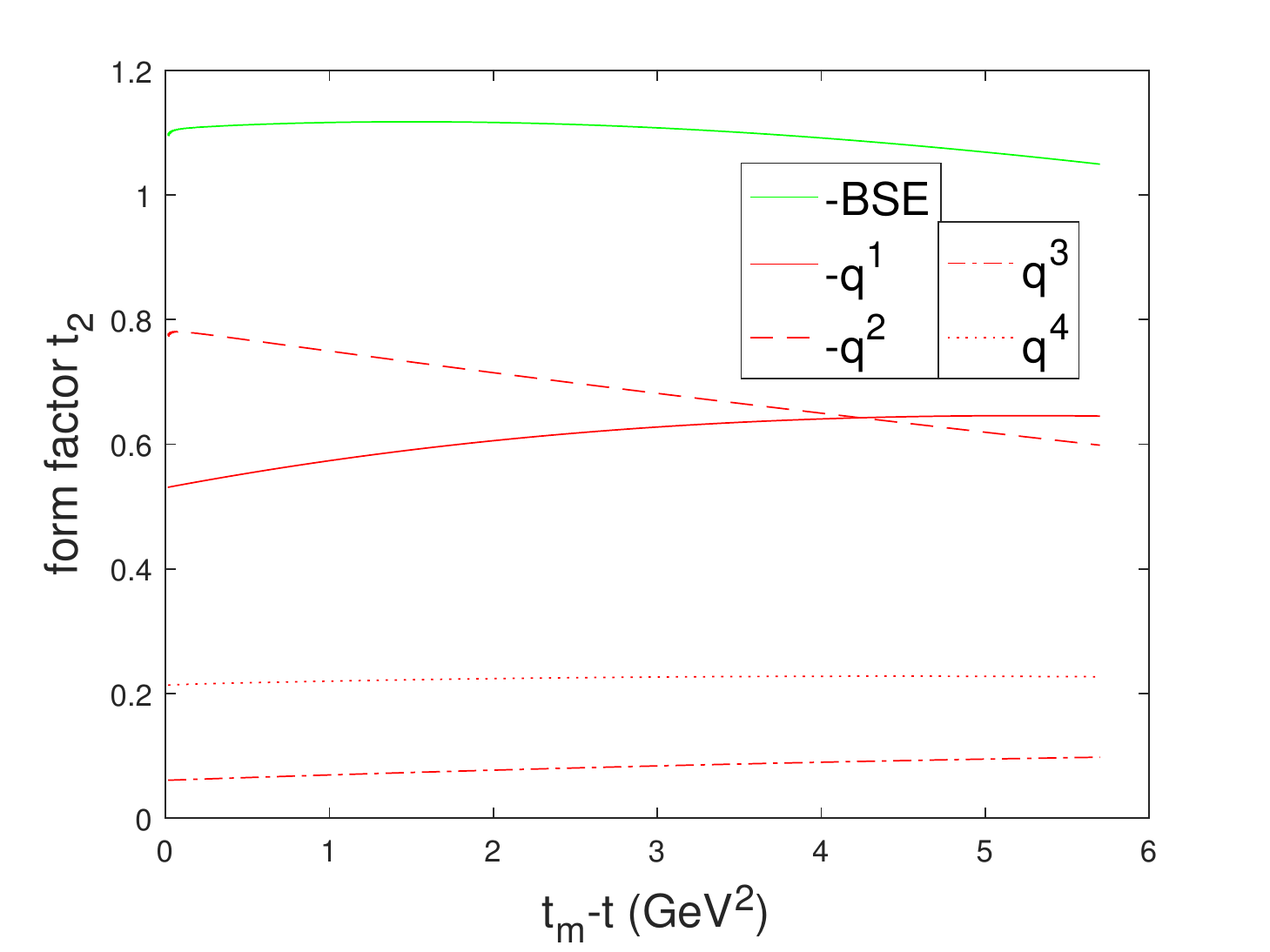}}
\subfigure[$B_c^+\to\chi_{c0}(2P)e^+\nu_e$]{
			      \includegraphics[width=0.4\textwidth]{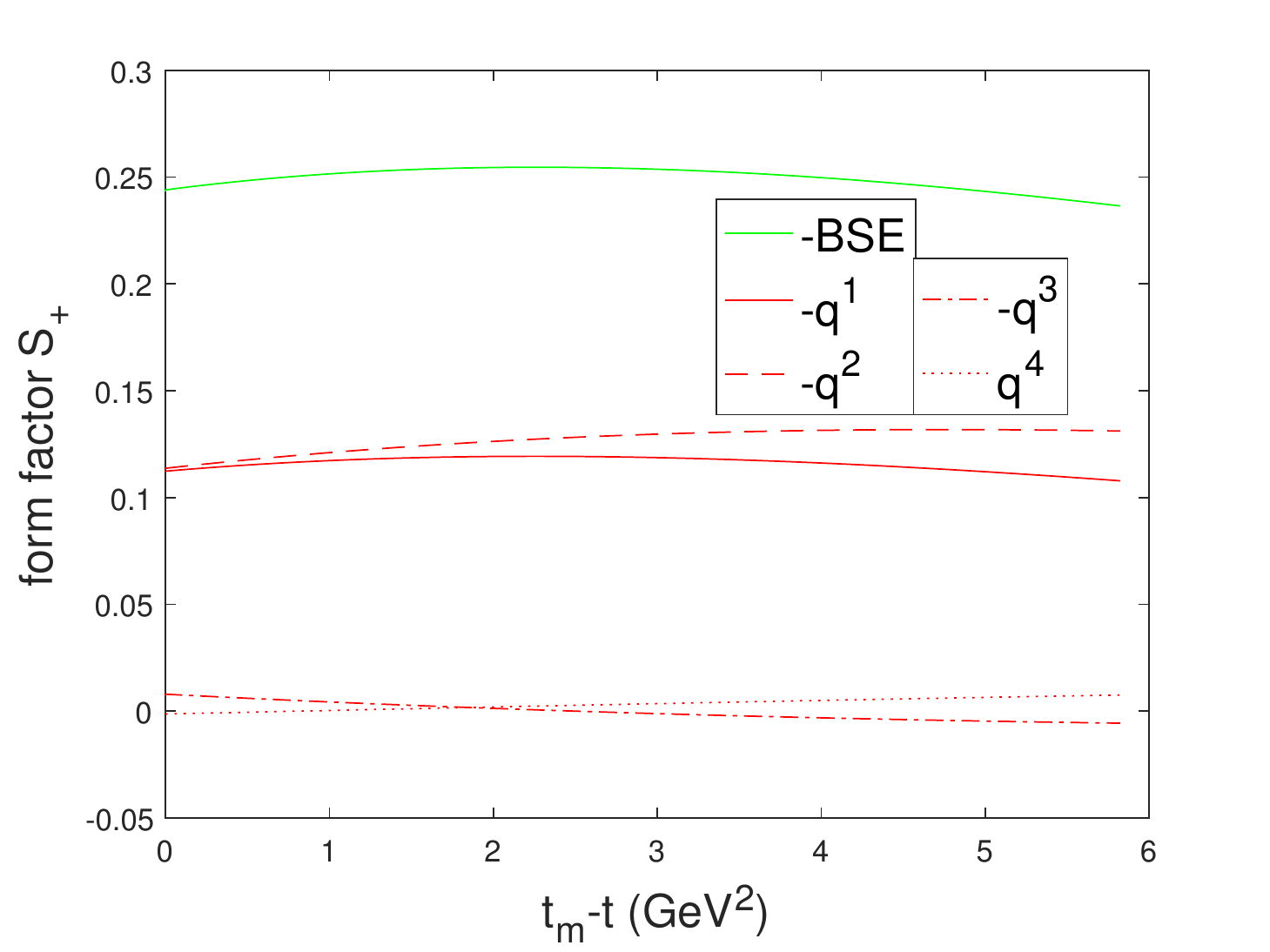}}
\subfigure[$B_c^+\to\chi_{c1}(2P)e^+\nu_e$]{
			      \includegraphics[width=0.4\textwidth]{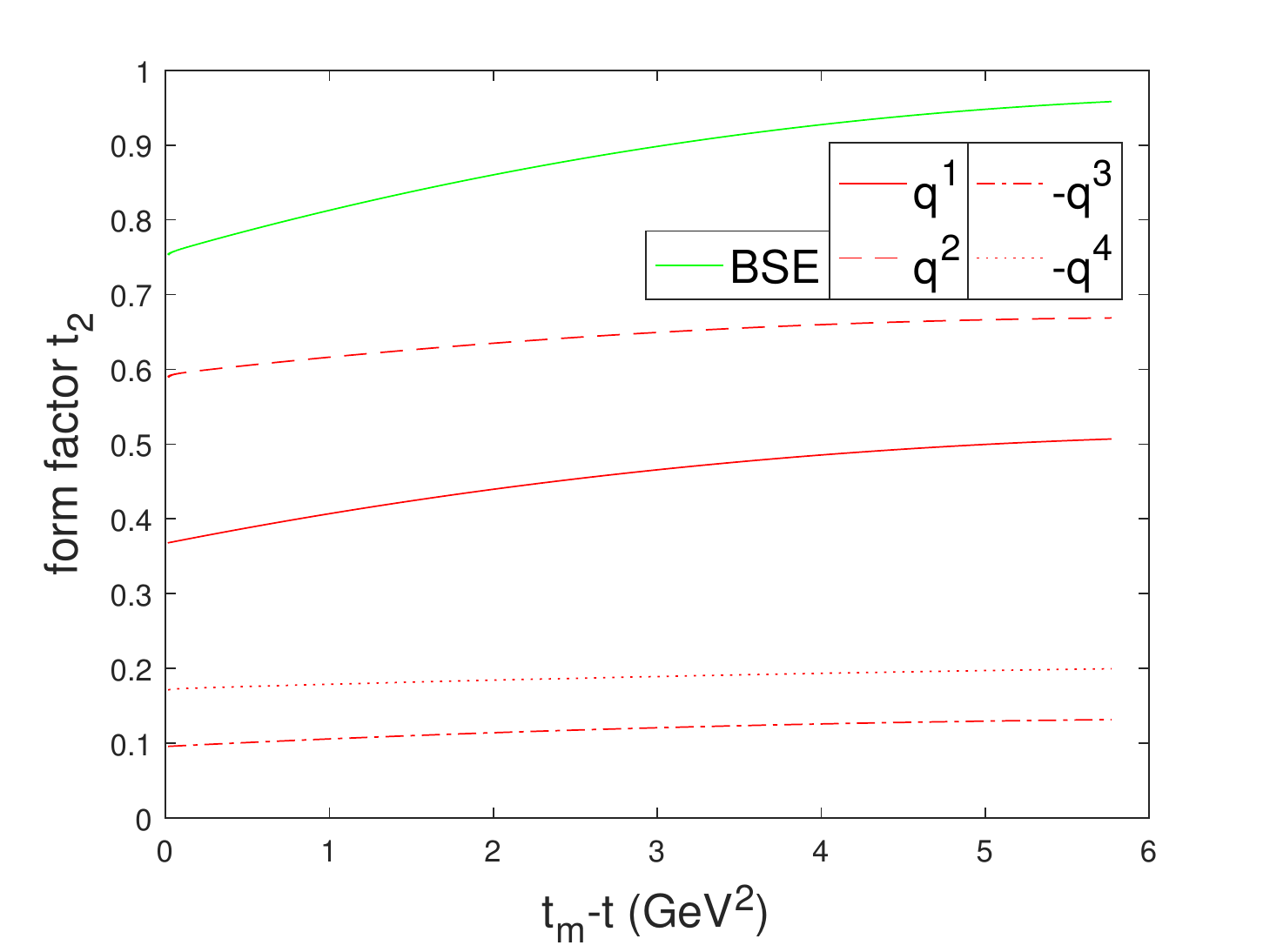}}

\caption{Form factors of the $B_c$ decays to $2S$ and $2P$ charmonium.}\label{fig:ffe2}
\end{figure}

\begin{figure}[tbp]
\centering
\subfigure[$B_c^+\to\eta_c(3S)e^+\nu_e$]{
			      \includegraphics[width=0.4\textwidth]{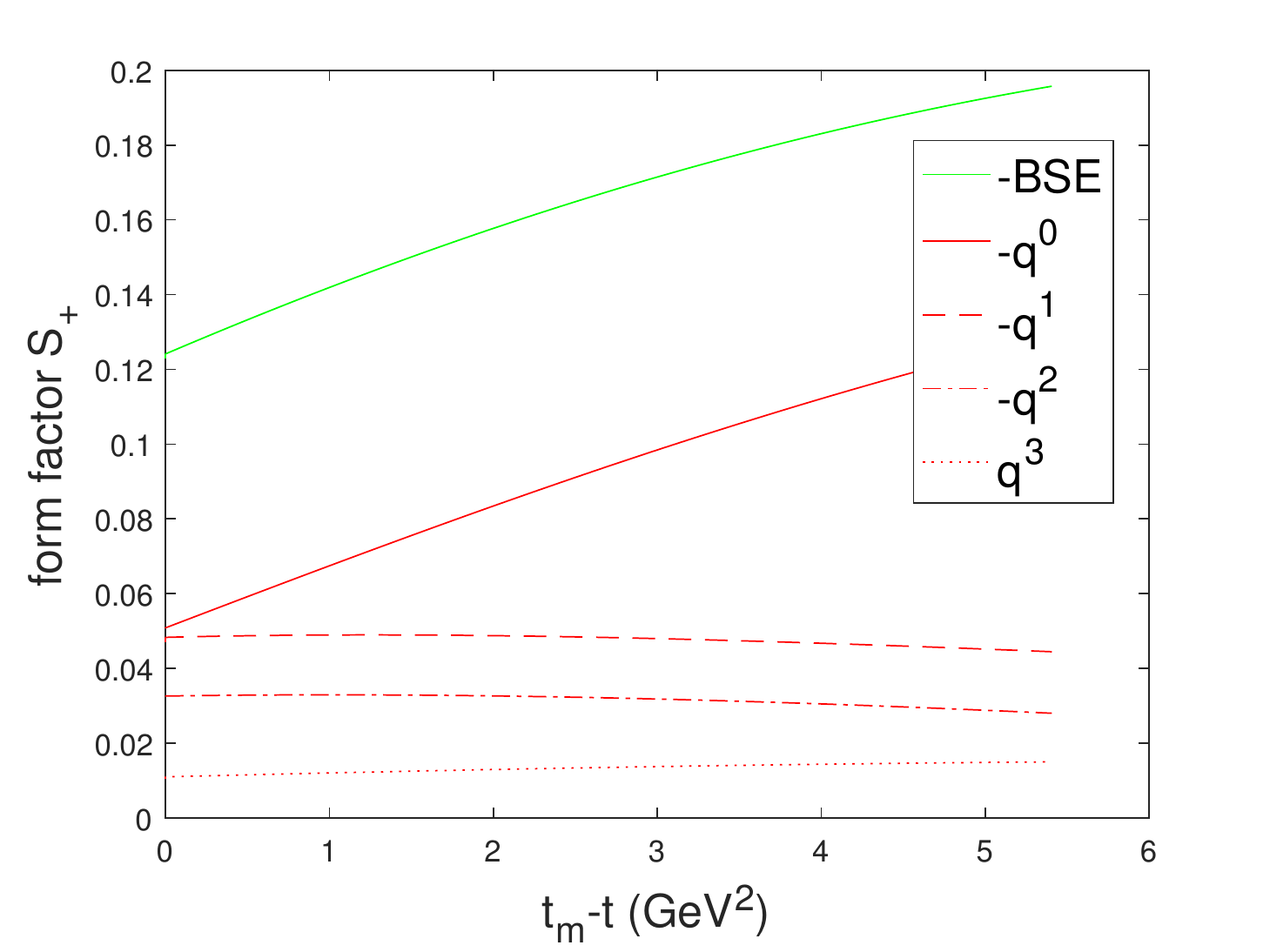}}
\subfigure[$B_c^+\to\psi(3S)e^+\nu_e$]{
			      \includegraphics[width=0.4\textwidth]{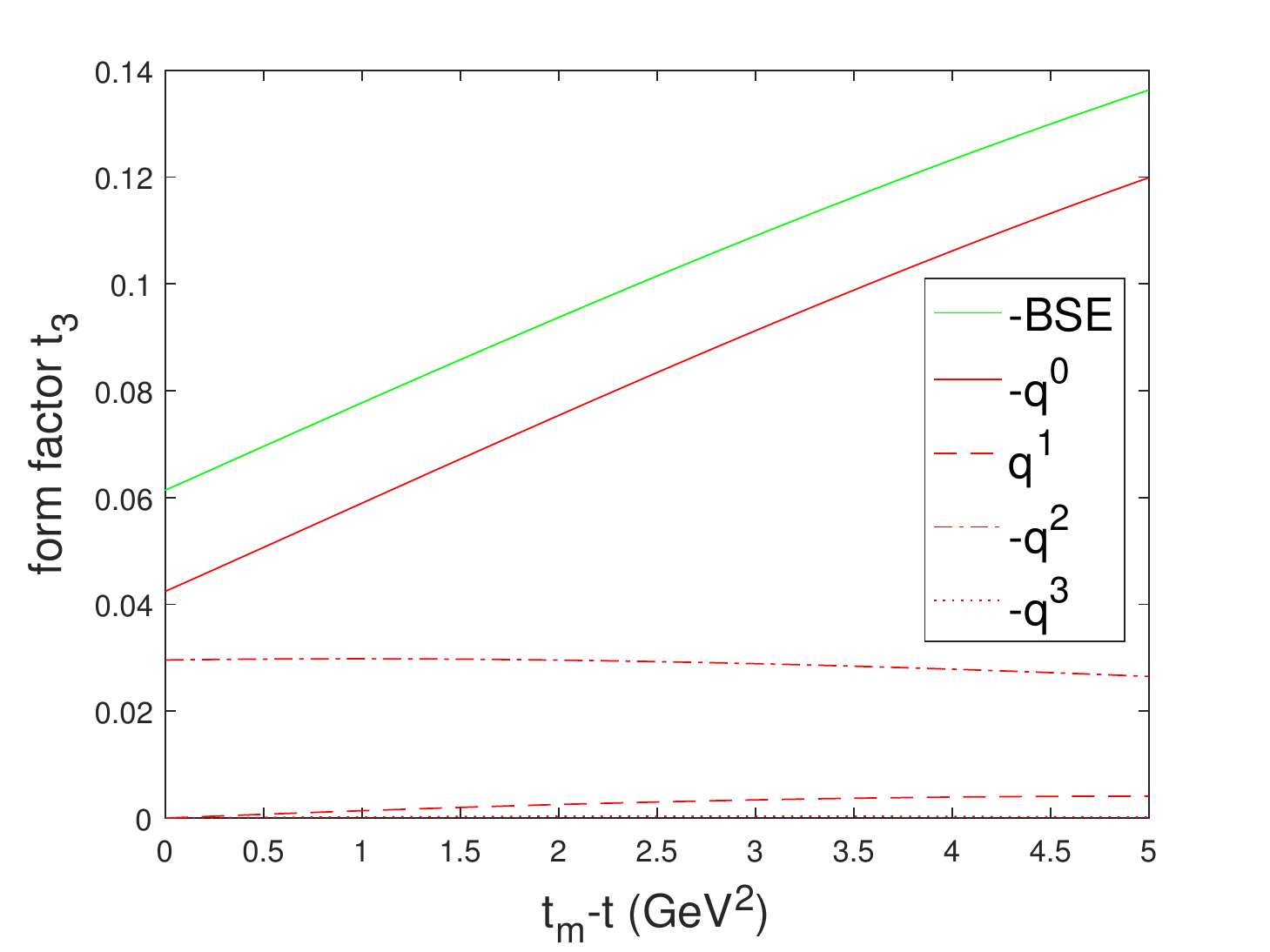}}
\subfigure[$B_c^+\to h_c(3P)e^+\nu_e$]{
			      \includegraphics[width=0.4\textwidth]{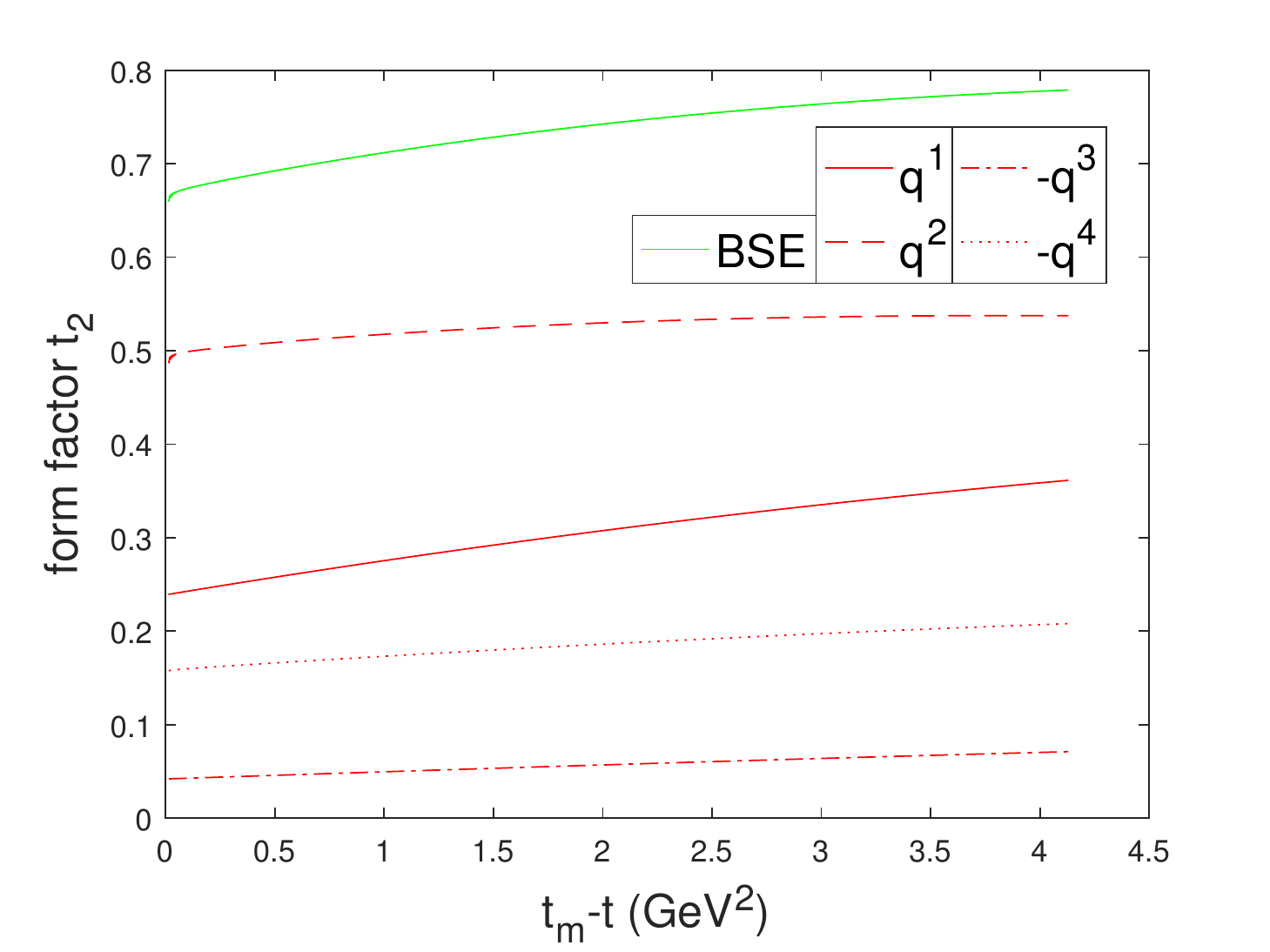}}
\subfigure[$B_c^+\to\chi_{c0}(3P)e^+\nu_e$]{
			      \includegraphics[width=0.4\textwidth]{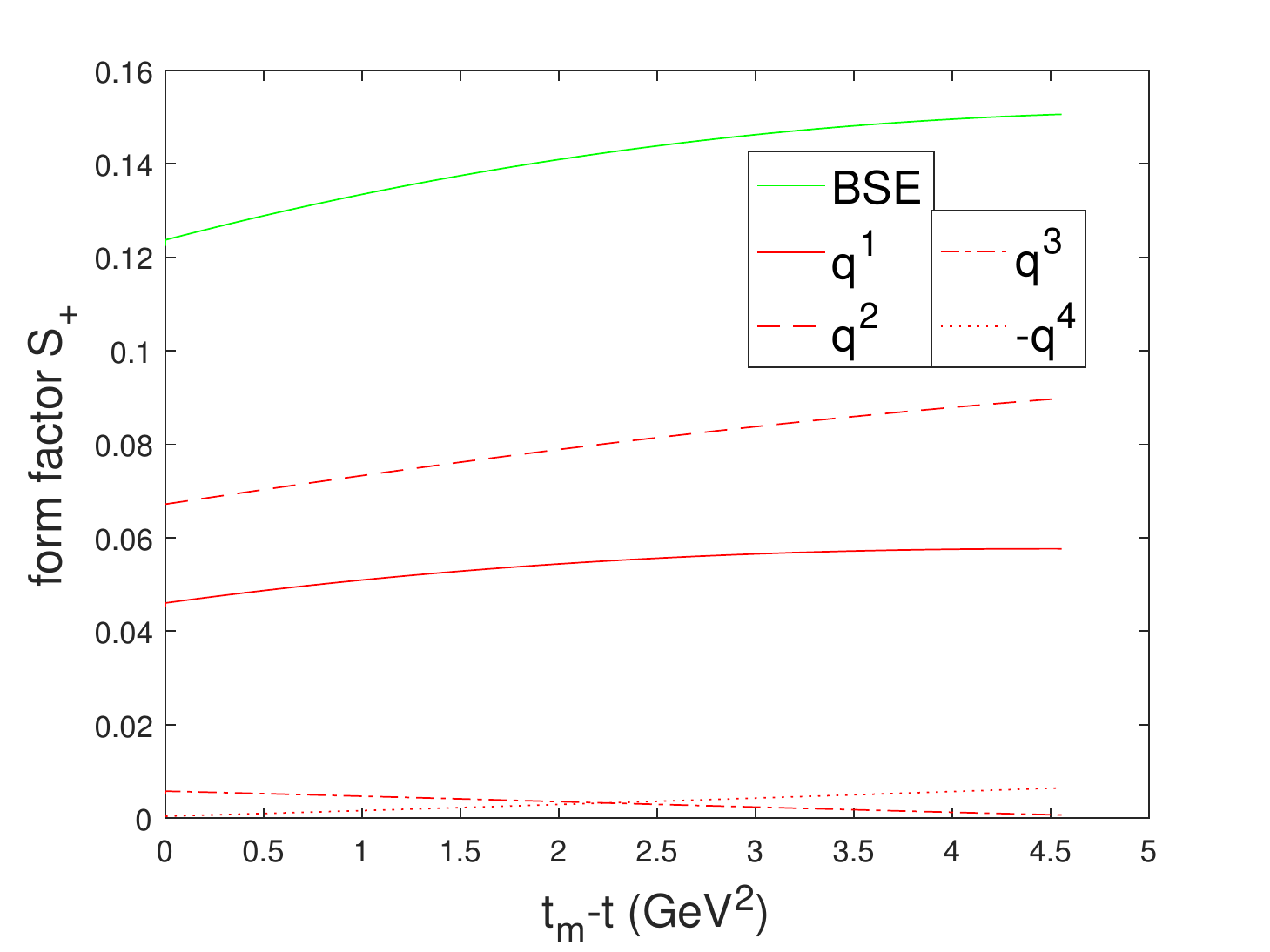}}
\subfigure[$B_c^+\to\chi_{c1}(3P)e^+\nu_e$]{
			      \includegraphics[width=0.4\textwidth]{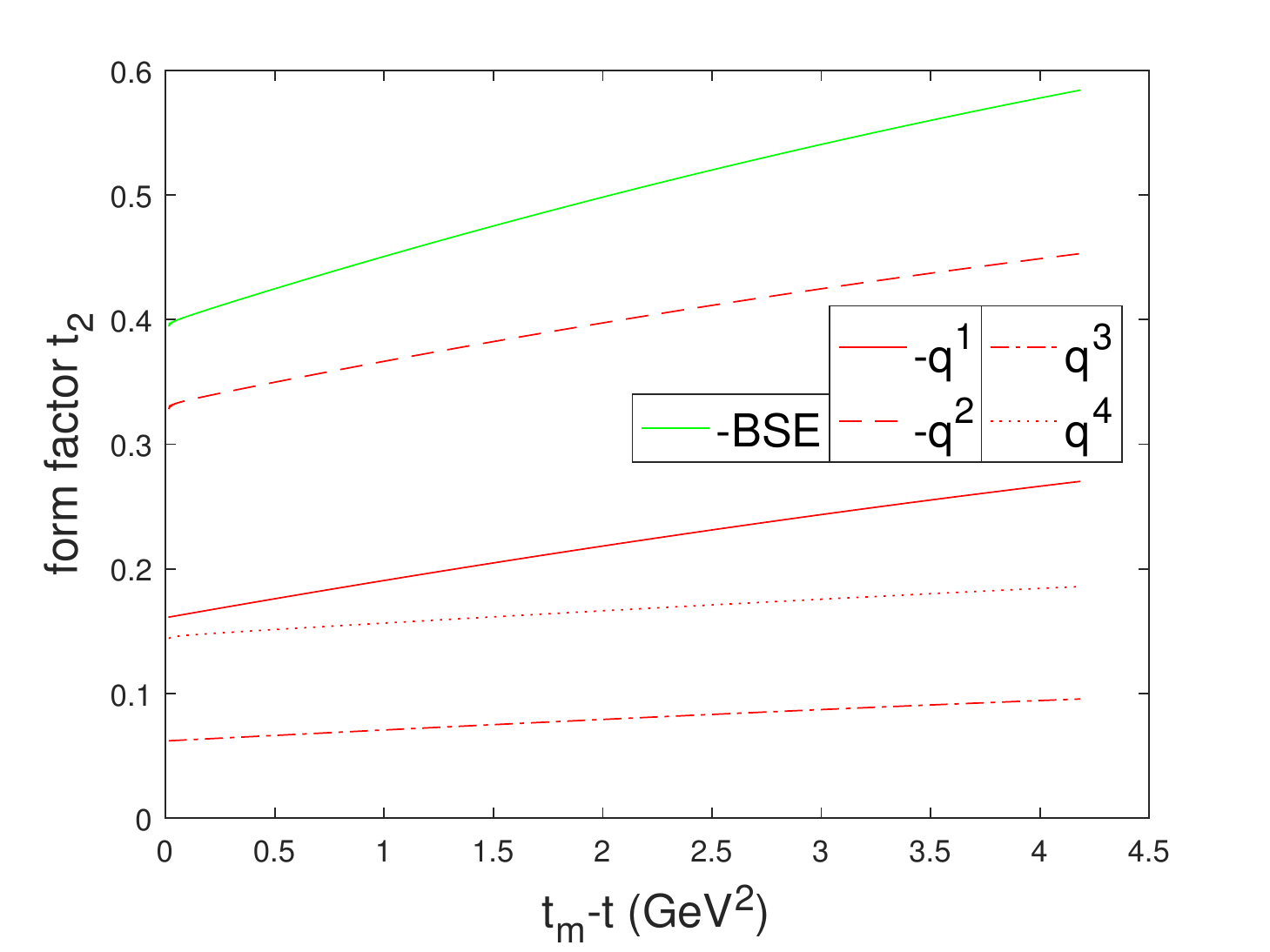}}

\caption{Form factors of the $B_c$ decays to $3S$ and $3P$ charmonium.}\label{fig:ffe3}
\end{figure}

The form factors of the $B_c$ decays to radially excited charmonium are plotted in figure~\ref{fig:ffe2}--\ref{fig:ffe3}. The 1st relativistic corrections are comparable to the leading order for the $B_c$ decays to $2S$ or $3S$ charmonium. The 1st relativistic corrections are larger than the leading order for the $B_c$ decays to $2P$ or $3P$ charmonium. The 2nd order contribution and the 3rd one may also appear reversal. We conclude that compared with the ground states, the relativistic corrections of the $B_c$ decays to corresponding excited charmonium are much larger.

\begin{table}[tbp]
\vspace{0.2cm}
\setlength{\tabcolsep}{0.2cm}
\centering
\begin{tabular*}{\textwidth}{@{}@{\extracolsep{\fill}}|c|ccccccc|}
\hline
Mode&${\vec{q}~}^0$ &${\vec{q}~}^1$&${\vec{q}~}^2$&${\vec{q}~}^3$&${\vec{q}~}^4$&${\vec{q}~}^5$&${\vec{q}~}^6$\\
\hline
{\phantom{\Large{l}}}\raisebox{+.2cm}{\phantom{\Large{j}}}
$\eta_c$  &  47.4  &  11.0 & 4.10 &  -3.49 &  -0.390  &  -0.135 & 0.0811\\
$J/\psi$  &  157  & 20.2 & 18.8 & 0.517 & 0.150 & -0.00270 & 0.0203\\
$\eta_c(2S)$  &  3.66  &  2.29 & 1.43 &  -0.294 &  -0.122  &  -0.0951 & 0.0284 \\
$\psi(2S)$  &  6.88  & 1.87  & 2.74 & 0.362 & 0.185 & 0.0247 & 0.00849\\
$\eta_c(3S)$  &  0.408  &  0.339 & 0.293 &  -0.0100 &  -0.0133  &  -0.0287 & 0.00675 \\
$\psi(3S)$  &  0.652  & 0.241  & 0.510 & 0.108 & 0.0809 & 0.0138 & 0.00282 \\
\hline
Mode&${\vec{q}~}^2$&${\vec{q}~}^3$&${\vec{q}~}^4$&${\vec{q}~}^5$&${\vec{q}~}^6$&${\vec{q}~}^7$ &${\vec{q}~}^8$\\
\hline

$h_c$   &  15.4 & 11.0 &  4.20 &  -0.420 & -0.142 & 0.00354 & 0.0100 \\
$\chi_{c0}$  &  5.13 & 7.90 &  1.85 & -1.12 & -0.0765 & 0.0224 & 0.00224 \\
$\chi_{c1}$  &  7.82 & 1.50 &  2.30 & 0.218 & -0.480 & -0.0189 & 0.0278 \\
$h_c(2P)$  &  1.75 & 1.86 & 1.11 &  -0.0110 & -0.0810 & -0.0168 & 0.00575 \\
$\chi_{c0}(2P)$    &  0.508 & 1.16 &  0.635 & -0.0776 & -0.0531 & 0.00158 & 0.00121 \\
$\chi_{c1}(2P)$   &  0.666 & 0.104 &  0.444 & 0.0265 & -0.157 & -0.00197 & 0.0159 \\
$h_c(3P)$  &  0.173 & 0.249 & 0.207&  0.0205 & -0.0136 & -0.00470 & 0.00103 \\
$\chi_{c0}(3P)$    &  0.0731 & 0.220 &  0.170 & -0.00436 & -0.0183 & -0.000377 & 0.000541 \\
$\chi_{c1}(3P)$   &  0.0580 & 0.00784 &  0.0758 & 0.00580 & -0.0379 & -0.000542 & 0.00539 \\
\hline
\end{tabular*}
\caption{\label{tab:method1e} The branch ratios of $B_c^+\to (c\bar{c})+e^++\nu_e$ in Method I according to the power ${\vec{q}~}^n$ (in $10^{-4}$).}
\end{table}

\begin{table}[tbp]
\vspace{0.2cm}
\setlength{\tabcolsep}{0.2cm}
\centering
\begin{tabular*}{\textwidth}{@{}@{\extracolsep{\fill}}|c|ccccc|c|}
\hline
Mode& LO & 1st & 2nd & 3rd & 4th  & Total(BS) \\
\hline
{\phantom{\Large{l}}}\raisebox{+.2cm}{\phantom{\Large{j}}}
$\eta_c$  &  44.1  &  8.24 & 7.32 &  0.650 &  0.279   & 60.7\\
$J/\psi$  &  158  & 18.2  & 15.2 & 1.94 & 0.219  & 193\\
$\eta_c(2S)$  &  3.24  &  1.81 & 1.71 &  0.420 &  0.166   & 7.34\\
$\psi(2S)$  &  6.96  & 1.66  & 2.00 & 0.353 & 0.108 &  11.1\\
$\eta_c(3S)$  &  0.355  &  0.272 & 0.311 &  0.101 &  0.0475   & 1.09\\
$\psi(3S)$  &  0.651  & 0.201  & 0.365 & 0.0834 & 0.0388 &  1.34\\
$h_c$   &  14.7 & 10.2  &  5.21 &  0.688 & 0.0822 &  30.9\\
$\chi_{c0}$  &  5.20 & 7.88 &  2.03  & -0.736 & 0.0453  & 14.4\\
$\chi_{c1}$  &  7.75 & 1.35  &  2.31 & 0.307 & 0.0292  & 11.8\\
$h_c(2P)$   &  1.61 & 1.67  &  1.27 &  0.288 & 0.0448  & 4.88\\
$\chi_{c0}(2P)$  &  0.523 & 1.16  &  0.664 & 0.0211 & 0.000490  & 2.37\\
$\chi_{c1}(2P)$  &  0.650 & 0.0804  &  0.406 & 0.0353 & 0.00421  & 1.18\\
$h_c(3P)$   &  0.159 & 0.228  &  0.222 &  0.0614 & 0.0111  & 0.682\\
$\chi_{c0}(3P)$  &  0.0760 & 0.221  &  0.175 & 0.0208 & 0.000717  & 0.493\\
$\chi_{c1}(3P)$  &  0.0562 & 0.00589  &  0.0615 & 0.00427 & 0.000765  & 0.129\\

\hline
\end{tabular*}
\caption{\label{tab:method2e} The Branch Ratios of $B_c^+\to (c\bar{c})+e^++\nu_e$ in Method II (in $10^{-4}$) according to eq.~(\ref{eq:tt}).}
\end{table}

Then we calculate the decay widths of the semileptonic decay $B_c^+\to (c\bar c)+e^++\bar\nu_e$. To see the relativistic effects, we give each order expansion of the decay fractions and their sums in two methods. The results from Method I are shown in table~\ref{tab:method1e}, and the results from Method II are shown in table~\ref{tab:method2e}. In Method I, the amplitude is expanded in powers of $\vec q$ before integral, and $\vec q$ relates directly to the relative velocity $\vec{v}$ between quark and antiquark in the meson, $\vec{q}=\frac{m_1 m_2}{m_1 + m_2}\vec{v}$. The average relative velocity is not very large in a double-heavy meson, so one can make such kind of expansion. $|\vec{q}|$ is actually not a fixed quantity, whose range is from zero to infinity. We argue when $|\vec{q}|$ is large, the contribution is suppressed by the values of wave functions. But from the diagrams of wave functions in figures~\ref{fig:wfs}--\ref{fig:wfs2}, these arguments may be not very good especially for highly excited states. It may lead to the problem of convergence. For $1S$ wave ($\eta_c$ and $J/\psi$) as the final states, the leading order ($\vec q~^0$) makes dominant contribution. For the channels $2S$ or $3S$ as the final states, the leading order makes largest contribution. Meanwhile, the first and second order relativistic corrections, $\vec{q}~^1$ and $\vec{q}~^2$ items, provide comparable contributions to the leading order. For some of $P$ wave final states, the leading order ($\vec q~^2$) which is non-relativistic contributions, may be not dominant. Instead, the first order relativistic corrections make the dominant contributions. For both the orbitally excited states and the radially ones as the final states, the relativistic effects are huge.

In table~\ref{tab:method1e1}, the comparisons of the branch ratios obtained by different ways are given, where {\bf sum} column means the sum of all of expansion orders; {\bf BS} column means the results by the BS method without expansion; {\bf NR} means the result by the non-relativistic wave function and the leading order expansion of the amplitude.  For the $B_c$ decays to the $1S$ charmonium, the non-relativistic results (NR) are very close to those by the BS method. But for the orbitally or radially excited charmonium as final state, the NR results are closer to the leading orders ($\vec q~^0$) rather than the BS results. Therefore, we conclude that the non-relativistic approach may be a good approximation for the ground heavy meson, but it is unreasonable for the excited state. The sum of all of expansion orders are not exactly equal to the BS results, which means the convergence rate is not fast enough. In the last column, we show the differences between them, and conclude that high order $\vec{q}~^n$ ($n>6$) contributions are still important for the radially excited states. We provide supplementary Method II to see the relativistic effects. The details are given in section~\ref{sec:method2}, and numerical results are shown in table~\ref{tab:method2e}. The main difference between two methods is the treatment of $\vec {q}~^2$ in $\omega_1$ and $\omega_2$. The amplitude is expanded in powers of $|\vec {q}|/m_i$ or $|\vec {q}|/M$ in Method I, which suffers from the problem of convergence. In the Method II, the relativistic correction is based on the wave functions. It is no problem with convergence, but suffers from the overlapping problem. 

\begin{table}[tbp]
\vspace{0.2cm}
\setlength{\tabcolsep}{0.2cm}
\centering
\begin{tabular*}{\textwidth}{@{}@{\extracolsep{\fill}}|c|cccc|c|}
\hline
Mode&${\vec{q}~}^0$&sum&BS&NR&$\mathrm{\frac{BS-sum}{BS}}$\\
\hline
{\phantom{\Large{l}}}\raisebox{+.2cm}{\phantom{\Large{j}}}
$\eta_c$  &  47.4  &  58.6  &  60.7 & 56.7 & 3.4\%\\
$J/\psi$  &  157  &197 & 193 & 188 & -1.8\%\\
$\eta_c(2S)$  &  3.66  &  6.90  &  7.34 & 4.48 & 6.0\%\\
$\psi(2S)$  &  6.88  & 12.1 & 11.1 & 8.40 & -8.8\%\\
$\eta_c(3S)$  &  0.408  &  0.995  &  1.09 & 0.509 & 8.7\%\\
$\psi(3S)$  &  0.652  & 1.61 & 1.34 & 0.806 & -20\%\\
\hline
Mode&${\vec{q}~}^2$ &sum&BS&NR&$\mathrm{\frac{BS-sum}{BS}}$\\
\hline

$h_c$   &  15.4  & 30.0 & 30.9 & 18.8 & 2.9\%\\
$\chi_{c0}$  &  5.13 & 13.7 & 14.4 & 6.28 & 4.8\%\\
$\chi_{c1}$  &  7.82 & 11.4 & 11.8 & 9.60 & 2.8\%\\
$h_c(2P)$  &  1.75  & 4.62 & 4.88 & 2.18 & 5.3\%\\
$\chi_{c0}(2P)$    &  0.508 & 2.17 & 2.37 & 0.633 & 8.4\%\\
$\chi_{c1}(2P)$   &  0.666  & 1.10 & 1.18 & 0.853 & 7.2\%\\
$h_c(3P)$  &  0.173 & 0.633 & 0.682 & 0.220 & 7.1\%\\
$\chi_{c0}(3P)$    &  0.0731 & 0.440 & 0.493 & 0.0923 & 11\%\\
$\chi_{c1}(3P)$   &  0.0580 & 0.114 & 0.129 & 0.0735 & 11\%\\

\hline
\end{tabular*}
\caption{\label{tab:method1e1} Comparisons of the branch ratios of $B_c^+\to (c\bar{c})+e^++\nu_e$ obtained by different ways, where  $\boldsymbol{\vec{q}~^0}$ means the leading order result; {\bf sum} means the sum of all of expansion orders; {\bf BS} means the result by BS method without expansion, and {\bf NR} means the result by the non-relativistic wave function and the leading order expansion of the amplitude (in $10^{-4}$ except the last column).}
\end{table}

For the highly excited charmonium as final states, the first order relativistic corrections may be larger than the leading order. This special behavior can be understood. In the transition of $B_c\to\chi_{c0}$, for example, the ratio of 1st order expansion over LO, $T_1/T_0\approx 80\%$ (figure~\ref{fig:ffg}) in the level of form factor. As a rough estimate, see eq.~(\ref{eq:tt}), $(T_0T_1^*+T_0^*T_1)/T_0^2\approx 160\%$ at most, so the large contribution of 1st order expansion is reasonable. For the process of $B_c\to\chi_{c1}$, the ratio of 1st order expansion form factor over the LO form factor is as high as 80\%. However, the 1st order expansion of decay width is much smaller than LO decay width. It means there is cancellation when calculating $T_0T_1^*+T_0^*T_1$. Therefore it is not enough to just provide the relativistic corrections of form factors. To investigate the relativistic effects of the decay processes, the relativistic corrections should be given not only in the level of form factors, but also in the level of decay widths.

To see the whole relativistic effects in the level of decay widths, we define the ratios $\frac{BS-LO}{BS}$ as the relativistic effects, and show them in tables~\ref{tab:comp1}--\ref{tab:comp2}. Except the $\eta_c$ case, two methods give consistent results of the relativistic effects. The Method I suffers from the problem of convergence of $|\vec {q}|$ expansion, but it dose not affect the leading order; the Method II suffers from the overlapping problem because $\omega_i$ ($i=1,~2$) has no expansion on $|\vec {q}|$. But in leading order, if we use the approximate formula $\omega_i=m_i+\vec{q}^{\:2}/{2m_i}$, the difference between two methods is left to the $\vec {q}^{~2}$ order. Therefore both methods give accurate estimate of the relativistic corrections. For the ground states, the relativistic effects are about $20\%$; for the excited states, we reach much larger relativistic effects. The smallest one is $\chi_{c1}$ case, which is $33\%$. The largest one is $\chi_{c0}(3P)$ case, which is $85\%$.

\begin{table}[tbp]
\vspace{0.2cm}
\setlength{\tabcolsep}{0.2cm}
\centering
\begin{tabular*}{\textwidth}{@{}@{\extracolsep{\fill}}|c|cccccc|}
\hline
Method & $\eta_c$ & $J/\psi$ & $\eta_c(2S)$ & $\psi(2S)$ & $\eta_c(3S)$ & $\psi(3S)$\\
\hline
~I~ &  21.9  & 18.8  & 50.2 & 38.0 & 62.5 & 51.3 \\
~II~ & 27.3  & 18.5 & 55.8 & 37.2 & 67.3 & 51.5 \\
\hline
\end{tabular*}
\caption{\label{tab:comp1} The relativistic effects of $B_c^+\to (c\bar{c})e^+\nu_e$: $\frac{BS-LO}{BS}$ from two methods (in $\%$).}
\end{table}

\begin{table}[tbp]
\vspace{0.2cm}
\setlength{\tabcolsep}{0.2cm}
\centering
\begin{tabular*}{\textwidth}{@{}@{\extracolsep{\fill}}|c|ccccccccc|}
\hline
Method &  $h_c$ & $\chi_{c0}$ & $\chi_{c1}$ & $h_c(2P)$ & $\chi_{c0}(2P)$ & $\chi_{c1}(2P)$ & $h_c(3P)$ & $\chi_{c0}(3P)$ & $\chi_{c1}(3P)$\\
\hline
~I~ &  50.2 & 64.4 & 33.7 & 64.1 & 78.5 & 43.3 & 74.6 & 85.2 & 54.9\\
~II~ & 52.5 & 63.9 & 34.0 & 67.0 & 77.9 & 44.7 & 76.7 & 84.6 & 56.3\\
\hline
\end{tabular*}
\caption{\label{tab:comp2} The relativistic effects of $B_c^+\to (c\bar{c})e^+\nu_e$: $\frac{BS-LO}{BS}$ from two methods (in $\%$).}
\end{table}

We also calculate the processes $B_c^+\to (c\bar{c})\tau^+\nu_\tau$, as tables~\ref{tab:method1t}--\ref{tab:comp2tau} shown.

\begin{table}[tbp]
\vspace{0.2cm}
\setlength{\tabcolsep}{0.2cm}
\centering
\begin{tabular*}{\textwidth}{@{}@{\extracolsep{\fill}}|c|ccccccc|}
\hline
Mode&${\vec{q}~}^0$ &${\vec{q}~}^1$&${\vec{q}~}^2$&${\vec{q}~}^3$&${\vec{q}~}^4$&${\vec{q}~}^5$&${\vec{q}~}^6$\\
\hline
{\phantom{\Large{l}}}\raisebox{+.2cm}{\phantom{\Large{j}}}
$\eta_c$  &  158  &  19.0 & 20.0 &  -5.16 &  0.109  &  -0.333 & 0.0832 \\
$J/\psi$  &  405  & 33.7 & 48.4 & 1.20 & 0.842 & 0.00264 & 0.0264 \\
$\eta_c(2S)$  &  2.01  &  0.870 & 1.35 &  0.0262 &  0.116  &  -0.0543 & 0.00792 \\
$\psi(2S)$  &  2.99  & 0.490  & 1.66 & 0.121 & 0.183 & 0.00833 & 0.00288 \\
$\eta_c(3S)$  &  0.0430  &  0.0187 & 0.0534 &  0.00462 &  0.0120  &  -0.00245 & 0.000284 \\
$\psi(3S)$  &  0.0493  & 0.00698  & 0.0571 & 0.00450 & 0.0149 & 0.000764 & 0.000151 \\
\hline
Mode&${\vec{q}~}^2$&${\vec{q}~}^3$&${\vec{q}~}^4$&${\vec{q}~}^5$&${\vec{q}~}^6$&${\vec{q}~}^7$ &${\vec{q}~}^8$\\
\hline

$h_c$   &  9.00 & 6.28 &  5.61 &  0.310 & 0.0620 & 0.00222 & 0.00155\\
$\chi_{c0}$  &  4.12 & 7.74 &  5.65 & -0.886 & -0.0110 & 0.00549 & 0.000419 \\
$\chi_{c1}$  &  6.84 & 0.915 &  3.14 & 0.226 & -0.522 & -0.0139 & 0.0232 \\
$h_c(2P)$  &  0.141 & 0.163 & 0.292 &  0.0292 & 0.00428 & -0.000102 & 0.000126 \\
$\chi_{c0}(2P)\times10^{-2}$    &  4.98 & 15.5 &  26.7 & 0.272 & -0.397 & -0.00241 & 0.00344 \\
$\chi_{c1}(2P)$   &  0.114 & 0.0124 &  0.157 & 0.00599 & -0.0520 & -0.000127 & 0.00480 \\
$h_c(3P)\times10^{-3}$  &  0.444 & 0.779 & 4.01& 0.270 & -0.000403 & 0.00639 & 0.00111 \\
$\chi_{c0}(3P)\times10^{-3}$    &  0.721 & 3.36 &  12.5 & 0.278 & -0.406 & -0.00608 & 0.00385 \\
$\chi_{c1}(3P)\times10^{-3}$   &  0.443 & 0.0467 &  2.36 & 0.0544 & -1.22 & -0.00180& 0.165 \\
\hline
\end{tabular*}
\caption{\label{tab:method1t} The branch ratios of $B_c^+\to (c\bar{c})+\tau^++\nu_\tau$ in Method I according to the power ${\vec{q}~}^n$ (in $10^{-5}$).}
\end{table}

\begin{table}[tbp]
\vspace{0.2cm}
\setlength{\tabcolsep}{0.2cm}
\centering
\begin{tabular*}{\textwidth}{@{}@{\extracolsep{\fill}}|c|ccccc|c|}
\hline
Mode& LO & 1st & 2nd & 3rd & 4th &  Total(BS) \\
\hline
{\phantom{\Large{l}}}\raisebox{+.2cm}{\phantom{\Large{j}}}
$\eta_c$  &  152  &  14.6 & 24.5 &  1.18 &  0.953 & 193\\
$J/\psi$  &  406  & 30.5 & 40.9 & 3.57 & 0.507  & 481\\
$\eta_c(2S)$  &  1.89  &  0.712 & 1.33 &  0.234 &  0.199 & 4.36\\
$\psi(2S)$  &  2.96  & 0.422 & 1.30 & 0.128 & 0.0991  & 4.91\\
$\eta_c(3S)$  &  0.0417  &  0.0159 & 0.0473 &  0.00865 &  0.0115 & 0.125\\
$\psi(3S)$  &  0.0464  & 0.00548 & 0.0418 & 0.00351 & 0.00707  & 0.104\\
$h_c$   &  8.85 & 6.16 &  5.99 &  0.523 & 0.0664  & 21.6\\
$\chi_{c0}$  &  4.15 & 7.78 &  5.86 & -0.657 & 0.0296 &  17.2\\
$\chi_{c1}$  &  6.82 & 0.850 &  3.01 & 0.233 & 0.0177 &  10.9\\
$h_c(2P)$   &  0.136 & 0.157 &  0.302 &  0.0374 & 0.00715  & 0.640\\
$\chi_{c0}(2P)$  &  0.0509 & 0.157 &  0.278 & 0.0125 & 0.000336 &  0.497\\
$\chi_{c1}(2P)$  &  0.112 & 0.0104 &  0.131 & 0.00457 & 0.000899 &  0.260\\
$h_c(3P)$   &  0.000446 & 0.000779 &  0.00408 &  0.000274 & 0.0000527  & 0.00563\\
$\chi_{c0}(3P)$  & 0.000748 & 0.00343 & 0.0127 & 0.000568 & 0.0000205 &  0.0175\\
$\chi_{c1}(3P)$  &  0.000432 & 0.0000405 &  0.00169 & 0.0000267 & 0.00000797 &  0.00220\\
\hline
\end{tabular*}
\caption{\label{tab:cctau} The decay fractions of the $B_c^+\to (c\bar{c})+\tau^++\nu_\tau$ in Method II according to eq.~(\ref{eq:tt}) (in $10^{-5}$).}
\end{table}

\begin{table}[tbp]
\vspace{0.2cm}
\setlength{\tabcolsep}{0.2cm}
\centering
\begin{tabular*}{\textwidth}{@{}@{\extracolsep{\fill}}|c|cccc|c|}
\hline
Mode&${\vec{q}~}^0$ &sum&BS&NR&$\mathrm{\frac{BS-sum}{BS}}$\\
\hline
{\phantom{\Large{l}}}\raisebox{+.2cm}{\phantom{\Large{j}}}
$\eta_c$  &  158  &  191  &  193 & 189 & 1.0\%\\
$J/\psi$  &  405  & 489 & 481 & 485 & -1.6\%\\
$\eta_c(2S)$  &  2.01 & 4.33 &  4.36 & 2.46 & 0.7\%\\
$\psi(2S)$  &  2.99  & 5.45 & 4.91  & 3.65 & -11\%\\
$\eta_c(3S)$  &  0.0430  & 0.130 &  0.125 & 0.0536 & -4.0\%\\
$\psi(3S)$  & 0.0495  & 0.134 & 0.104 & 0.0609 & -29\%\\
\hline
Mode&${\vec{q}~}^2$&sum&BS&NR&$\mathrm{\frac{BS-sum}{BS}}$\\
\hline

$h_c$   &  9.00 & 21.3 & 21.6 & 11.0 & 1.4\%\\
$\chi_{c0}$  &  4.12 & 16.6 & 17.2 & 5.04 & 3.1\%\\
$\chi_{c1}$  &  6.86 & 10.6 & 10.9 & 8.39 & 2.9\%\\
$h_c(2P)$  &  0.141 & 0.628 & 0.640 & 0.175 & 1.9\%\\
$\chi_{c0}(2P)$    &  0.0498 & 0.470 & 0.497 & 0.0621 & 5.3\%\\
$\chi_{c1}(2P)$   &  0.114 &  0.241 & 0.260 & 0.153 & 7.0\%\\
$h_c(3P)$  &  0.000444 & 0.00551 & 0.00563 & 0.000562 & 2.1\%\\
$\chi_{c0}(3P)$    &  0.000721 & 0.0164 & 0.0175 & 0.000911 & 6.1\%\\
$\chi_{c1}(3P)$   &  0.000443 & 0.00185 & 0.00220 & 0.000562 & 16\%\\

\hline
\end{tabular*}
\caption{\label{tab:corr} Comparisons of the branch ratios of $B_c^+\to (c\bar{c})+\tau^++\nu_\tau$ obtained by different ways, where $\boldsymbol{\vec{q}~^0}$ means the leading order result; {\bf sum} means the summed values of expanded items; {\bf BS} means the result by BS method without expansion, and {\bf NR} means the result by the non-relativistic wave function and the leading order expansion of the amplitude (in $10^{-5}$ except the last column).}
\end{table}

\begin{table}[tbp]
\vspace{0.2cm}
\setlength{\tabcolsep}{0.2cm}
\centering
\begin{tabular*}{\textwidth}{@{}@{\extracolsep{\fill}}|c|cccccc|}
\hline
Method & $\eta_c$ & $J/\psi$ & $\eta_c(2S)$ & $\psi(2S)$ & $\eta_c(3S)$ & $\psi(3S)$ \\
\hline
~I~ &  18.4  & 15.8  & 53.8 & 39.1 & 65.6 & 52.6 \\
~II~ & 21.3  & 15.7 & 56.7 & 39.7 & 66.6 & 55.3 \\
\hline
\end{tabular*}
\caption{\label{tab:comp1tau} The relativistic effects of $B_c^+\to (c\bar{c})\tau^+\nu_\tau$ : $\frac{BS-LO}{BS}$ from two methods (in $\%$).}
\end{table}

\begin{table}[tbp]
\vspace{0.2cm}
\setlength{\tabcolsep}{0.2cm}
\centering
\begin{tabular*}{\textwidth}{@{}@{\extracolsep{\fill}}|c|ccccccccc|}
\hline
Method &  $h_c$ & $\chi_{c0}$ & $\chi_{c1}$ & $h_c(2P)$ & $\chi_{c0}(2P)$ & $\chi_{c1}(2P)$ & $h_c(3P)$ & $\chi_{c0}(3P)$ & $\chi_{c1}(3P)$\\
\hline
~I~ &  58.3 & 76.0 & 37.2 & 78.0 & 90.0 & 56.2 & 92.1 & 95.9 & 79.8\\
~II~ & 59.0 & 75.8 & 37.6 & 78.8 & 89.8 & 56.8 & 92.1 & 95.7 & 80.3\\
\hline
\end{tabular*}
\caption{\label{tab:comp2tau} The relativistic effects of $B_c^+\to (c\bar{c})\tau^+\nu_\tau$ : $\frac{BS-LO}{BS}$ from two methods (in $\%$).}
\end{table}

\section{\label{sec:conclude} Conclusion}

In this paper, we choose the instantaneous BS method to calculate the semileptonic $B_c$ decays to charmonium, whose final states include $1S$, $2S$, $3S$, $1P$, $2P$ and $3P$. We focus on the relativistic effects. Two methods are provided to see the relativistic corrections. In the first method, the amplitude is expanded in powers of $\vec q$ which is the relative momentum between the quark and the antiquark. We find this widely used method suffers from the problem of convergence though the quark and antiquark are heavy, and the high order $\vec{q}~^n$ items still have sizable contributions. The other method is based on the relativistic wave functions. In this method there is no problem of convergence, while it suffers from the overlapping problem. Both methods give accurate and consistent leading order (non-relativistic) contributions. In another words, both methods provide accurate relativistic corrections. We find that for the semileptonic $B_c$ decays, the relativistic effects are about $20\%$ when final states are $1S$ charmonium ($\eta_c$ and $J/\psi$), but the relativistic effects are much higher for the excited final states. First of all, the $nP$ final state has larger relativistic corrections than the corresponding $nS$ state; secondly, the $(n+1)S$ state has larger relativistic corrections than the corresponding $nS$ state; thirdly, the $(n+1)P$ state has larger relativistic corrections than the corresponding $nP$ state. Therefore, we conclude that even though the higher excited state has higher mass, its relativistic effect is larger. A relativistic method is needed to deal with a problem including a excited state, though the corresponding quark and antiquark are heavy.

\appendix

\section{\label{appendix} Equation and solution for heavy mesons}
BS equation for a quark-antiquark bound state generally is written as \cite{Chang:2014jca}
\begin{equation}
(\slashed p_1-m_1)\chi_P(q)(\slashed p_2+m_2)=\mathrm i\int\frac{\mathrm d^4k}{(2\pi)^4}V(P,k,q)\chi_P(k),
\end{equation}
where $p_1,p_2;m_1,m_2$ are the momenta and masses of the quark and antiquark, respectively; $\chi_P(q)$ is the BS wave function with the total momentum $P$ and relative momentum $q$; $V(P,k,q)$ is the kernel between the quark-antiquark in the bound state. $P$ and $q$ are defined as
\begin{equation}
\begin{aligned}
&\vec p_1=\alpha_1\vec P+\vec q,\quad\alpha_1=\frac{m_1}{m_1+m_2},\\
&\vec p_2=\alpha_2\vec P+\vec q,\quad\alpha_2=\frac{m_2}{m_1+m_2}.
\end{aligned}
\end{equation}
The instantaneous kernel has the following form
\begin{equation}
V(P,k,q)\sim V(|k-q|),
\end{equation}
especially when the two constituents of meson are very heavy.
We divide the relative momentum $q$ into two parts, $q_{P_{||}}$ and $q_{P_\perp}$, a parallel part and an orthogonal one to $P$, respectively
\begin{equation}
q^\mu=q_{P_{||}}^\mu+q_{P_\perp}^\mu,
\end{equation}
where $q_{P_{||}}^\mu\equiv(P\cdot q/M^2)P^\mu,~q_{P_\perp}^\mu\equiv q^\mu-q_{P_{||}}^\mu$, and $M$ is the mass of the relevant meson. Correspondingly, we have two Lorentz-invariant variables
\begin{equation}
q_P=\frac{P\cdot q}{M},~q_{P_T}=\sqrt{q_P^2-q^2}=\sqrt{-q_{P_\perp}^2}.
\end{equation}
If we introduce two notations as below
\begin{equation}
\begin{aligned}
&\eta(q_{P_\perp}^\mu)\equiv \int\frac{k_{P_T}^2\ud k_{P_T}\ud s}{(2\pi)^2}V(k_{P_\perp},s,q_{P_\perp})\varphi(k_{p_\perp}^\mu),\\
&\varphi(q_{p_\perp}^\mu)\equiv\mathrm i\int\frac{\mathrm dq_P}{2\pi}\chi_P(q_{P_{||}}^\mu,q_{P_\perp}^\mu).
\end{aligned}
\end{equation}
Then the BS equation can take the form as follow
\begin{equation}
\chi_P(q_{P_{||}}^\mu,q_{P_\perp}^\mu)=S_1(p_1^\mu)\eta(q_{P_\perp}^\mu)S_2(p_2^\mu).
\end{equation}
The propagators of the relevant particles with masses $m_1$ and $m_2$ can be decomposed as
\begin{equation}
S_i(p_i^\mu)=\frac{\Lambda_{i_P}^+(q_{P_\perp}^\mu)}{J(i)q_P+\alpha_iM-\omega_{i_P}+\mathrm i\varepsilon}+\frac{\Lambda_{i_P}^-(q_{P_\perp}^\mu)}{J(i)q_P+\alpha_iM+\omega_{i_P}-\mathrm i\varepsilon},
\end{equation}
with
\begin{equation}
\begin{aligned}
\omega_{i_P}&=\sqrt{m_i^2+q_{P_T}^2},\\
\Lambda_{i_P}^{\pm}(q_{P_\perp}^\mu)&=\frac{1}{2\omega_{i_P}}\left[\frac{\slashed P}{M}\omega_{i_P}\pm J(i)(\slashed q_{P_\perp}+m_i)\right],
\end{aligned}
\end{equation}
where $i=1,2$ for quark and antiquark, respectively, and $J(i)=(-1)^{i+1}$.

Then the instantaneous Bethe-Salpeter equation can be decomposed into the coupled equations
\begin{equation}
\begin{aligned}
(M-\omega_{1p}-\omega_{2p})\varphi^{++}(q_{P_\perp})&=\Lambda_1^+(P_{1p_\perp})\eta(q_{P_\perp})\Lambda_2^+(P_{2p_\perp}),\\
(M+\omega_{1p}+\omega_{2p})\varphi^{--}(q_{P_\perp})&=-\Lambda_1^-(P_{1p_\perp})\eta(q_{P_\perp})\Lambda_2^-(P_{2p_\perp}),\\
\varphi^{+-}(q_{P_\perp})=0,\quad&\qquad\varphi^{-+}(q_{P_\perp})=0.
\end{aligned}\label{eq:phi}
\end{equation}

The instantaneous Bethe-Salpeter wave function for $1^-$ states mesons has the general form \cite{Wang:2005qx}
\begin{equation}
\begin{aligned}
\varphi_{1^-}(q_\perp)&=(q_\perp\cdot\epsilon)\left[g_1(q_\perp)+\frac{\slashed{P}}{M}g_2(q_\perp)
+\frac{\slashed{q}_\perp}{M}g_3(q_\perp)+\frac{\slashed{P}\slashed{q}_\perp}{M^2}g_4(q_\perp)\right]+\\
&\quad+M\slashed\epsilon\left[g_5(q_\perp)+\frac{\slashed{P}}{M}g_6(q_\perp)
+\frac{\slashed{q}_\perp}{M}g_7(q_\perp)+\frac{\slashed{P}\slashed{q}_\perp}{M^2}g_8(q_\perp)\right],
\end{aligned}
\end{equation}
with
\begin{equation}
\begin{aligned}
g_1(q_\perp)&=\frac{q_\perp^2g_3(\omega_1+\omega_2)+2M^2g_5\omega_2}{M(m_1\omega_2+m_2\omega_1)},\\
g_2(q_\perp)&=\frac{q_\perp^2g_4(\omega_1+\omega_2)+2M^2g_6\omega_2}{M(m_1\omega_2+m_2\omega_1)},\\
g_7(q_\perp)&=\frac{M(\omega_1-\omega_2)}{m_1\omega_2+m_2\omega_1}g_5,\\
g_8(q_\perp)&=\frac{M(\omega_1+\omega_2)}{m_1\omega_2+m_2\omega_1}g_6.\label{eq:1-}
\end{aligned}
\end{equation}
The wave function corresponding to the positive projection has the form
\begin{equation}
\begin{aligned}
\varphi_{1^-}^{++}(q_\perp)&=(q_\perp\cdot\epsilon)\left[B_1(q_\perp)+\frac{\slashed P}{M}B_2(q_\perp)+\frac{\slashed q_\perp}{M}B_3(q_\perp)+\frac{\slashed P\slashed q_\perp}{M^2}B_4(q_\perp)\right]+\\
&\quad+M\slashed\epsilon\left[B_5(q_\perp)+\frac{\slashed P}{M}B_6(q_\perp)+\frac{\slashed q_\perp}{M}B_7(q_\perp)+\frac{\slashed P\slashed q_\perp}{M^2}B_8(q_\perp)\right],
\end{aligned}
\end{equation}
where
\begin{equation}
\begin{aligned}
B_1&=\frac{1}{2M(m_1\omega_2+m_2\omega_1)}[(\omega_1+\omega_2)q_{\perp}^2g_3+(m_1+m_2)q_{\perp}^2g_4+2M^2\omega_2g_5-2M^2m_2g_6],\\
B_2&=\frac{1}{2M(m_1\omega_2+m_2\omega_1)}[(m_1-m_2)q_{\perp}^2g_3+(\omega_1-\omega_2)q_{\perp}^2g_4+2M^2\omega_2g_6-2M^2m_2g_5],\\
B_3&=\frac{1}{2}\left[g_3+\frac{m_1+m_2}{\omega_1+\omega_2}g_4-\frac{2M^2}{m_1\omega_2+m_2\omega_1}g_6\right],\\
B_4&=\frac{1}{2}\left[\frac{\omega_1+\omega_2}{m_1+m_2}g_3+g_4-\frac{2M^2}{m_1\omega_2+m_2\omega_1}g_5\right],\\
B_5&=\frac{1}{2}\left[g_5-\frac{\omega_1+\omega_2}{m_1+m_2}g_6\right],\qquad B_6=\frac{1}{2}\left[-\frac{m_1+m_2}{\omega_1+\omega_2}g_5+g_6\right],\\
B_7&=\frac{M}{2}\frac{\omega_1-\omega_2}{m_1\omega_2+m_2\omega_1}\left[g_5-\frac{\omega_1+\omega_2}{m_1+m_2}g_6\right],\\
B_8&=\frac{M}{2}\frac{m_1+m_2}{m_1\omega_2+m_2\omega_1}\left[-g_5+\frac{\omega_1+\omega_2}{m_1+m_2}g_6\right].
\end{aligned}
\end{equation}
If the masses of the quark and antiquark are equal, the normalization condition reads as
\begin{equation}
\int\frac{\ud\vec q}{(2\pi)^3}\frac{16\omega_1\omega_2}{3}\left\{3g_5g_6\frac{M^2}{2m_1\omega_2}+\frac{\vec q^2}{2m_1\omega_2}\left[g_4g_5-g_3\left(g_4\frac{\vec q^2}{M^2}+g_6\right)\right]\right\}=2M.
\end{equation}

The instantaneous Bethe-Salpeter wave function for $1^{+-}$ states mesons has the form \cite{Wang:2007av}
\begin{equation}
\varphi_{1^{+-}}(q_\perp)=(q_\perp\cdot\epsilon)\left[h_1(q_\perp)+\frac{\slashed P}{M}h_2(q_\perp)+\frac{\slashed q_\perp}{M}h_3(q_\perp)+\frac{\slashed P\slashed q_\perp}{M^2}h_4(q_\perp)\right]\gamma_5,\label{eq:wf1+-}
\end{equation}
with
\begin{equation}
\begin{aligned}
h_3(q_\perp)&=-\frac{M(\omega_1-\omega_2)}{m_1\omega_2+m_2\omega_1}h_1,\\
h_4(q_\perp)&=-\frac{M(\omega_1+\omega_2)}{m_1\omega_2+m_2\omega_1}h_2.
\end{aligned}
\end{equation}
The wave function corresponding to the positive projection has the form
\begin{equation}
\varphi_{1^{+-}}^{++}(q_{\perp})=q_\perp\cdot\epsilon\left[C_1(q_{\perp})+\frac{\slashed P}{M}C_2(q_{\perp})+\frac{\slashed q_{\perp}}{M}C_3(q_{\perp})+\frac{\slashed P\slashed q_{\perp}}{M^2}C_4(q_{\perp})\right]\gamma^5,
\end{equation}
where
\begin{equation}
\begin{aligned}
C_1&=\frac{1}{2}\left[h_1+\frac{\omega_1+\omega_2}{m_1+m_2}h_2\right],\\
C_2&=\frac{1}{2}\left[\frac{m_1+m_2}{\omega_1+\omega_2}h_1+h_2\right],\\
C_3&=-\frac{M(\omega_1-\omega_2)}{m_1\omega_2+m_2\omega_1}C_1,\\
C_4&=-\frac{M(m_1+m_2)}{m_1\omega_2+m_2\omega_1}C_1.
\end{aligned}
\end{equation}
If the masses of the quark and antiquark are equal, the normalization condition reads as
\begin{equation}
\int\frac{\ud\vec q}{(2\pi)^3}\frac{4h_1h_2\omega_1\vec q^2}{3m_1}=M.
\end{equation}

The instantaneous Bethe-Salpeter wave function for $0^{+}$ states mesons has the form \cite{Wang:2007av}
\begin{equation}
\varphi_{0^{+}}(q_\perp)=M\left[\frac{\slashed q_\perp}{M}\phi_1(q_\perp)+\frac{\slashed P\slashed q_\perp}{M^2}\phi_2(q_\perp)+\phi_3(q_\perp)+\frac{\slashed P}{M}\phi_4(q_\perp)\right]\label{eq:wf0++},
\end{equation}
with
\begin{equation}
\begin{aligned}
\phi_3(q_\perp)&=\frac{q_\perp^2(\omega_1+\omega_2)}{M(m_1\omega_2+m_2\omega_1)}\phi_1,\\
\phi_4(q_\perp)&=\frac{q_\perp^2(\omega_1-\omega_2)}{M(m_1\omega_2+m_2\omega_1)}\phi_2.
\end{aligned}
\end{equation}
The wave function corresponding to the positive projection has the form
\begin{equation}
\varphi_{0^{+}}^{++}(q_{\perp})=D_1(q_{\perp})+\frac{\slashed P}{M}D_2(q_{\perp})+\frac{\slashed q_{\perp}}{M}D_3(q_{\perp})+\frac{\slashed P\slashed q_{\perp}}{M^2}D_4(q_{\perp}),
\end{equation}
where
\begin{equation}
\begin{aligned}
D_1&=\frac{(\omega_1+\omega_2)q_\perp^2}{2(m_1\omega_2+m_2\omega_1)}\left[\phi_1+\frac{m_1+m_2}{\omega_1+\omega_2}\phi_2\right],\\
D_2&=\frac{(m_1-m_2)q_\perp^2}{2(m_1\omega_2+m_2\omega_1)}\left[\phi_1+\frac{m_1+m_2}{\omega_1+\omega_2}\phi_2\right],\\
D_3&=\frac{M}{2}\left[\phi_1+\frac{m_1+m_2}{\omega_1+\omega_2}\phi_2\right],\\
D_4&=\frac{M}{2}\left[\frac{\omega_1+\omega_2}{m_1+m_2}\phi_1+\phi_2\right].
\end{aligned}
\end{equation}
If the masses of the quark and antiquark are equal, the normalization condition reads as
\begin{equation}
\int\frac{\ud\vec q}{(2\pi)^3}\frac{4\phi_1\phi_2\omega_1\vec q^2}{m_1}=M.
\end{equation}

The instantaneous Bethe-Salpeter wave function for $1^{++}$ states mesons has the form \cite{Wang:2007av}
\begin{equation}
\varphi_{1^{++}}(q_\perp)=\mathrm i\varepsilon_{\mu\nu\alpha\beta}\frac{P^\nu}{M}q_\perp^\alpha\epsilon^\beta\gamma^\mu\left[\psi_1(q_\perp)+\frac{\slashed P}{M}\psi_2(q_\perp)+\frac{\slashed q_\perp}{M}\psi_3(q_\perp)+\frac{\slashed P\slashed q_\perp}{M^2}\psi_4(q_\perp)\right]\label{eq:wf1++},
\end{equation}
with
\begin{equation}
\begin{aligned}
\psi_3(q_\perp)&=-\frac{M(\omega_1-\omega_2)}{m_1\omega_2+m_2\omega_1}\psi_1,\\
\psi_4(q_\perp)&=-\frac{M(\omega_1+\omega_2)}{m_1\omega_2+m_2\omega_1}\psi_2.
\end{aligned}
\end{equation}
The wave function corresponding to the positive projection has the form
\begin{equation}
\varphi_{1^{++}}^{++}(q_{\perp})=\mathrm i\varepsilon_{\mu\nu\alpha\beta}\frac{P^\nu}{M}q_\perp^\alpha\epsilon^\beta\gamma^\mu\left[F_1(q_{\perp})+\frac{\slashed P}{M}F_2(q_{\perp})+\frac{\slashed q_{\perp}}{M}F_3(q_{\perp})+\frac{\slashed P\slashed q_{\perp}}{M^2}F_4(q_{\perp})\right],
\end{equation}
where
\begin{equation}
\begin{aligned}
F_1&=\frac{1}{2}\left[\psi_1+\frac{\omega_1+\omega_2}{m_1+m_2}\psi_2\right],\\
F_2&=-\frac{1}{2}\left[\frac{m_1+m_2}{\omega_1+\omega_2}\psi_1+\psi_2\right],\\
F_3&=\frac{M(\omega_1-\omega_2)}{m_1\omega_2+m_2\omega_1}F_1,\\
F_4&=-\frac{M(m_1+m_2)}{m_1\omega_2+m_2\omega_1}F_1.
\end{aligned}
\end{equation}
If the masses of the quark and antiquark are equal, the normalization condition reads as
\begin{equation}
\int\frac{\ud\vec q}{(2\pi)^3}\frac{8\psi_1\psi_2\omega_1\vec q^2}{3m_1}=M.
\end{equation}

\acknowledgments
This work was supported in part by the National Natural Science
Foundation of China (NSFC) under Grant No.~11405037, No.~11575048 and No.~11505039.


\bibliographystyle{JHEP}
\bibliography{geng1.bib} 

\providecommand{\href}[2]{#2}\begingroup\raggedright\begin{thebibliography}{10}

\bibitem{Zhu:2017lqu}
R.~Zhu, Y.~Ma, X.-L. Han and Z.-J. Xiao, \emph{{Relativistic corrections to the
  form factors of $B_c$ into $S$-wave Charmonium}},
  \href{https://doi.org/10.1103/PhysRevD.95.094012}{\emph{Phys. Rev.}
  {\bfseries D95} (2017) 094012}
  [\href{https://arxiv.org/abs/1703.03875}{{\ttfamily 1703.03875}}].

\bibitem{Wang:2017bgv}
W.~Wang, J.~Xu, D.~Yang and S.~Zhao, \emph{{Relativistic corrections to
  light-cone distribution amplitudes of S-wave B$_{c}$ mesons and heavy
  quarkonia}}, \href{https://doi.org/10.1007/JHEP12(2017)012}{\emph{JHEP}
  {\bfseries 12} (2017) 012}
  [\href{https://arxiv.org/abs/1706.06241}{{\ttfamily 1706.06241}}].

\bibitem{Braaten:2002fi}
E.~Braaten and J.~Lee, \emph{{Exclusive double charmonium production from e+ e-
  annihilation into a virtual photon}},
  \href{https://doi.org/10.1103/PhysRevD.72.099901,
  10.1103/PhysRevD.67.054007}{\emph{Phys. Rev.} {\bfseries D67} (2003) 054007}
  [\href{https://arxiv.org/abs/hep-ph/0211085}{{\ttfamily hep-ph/0211085}}].

\bibitem{Liu:2002wq}
K.-Y. Liu, Z.-G. He and K.-T. Chao, \emph{{Problems of double charm production
  in e+ e- annihilation at s**(1/2) = 10.6-GeV}},
  \href{https://doi.org/10.1016/S0370-2693(03)00176-X}{\emph{Phys. Lett.}
  {\bfseries B557} (2003) 45}
  [\href{https://arxiv.org/abs/hep-ph/0211181}{{\ttfamily hep-ph/0211181}}].

\bibitem{Abe:2002rb}
{\scshape Belle} collaboration, K.~Abe et~al., \emph{{Observation of double c
  anti-c production in e+ e- annihilation at s**(1/2) approximately 10.6-GeV}},
  \href{https://doi.org/10.1103/PhysRevLett.89.142001}{\emph{Phys. Rev. Lett.}
  {\bfseries 89} (2002) 142001}
  [\href{https://arxiv.org/abs/hep-ex/0205104}{{\ttfamily hep-ex/0205104}}].

\bibitem{Aubert:2005tj}
{\scshape BaBar} collaboration, B.~Aubert et~al., \emph{{Measurement of double
  charmonium production in $e^+e^-$ annihilations at $\sqrt{s}=10.6$ GeV}},
  \href{https://doi.org/10.1103/PhysRevD.72.031101}{\emph{Phys. Rev.}
  {\bfseries D72} (2005) 031101}
  [\href{https://arxiv.org/abs/hep-ex/0506062}{{\ttfamily hep-ex/0506062}}].

\bibitem{Bodwin:2006ke}
G.~T. Bodwin, D.~Kang, T.~Kim, J.~Lee and C.~Yu, \emph{{Relativistic
  Corrections to e+ e- ---> J/psi + eta(c) in a Potential Model}},
  \href{https://doi.org/10.1063/1.2714404}{\emph{AIP Conf. Proc.} {\bfseries
  892} (2007) 315} [\href{https://arxiv.org/abs/hep-ph/0611002}{{\ttfamily
  hep-ph/0611002}}].

\bibitem{Bodwin:2007ga}
G.~T. Bodwin, J.~Lee and C.~Yu, \emph{{Resummation of Relativistic Corrections
  to e+ e- ---> J/psi + eta(c)}},
  \href{https://doi.org/10.1103/PhysRevD.77.094018}{\emph{Phys. Rev.}
  {\bfseries D77} (2008) 094018}
  [\href{https://arxiv.org/abs/0710.0995}{{\ttfamily 0710.0995}}].

\bibitem{Patrignani:2016xqp}
{\scshape Particle Data Group} collaboration, C.~Patrignani et~al.,
  \emph{{Review of Particle Physics}},
  \href{https://doi.org/10.1088/1674-1137/40/10/100001}{\emph{Chin. Phys.}
  {\bfseries C40} (2016) 100001}.

\bibitem{Chen:2018obq}
G.~Chen, C.-H. Chang and X.-G. Wu, \emph{{$B_c (B_c^*)$ meson production via
  the proton-nucleus and the nucleus-nucleus collision modes at the colliders
  RHIC and LHC}}, \href{https://doi.org/10.1103/PhysRevD.97.114022}{\emph{Phys.
  Rev.} {\bfseries D97} (2018) 114022}
  [\href{https://arxiv.org/abs/1803.11447}{{\ttfamily 1803.11447}}].

\bibitem{Qiao:2012hp}
C.-F. Qiao, P.~Sun, D.~Yang and R.-L. Zhu, \emph{{B$_c$ exclusive decays to
  charmonium and a light meson at next-to-leading order accuracy}},
  \href{https://doi.org/10.1103/PhysRevD.89.034008}{\emph{Phys. Rev.}
  {\bfseries D89} (2014) 034008}
  [\href{https://arxiv.org/abs/1209.5859}{{\ttfamily 1209.5859}}].

\bibitem{Shen:2014msa}
J.-M. Shen, X.-G. Wu, H.-H. Ma and S.-Q. Wang, \emph{{QCD corrections to the
  $B_c$ to charmonia semileptonic decays}},
  \href{https://doi.org/10.1103/PhysRevD.90.034025}{\emph{Phys. Rev.}
  {\bfseries D90} (2014) 034025}
  [\href{https://arxiv.org/abs/1407.7309}{{\ttfamily 1407.7309}}].

\bibitem{Zhu:2017lwi}
R.~Zhu, \emph{{Relativistic corrections to the form factors of $B_c$ into
  $P$-wave orbitally excited charmonium}},
  \href{https://doi.org/10.1016/j.nuclphysb.2018.04.018}{\emph{Nucl. Phys.}
  {\bfseries B931} (2018) 359}
  [\href{https://arxiv.org/abs/1710.07011}{{\ttfamily 1710.07011}}].

\bibitem{Du:1988ws}
D.-s. Du and Z.~Wang, \emph{{Predictions of the Standard Model for $B_c^\pm$
  Weak Decays}}, \href{https://doi.org/10.1103/PhysRevD.39.1342}{\emph{Phys.
  Rev.} {\bfseries D39} (1989) 1342}.

\bibitem{Sun:2008ew}
J.-F. Sun, D.-S. Du and Y.-L. Yang, \emph{{Study of $B_c \to J/\psi \pi$,
  $\eta_c \pi$ decays with perturbative QCD approach}},
  \href{https://doi.org/10.1140/epjc/s10052-009-0872-y}{\emph{Eur. Phys. J.}
  {\bfseries C60} (2009) 107}
  [\href{https://arxiv.org/abs/0808.3619}{{\ttfamily 0808.3619}}].

\bibitem{Wen-Fei:2013uea}
W.-F. Wang, Y.-Y. Fan and Z.-J. Xiao, \emph{{Semileptonic decays
  $B_c\to(\eta_c,J/\Psi)l\nu$ in the perturbative QCD approach}},
  \href{https://doi.org/10.1088/1674-1137/37/9/093102}{\emph{Chin. Phys.}
  {\bfseries C37} (2013) 093102}
  [\href{https://arxiv.org/abs/1212.5903}{{\ttfamily 1212.5903}}].

\bibitem{Rui:2014tpa}
Z.~Rui and Z.-T. Zou, \emph{{S-wave ground state charmonium decays of $B_c$
  mesons in the perturbative QCD approach}},
  \href{https://doi.org/10.1103/PhysRevD.90.114030}{\emph{Phys. Rev.}
  {\bfseries D90} (2014) 114030}
  [\href{https://arxiv.org/abs/1407.5550}{{\ttfamily 1407.5550}}].

\bibitem{Nobes:2000pm}
M.~A. Nobes and R.~M. Woloshyn, \emph{{Decays of the $B_c$ meson in a
  relativistic quark meson model}},
  \href{https://doi.org/10.1088/0954-3899/26/7/308}{\emph{J. Phys.} {\bfseries
  G26} (2000) 1079} [\href{https://arxiv.org/abs/hep-ph/0005056}{{\ttfamily
  hep-ph/0005056}}].

\bibitem{Ebert:2003cn}
D.~Ebert, R.~N. Faustov and V.~O. Galkin, \emph{{Weak decays of the $B_c$ meson
  to charmonium and $D$ mesons in the relativistic quark model}},
  \href{https://doi.org/10.1103/PhysRevD.68.094020}{\emph{Phys. Rev.}
  {\bfseries D68} (2003) 094020}
  [\href{https://arxiv.org/abs/hep-ph/0306306}{{\ttfamily hep-ph/0306306}}].

\bibitem{Ivanov:2005fd}
M.~A. Ivanov, J.~G. Korner and P.~Santorelli, \emph{{Semileptonic decays of
  $B_c$ mesons into charmonium states in a relativistic quark model}},
  \href{https://doi.org/10.1103/PhysRevD.75.019901,
  10.1103/PhysRevD.71.094006}{\emph{Phys. Rev.} {\bfseries D71} (2005) 094006}
  [\href{https://arxiv.org/abs/hep-ph/0501051}{{\ttfamily hep-ph/0501051}}].

\bibitem{Ebert:2010zu}
D.~Ebert, R.~N. Faustov and V.~O. Galkin, \emph{{Semileptonic and Nonleptonic
  Decays of $B_c$ Mesons to Orbitally Excited Heavy Mesons in the Relativistic
  Quark Model}}, \href{https://doi.org/10.1103/PhysRevD.82.034019}{\emph{Phys.
  Rev.} {\bfseries D82} (2010) 034019}
  [\href{https://arxiv.org/abs/1007.1369}{{\ttfamily 1007.1369}}].

\bibitem{Huang:2007kb}
T.~Huang and F.~Zuo, \emph{{Semileptonic $B_c$ decays and charmonium
  distribution amplitude}},
  \href{https://doi.org/10.1140/epjc/s10052-007-0333-4}{\emph{Eur. Phys. J.}
  {\bfseries C51} (2007) 833}
  [\href{https://arxiv.org/abs/hep-ph/0702147}{{\ttfamily hep-ph/0702147}}].

\bibitem{Hernandez:2006gt}
E.~Hernandez, J.~Nieves and J.~M. Verde-Velasco, \emph{{Study of exclusive
  semileptonic and non-leptonic decays of $B_c$ - in a nonrelativistic quark
  model}}, \href{https://doi.org/10.1103/PhysRevD.74.074008}{\emph{Phys. Rev.}
  {\bfseries D74} (2006) 074008}
  [\href{https://arxiv.org/abs/hep-ph/0607150}{{\ttfamily hep-ph/0607150}}].

\bibitem{Colangelo:1992cx}
P.~Colangelo, G.~Nardulli and N.~Paver, \emph{{QCD sum rules calculation of
  B(c) decays}}, \href{https://doi.org/10.1007/BF01555737}{\emph{Z. Phys.}
  {\bfseries C57} (1993) 43}.

\bibitem{Kiselev:1993ea}
V.~V. Kiselev and A.~V. Tkabladze, \emph{{Semileptonic B(c) decays from QCD sum
  rules}}, \href{https://doi.org/10.1103/PhysRevD.48.5208}{\emph{Phys. Rev.}
  {\bfseries D48} (1993) 5208}.

\bibitem{Kiselev:1999sc}
V.~V. Kiselev, A.~K. Likhoded and A.~I. Onishchenko, \emph{{Semileptonic $B_c$
  meson decays in sum rules of QCD and NRQCD}},
  \href{https://doi.org/10.1016/S0550-3213(99)00505-2}{\emph{Nucl. Phys.}
  {\bfseries B569} (2000) 473}
  [\href{https://arxiv.org/abs/hep-ph/9905359}{{\ttfamily hep-ph/9905359}}].

\bibitem{Azizi:2009ny}
K.~Azizi, H.~Sundu and M.~Bayar, \emph{{Semileptonic B(c) to P-Wave Charmonia
  (X(c0), X(c1), h(c)) Transitions within QCD Sum Rules}},
  \href{https://doi.org/10.1103/PhysRevD.79.116001}{\emph{Phys. Rev.}
  {\bfseries D79} (2009) 116001}
  [\href{https://arxiv.org/abs/0902.1467}{{\ttfamily 0902.1467}}].

\bibitem{Salpeter:1951sz}
E.~E. Salpeter and H.~A. Bethe, \emph{{A Relativistic equation for bound state
  problems}}, \href{https://doi.org/10.1103/PhysRev.84.1232}{\emph{Phys. Rev.}
  {\bfseries 84} (1951) 1232}.

\bibitem{Salpeter:1952ib}
E.~E. Salpeter, \emph{{Mass corrections to the fine structure of hydrogen -
  like atoms}}, \href{https://doi.org/10.1103/PhysRev.87.328}{\emph{Phys. Rev.}
  {\bfseries 87} (1952) 328}.

\bibitem{Wang:2007nb}
G.-L. Wang, \emph{{Annihilation rate of heavy 0++ P-wave quarkonium in
  relativistic Salpeter method}},
  \href{https://doi.org/10.1016/j.physletb.2007.08.017}{\emph{Phys. Lett.}
  {\bfseries B653} (2007) 206}
  [\href{https://arxiv.org/abs/0708.3516}{{\ttfamily 0708.3516}}].

\bibitem{Mandelstam:1955sd}
S.~Mandelstam, \emph{{Dynamical variables in the Bethe-Salpeter formalism}},
  \href{https://doi.org/10.1098/rspa.1955.0261}{\emph{Proc. Roy. Soc. Lond.}
  {\bfseries A233} (1955) 248}.

\bibitem{Cvetic:2004qg}
G.~Cvetic, C.~S. Kim, G.-L. Wang and W.~Namgung, \emph{{Decay constants of
  heavy meson of 0- state in relativistic Salpeter method}},
  \href{https://doi.org/10.1016/j.physletb.2004.06.092}{\emph{Phys. Lett.}
  {\bfseries B596} (2004) 84}
  [\href{https://arxiv.org/abs/hep-ph/0405112}{{\ttfamily hep-ph/0405112}}].

\bibitem{Chang:2004im}
C.-H. Chang, J.-K. Chen, X.-Q. Li and G.-L. Wang, \emph{{Instantaneous
  Bethe-Salpeter equation and its exact solution}},
  \href{https://doi.org/10.1088/0253-6102/43/1/023}{\emph{Commun. Theor. Phys.}
  {\bfseries 43} (2005) 113}
  [\href{https://arxiv.org/abs/hep-ph/0406050}{{\ttfamily hep-ph/0406050}}].

\bibitem{Wang:2007av}
G.-L. Wang, \emph{{Decay constants of P-wave mesons}},
  \href{https://doi.org/10.1016/j.physletb.2007.05.001}{\emph{Phys. Lett.}
  {\bfseries B650} (2007) 15}
  [\href{https://arxiv.org/abs/0705.2621}{{\ttfamily 0705.2621}}].

\bibitem{Chang:2006tc}
C.-H. Chang, J.-K. Chen and G.-L. Wang, \emph{{Instantaneous formulation for
  transitions between two instantaneous bound states and its gauge
  invariance}}, \href{https://doi.org/10.1088/0253-6102/46/3/017}{\emph{Commun.
  Theor. Phys.} {\bfseries 46} (2006) 467}.

\bibitem{Wang:2016enc}
T.~Wang, Z.-H. Wang, Y.~Jiang, L.~Jiang and G.-L. Wang, \emph{{Strong decays of
  $D_{3}^{*}(2760)$ , $D_{s3}^{*}(2860)$ , $B_{3}^{*}$ , and $B_{s3}^{*}$}},
  \href{https://doi.org/10.1140/epjc/s10052-017-4611-5}{\emph{Eur. Phys. J.}
  {\bfseries C77} (2017) 38}
  [\href{https://arxiv.org/abs/1610.04991}{{\ttfamily 1610.04991}}].

\bibitem{Chang:1992pt}
C.-H. Chang and Y.-Q. Chen, \emph{{The Decays of B(c) meson}},
  \href{https://doi.org/10.1103/PhysRevD.49.3399}{\emph{Phys. Rev.} {\bfseries
  D49} (1994) 3399}.

\bibitem{Kim:2003ny}
C.~S. Kim and G.-L. Wang, \emph{{Average kinetic energy of heavy quark
  (mu(pi)**2) inside heavy meson of 0- state by Bethe-Salpeter method}},
  \href{https://doi.org/10.1016/j.physletb.2004.01.058,
  10.1016/j.physletb.2006.01.053}{\emph{Phys. Lett.} {\bfseries B584} (2004)
  285} [\href{https://arxiv.org/abs/hep-ph/0309162}{{\ttfamily
  hep-ph/0309162}}].

\bibitem{Bodwin:1994jh}
G.~T. Bodwin, E.~Braaten and G.~P. Lepage, \emph{{Rigorous QCD analysis of
  inclusive annihilation and production of heavy quarkonium}},
  \href{https://doi.org/10.1103/PhysRevD.55.5853,
  10.1103/PhysRevD.51.1125}{\emph{Phys. Rev.} {\bfseries D51} (1995) 1125}
  [\href{https://arxiv.org/abs/hep-ph/9407339}{{\ttfamily hep-ph/9407339}}].

\bibitem{Chang:2014jca}
C.~Chang, H.-F. Fu, G.-L. Wang and J.-M. Zhang, \emph{{Some of semileptonic and
  nonleptonic decays of $B_c$ meson in a Bethe-Salpeter relativistic quark
  model}}, \href{https://doi.org/10.1007/s11433-015-5671-x}{\emph{Sci. China
  Phys. Mech. Astron.} {\bfseries 58} (2015) 071001}
  [\href{https://arxiv.org/abs/1411.3428}{{\ttfamily 1411.3428}}].

\bibitem{Wang:2005qx}
G.-L. Wang, \emph{{Decay constants of heavy vector mesons in relativistic
  Bethe-Salpeter method}},
  \href{https://doi.org/10.1016/j.physletb.2005.12.005}{\emph{Phys. Lett.}
  {\bfseries B633} (2006) 492}
  [\href{https://arxiv.org/abs/math-ph/0512009}{{\ttfamily math-ph/0512009}}].

\end{thebibliography}\endgroup


\end{document}